\documentclass{article}
\usepackage{graphicx,epsfig}
\usepackage{graphicx,subfigure}
\usepackage{arxiv}
\usepackage{amsmath}
\usepackage[utf8]{inputenc} 
\usepackage[T1]{fontenc}    
\usepackage{hyperref}       
\usepackage{url}            
\usepackage{booktabs}       
\usepackage{amsfonts}       
\usepackage{nicefrac}       
\usepackage{microtype}      
\usepackage{lipsum}

\title{Effect of floating flexible plate on the dynamics of flow over a mild-slope}

\author{
  Siluvai Antony Selvan\\
  Department of Mathematics\\
  SRM Institute of Science and Technology\\
   Kattankulathur-603203, India\\
  \texttt{antony.selvan22@gmail.com} \\
   \And
 Sukhendu Ghosh*\\
  Department of Mathematics\\
  Indian Institute of Technology\\
  Santa Narimana, Levand \\
  \texttt{sukhendu.math@gmail.com} \\
     \And
  Harekrushna Behera*\\
  Department of Mathematics\\
  SRM Institute of Science and Technology\\
  Kattankulathur-603203, India\\
  \texttt{sukhendu.math@gmail.com} \\
     \And
  Michael H Meylan\\
  School of Mathematical and Physical Sciences\\
 University of Newcastle\\
 NSW 2308, Australia\\
  \texttt{mike.meylan@newcastle.edu.au} \\
}

\begin{document}
\maketitle

\begin{abstract}
Hydrodynamic instability of a gravity-driven flow down an inclined plane is investigated in the presence of a floating elastic plate which rests on the top surface of the flow. Linear instability of the system with respect to infinitesimal disturbances is captured using normal-mode analysis. The critical conditions for instability are obtained analytically utilizing the long-wave approximation and small film aspect ratio. Further, the bifurcation of the nonlinear evolution equation is analyzed using weakly nonlinear stability analysis. The Orr-Sommerfeld system for the perturbed flow is derived, and it is solved numerically using the spectral collocation method.     The behaviour of the marginal stability curves and temporal growth of the unstable waves are portrayed for a range of dimensionless flow parameters. Moreover, the pressure acting on the surface and shearing stress are calculated and analysed for various structural and flow configurations. The study reveals that critical structural parameters such as rigidity and mass per unit length play a crucial role in suppressing and facilitating the unstable surface waves of the flow.  Numerical observations imply that the floating elastic plate helps to stabilize the surface flow and to damp the high amplitude waves. 
\end{abstract}


\section{Introduction}
The study of hydrodynamic instability of fluid flow in bounded and semi-bounded domains is useful in applications ranging from coating processes to pulmonary fluid mechanics and biomedical engineering (\cite{grad1967unified,anshus1967effect} and \cite{kistler1997liquid}). The onset/offset characteristics of the instability can both be employed usefully depending on the application. A flow bounded below by a rigid/slippery, flat/wavy bottom with a free surface is a kind of semi-bounded flow. There are several studies in the literature concerning the instability of free surface flow down an inclined plane (\cite{benjamin1957wave}, \cite{yih1963stability}, \cite{chin1986gravity}). The gravity-driven semi-bounded flow becomes unstable at low to moderate Reynolds number due to the development of unstable surface mode (\cite{chin1986gravity}). Further, the presence of a slippery bottom enhances the instability of fluid flow, which destabilizes the fluid by reducing the critical Reynolds number (\cite{samanta2011falling}). 

Numerous active and passive techniques have been employed to suppress or incite the instability of such flow depending on the diverse applications where such flows arise. One of the methods to control the surface instability is by adding an insoluble surfactant to the fluid surface.  The insoluble surfactant does not react with fluid, and fluid properties remain unaltered, but the surfactant significantly alters the surface properties of the flow. In addition, there will be two normal modes due to the presence of insoluble surfactants on the surface of the fluid. The first is due to the free surface flow known as a Yih mode (\cite{yih1963stability}), and the other is due to the insoluble surfactant known as a Marangoni mode (\cite{pozrikidis2003effect}).  Further, the decay rate of the Yih mode is found to be higher than that of Marangoni mode (\cite{blyth2004effect}). There is a parallel interest in analyzing the effect of shear stress on the film flow down an inclined plane in the presence of an insoluble surfactant. \cite{wei2005effect} showed that the Marangoni effect due to the shear stress destabilizes the fluid flow for smaller Reynolds numbers. Later, \cite{frenkel2005effect} extended the work of \cite{wei2005effect} and studied the consequence of inertia on the instability of the shear-induced flow having insoluble surfactant. It is observed that the application of shear stress to the fluid flow destabilizes the stable Marangoni-mode, which explains the behaviour of surface-laden lining liquid flow in the airways occlusion process. The effect of insoluble surfactant on the film flow down the porous inclined plane is investigated by \cite{anjalaiah2013thin}, and it is observed that the film flow over a thicker porous layer is more unstable than for a thinner layer. For a two-layer fluid, the effect of the mono-layer insoluble surfactants on the film flow in both the layers was analyzed by \cite{samanta2014effect} using the long-wave approximation. Recently, \cite{bhat2018linear} analyzed the effect of insoluble surfactant on the film down a slippery bottom using normal mode analysis.

An analogous way of controlling instability in a free-surface flow is the addition of a surface-active product, which alters the surface-elasticity due to the mass transfer at the interface between product/surfactant and the fluid. There are also many studies in literature regarding the effect of soluble surfactant on the fluid flow (\cite{benjamin1964effects,whitaker1964effect, whitaker1966stability} and \cite{anshus1967effect}). In addition to surface-elasticity, properties like  diffusion coefficient and concentration of surfactant also play an influential role in enhancing the stability of the fluid flow (\cite{lin1970stabilizing}). The existence of unstable Marangoni mode due to the soluble surfactant on the vertical film flow was first observed by  \cite{ji1994instabilities}. 
Recently, the role of soluble surfactant on the stability of two-layer fluid in the channel was investigated by \cite{kalogirou2019role}.

Apart from the linear theory describing the dynamics of the fluid flow down an inclined plane,  several attempts have made to analyze the problem using nonlinear theory. The first attempt was made by \cite{benney1966long}, who developed an asymptotic solution to the nonlinear problem of a thin film flow down the inclined plane. The dynamics of film flow is described by the nonlinear evolution equation known as the Benny equation (BE). But, this BE solution has a grave drawback due to the incorrect scaling of the surface tension (\cite{lin1969finite}). This incorrect scaling leads to the non-existence of finite-amplitude traveling waves around the neighbourhood of the critical region. There have been extensive studies made on the BE. \cite{lin1974finite} analyzed the bifurcation of first and second-order \textsc{be} and found that there is a predominant supercritical bifurcation. However, the solution blow-up for the BE corresponding to two-dimensional film flow is investigated extensively by \cite{rosenau1992bounded}. In the case of first-order BE, \cite{oron2004subcritical} disagrees with the result of \cite{lin1974finite}, and they found that there is predominant supercritical bifurcation till certain Reynolds number. Further, beyond the certain limiting values of the Reynolds number, the predominant bifurcation is sub-critical. \cite{sadiq2008thin} extended the idea of BE to the film flow down the inclined plane and observed that there exist both a supercritical and sub-critical region for a certain range of wavenumbers. 

In the past decades, many studies have been proposed on characteristics of flexural gravity wave motion employing the  Euler-Bernoulli plate equation (\cite{norris1995scattering}, \cite{vemula1997flexural}, \cite{williams2004oblique} and \cite{das2018flexural}). Consequently, several procedures have been used to investigate wave action on floating platforms of complex geometry (\cite{meylan1997forced},  \cite{wang2004higher}, \cite{porter2018trapping}).  In these investigations, the fluid flow was assumed to be incompressible and inviscid, which is not always realistic.

In the present work, the authors study the influences of floating flexible plate on the hydrodynamic instability of gravity-driven flow down an inclined surface. Inclusion of a floating elastic plate on the free surface is a passive way of controlling the surface instability of the flow system through changing the surface-elasticity. A viscous incompressible Newtonian fluid flow is considered, and the flexible structure is modelled using the Euler-Timenshenko plate theory (\cite{magrab2012vibrations}). Further, the plate length is multiple of the disturbance produced by the wavelength. The method of normal mode analysis is implemented to derive the linear stability equations. The resulting Orr-Sommerfeld system is solved numerically using the spectral collocation method. To analytically obtain the critical limit of the flow parameters which govern the instability, the long-wave approximation is used together with the small film aspect ratio. In addition to this, the primary bifurcation of the nonlinear evolution equation is investigated using weakly nonlinear stability analysis. The manuscript is organized as follows: Section \ref{MF} contains the Mathematical formulation for linear stability analysis, the long-wave approximation of the Orr-Sommerfeld system is presented in Section \ref{LWA},  the problem analysis through small film aspect ratio is given in Section \ref{SAR}, the weakly nonlinear stability analysis is studied in Section \ref{WN} numerical results are discussed in Section \ref{NRD} and Section \ref{con} contains the concluding remarks.

\section{Mathematical formulation}\label{MF}
\begin{figure}
	\centering
	{\includegraphics[width=10cm]{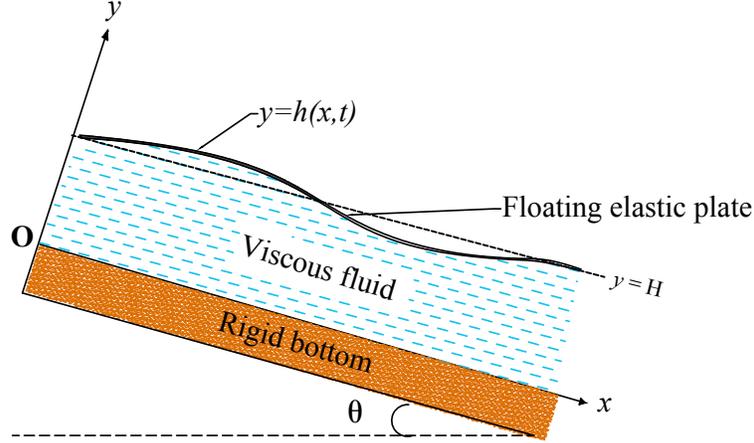}}
	\caption{Schematic diagram of a falling Newtonian fluid with a floating flexible plate at the top surface.} 
	\label{f1}
\end{figure}

A viscous, incompressible and irrotational Newtonian fluid flow over a rigid plate having an angle $\theta$ with horizontal is considered. There is a floating flexible plate of infinite length along the mean free surface at $y=H$ as shown in Fig. \ref{f1}. The problem is formulated under the two-dimensional Cartesian coordinate system with positive $x$-axis along the direction of the fluid flow and positive $y$-axis pointed perpendicular to inclined plate. The deflection of floating elastic plate is denoted by $y=h\,\big(x,t\big)$. The governing Navier-Stokes equations for the fluid flow is given by

\begin{subequations}
	\begin{eqnarray}
	\frac{\partial u}{\partial x}+\frac{\partial v}{\partial y}&=&0,\label{e1}\\
	\rho\,\bigg(\frac{\partial u}{\partial t}+u\,\frac{\partial u}{\partial x}+v\,\frac{\partial u}{\partial y}\bigg)&=&-\frac{\partial p}{\partial x}+\mu\, \bigg(\frac{\partial^2 u}{\partial x^2}+\frac{\partial^2 u}{\partial y^2}\bigg)+\rho\,g\sin\theta,\label{e2}
	\\
	\rho\,\bigg(\frac{\partial v}{\partial t}+u\,\frac{\partial v}{\partial x}+v\,\frac{\partial v}{\partial y}\bigg)&=&-\frac{\partial p}{\partial y}+\mu\, \bigg(\frac{\partial^2 v}{\partial x^2}+\frac{\partial^2 v}{\partial y^2}\bigg)-\rho\,g\cos\theta,\label{e3}
	\end{eqnarray} 
	\label{eq1}
\end{subequations}
\noindent where $\rho$, $p$ and $\mu$ are the density, pressure and viscosity of the fluid, respectively. The components of velocity along the $x$ and $y$ directions are given by $u$ and $v$, respectively, and $g$ is the gravitational acceleration. At the flexible plate covered surface, the kinematic boundary condition is given by
\begin{eqnarray}
v&=&\frac{\partial h}{\partial t}+u\,\frac{\partial h}{\partial x}
\quad \mbox{at} \quad y=h\,\big(x,t\big).\label{e4}
\end{eqnarray}
\noindent The tangential stress balance is same as that  of free-surface due to the uniform compressive force acting on the floating elastic plate. In the case of normal stress balance, the boundary condition is derived by coupling the Euler-Bernoulli's plate equation (\cite{magrab2012vibrations}) with the pressure due to the both fluid domain ($p$) and atmosphere ($p_\infty$). Further, the balance of tangential and normal stress at that surface is taken as, 
\begin{eqnarray}
\mu\,\bigg[\bigg(\frac{\partial u}{\partial y}+\frac{\partial v}{\partial x}\bigg)\,\bigg(1-\bigg\{\frac{\partial h}{\partial x}\bigg\}^2\bigg)-4\,\frac{\partial u}{\partial x}\frac{\partial h}{\partial x}\bigg]= 0~\mbox{at}~y=h\big(x,t\big), \label{e5}
\\
p-p_\infty=\frac{2\,\mu}{1+\displaystyle\big(\partial h/\partial x\big)^2}\,\left[\bigg(\frac{\partial u}{\partial x}\bigg)\bigg(\frac{\partial h}{\partial x}\bigg)^2-\bigg(\frac{\partial u}{\partial y}+\frac{\partial v}{\partial x}\bigg)\frac{\partial h}{\partial x}+\frac{\partial v}{\partial y}\right]+EI\,\frac{\partial^2}{\partial x^2}\nonumber\\
~\left[\frac{\displaystyle\partial^2h/\partial x^2}{\big(1+\{\partial h/\partial x\}^{2}\big)^{3/2}}\right]+N\,\frac{\partial^2 h/\partial x^2}{\big(1+\{\partial h/\partial x\}^{2}\big)^{3/2}}-m_p\,\frac{\partial^2 h}{\partial t^2}~~\text{at}~~y=h\,\big(x,t\big),\label{e6}
\end{eqnarray}
\noindent where $EI$ and $N$ are the structural rigidity and the compressive force acing on the flexible plate, respectively. The variables $m_p$ and $p_\infty$ are the mass per unit length of plate and the atmospheric pressure, respectively. Due to the uniform compressive force acting along the $x$-axis, the right hand side of Eq. \eqref{e5} will be zero ($N_x=0$). Letting, $EI \to 0$, $N \to -\sigma$ and $m_p \to 0$, the equation \eqref{e6} becomes normal stress balance between a usual free surface flow and atmosphere, where $\sigma$ be the surface tension of the fluid flow. At the bottom plane $y=0$, it is assumed that there is no velocity slip and the corresponding boundary conditions are given as
\begin{equation}
u=0~~\mbox{and}~~v=0 \quad \mbox{at} \quad y=0.\label{e7}
\end{equation}
\noindent The governing equations and boundary conditions are made non-dimensional and the dimensionless variables are,
\begin{align*}
&\bar{x}=x/H, \quad \bar{y}=y/H, \quad \bar{u}=u/V, \quad \bar{v}=v/V, \quad \bar{t}=tV/H, \quad \bar{p}=p/\rho V^2, \quad \bar{h}=h/H,
\\
&\alpha=\frac{2EI}{\sin\theta\,\rho gH^4}, \quad \beta=\frac{2N}{\sin\theta\,\rho gH^2}, \quad \gamma=\frac{m_p}{\rho H},\quad G=\frac{gH\sin\theta}{V^2}.
\end{align*}
The maximum velocity of the fluid film in the presence of elastic plate is considered as the characteristic velocity. It is given as $\displaystyle V=\frac{g\,H^2\sin\theta}{2\nu}$, with $H=({EI/\rho g\sin\theta})^{1/4}$ be the characteristic length scale as given in \cite{bonnefoy2009nonlinear}. Further, $\alpha$, $\beta$, $\gamma$ and $G$ denote the non-dimensional form of the structural rigidity, compressive force, mass per unit length and gravitational acceleration, respectively. The dimensionless equations and corresponding boundary conditions are given as (after suppressing the over-bars),
\begin{subequations}
	\begin{eqnarray}
	\frac{\partial u}{\partial x}+\frac{\partial v}{\partial y}&=&0,\label{e8}
	\\
	\frac{\partial u}{\partial t}+u\,\frac{\partial u}{\partial x}+v\,\frac{\partial u}{\partial y}&=&-\frac{\partial p}{\partial x}+\frac{1}{Re} \left(\frac{\partial^2 u}{\partial x^2}+\frac{\partial^2 u}{\partial y^2}\right)+G, \label{e9}
	\\
	\frac{\partial v}{\partial t}+u\,\frac{\partial v}{\partial x}+v\,\frac{\partial v}{\partial y}&=&-\frac{\partial p}{\partial y}+\frac{1}{Re} \left(\frac{\partial^2 v}{\partial x^2}+\frac{\partial^2 v}{\partial y^2}\right)-G\cot\theta,\label{e10}\\
	v&=&\frac{\partial h}{\partial t}+u\,\frac{\partial h}{\partial x} \quad \mbox{at} \quad y=h(x,t), \label{e11}
	\end{eqnarray}\vspace{-0.5cm}
	\begin{eqnarray}
	\bigg(\frac{\partial u}{\partial y}+\frac{\partial v}{\partial x}\bigg)\,\bigg(1-\bigg\{\frac{\partial h}{\partial x}\bigg\}^2\bigg)-4\,\frac{\partial u}{\partial x}\frac{\partial h}{\partial x}= 0 \quad \mbox{at} \quad y=h\,\big(x,t\big), \label{e12}
	\\
	Re\,p=\frac{2}{1+\displaystyle\big(\partial h/\partial x\big)^2}\,\left[\bigg(\frac{\partial u}{\partial x}\bigg)\bigg(\frac{\partial h}{\partial x}\bigg)^2-\bigg(\frac{\partial u}{\partial y}+\frac{\partial v}{\partial x}\bigg)\frac{\partial h}{\partial x}+\frac{\partial v}{\partial y}\right]+\alpha\,\frac{\partial^2}{\partial x^2}\nonumber\\
	\left[\frac{\displaystyle\partial^2h/\partial x^2}{\big(1+\{\partial h/\partial x\}^{2}\big)^{3/2}}\right]+\beta\,\frac{\partial^2 h/\partial x^2}{\big(1+\{\partial h/\partial x\}^{2}\big)^{3/2}}-\gamma\,Re\,\frac{\partial^2 h}{\partial t^2}
	\quad \text{at} \,\, y=h\,\big(x,t\big),\label{e13}
	\end{eqnarray}\vspace{-0.4cm}
	\begin{eqnarray}
	u=0 \quad \text{and} \quad v=0 \quad \mbox{at} \quad y=0, \label{e14}
	\end{eqnarray}
	\label{eq6}
\end{subequations}
where $Re$ denotes the Reynolds number, which is defined as $Re=VH/\nu$. Note, the effect of the plate curvature is accounted in the stress balance equation \eqref{e13}. Under the non-dimensionalization process $GRe=2$, which is used in the calculation of base flow in the following section.
\subsection{Basic state and linear stability equations}
We first determine the steady fully developed linear flow. 
The mean velocity of the fully developed flow is
$\mathbf{U}=(U_B(y),0)$ and $P_B$ is the corresponding pressure, which satisfy Eqs.~\eqref{e8} to \eqref{e10} along with the boundary conditions.
By solving these dimensionless equations and boundary conditions for a unidirectional steady flow, the following base flow solution is obtained:
\begin{subequations}
	\begin{eqnarray}
	U_B(y)&=&\frac{GRe}{2}(-y^2+2\,y)=(-y^2+2\,y), \label{e15}
	\\
	P_B(y)&=&G\,\cot\theta(1-y), \label{e16}    
	\end{eqnarray}
	\label{eq7}
\end{subequations}
\noindent since $GRe=2$. The presence of flexible plate in free surface does not alter the behavior of base flow. To check the linear instability of the base flow with respect to small disturbances, the flow variables are re-framed as the sum of contribution from the base flow and from the perturbed flow, which are given by, 
\begin{subequations}
	\begin{eqnarray}
	u(x,y,t)&=&U_B\,(y)+\tilde{u}\,(x,y,t),\quad \label{e17}\\ 
	v(x,y,t)&=&\tilde{v}\,(x,y,t), \quad  \label{e18} \\
	p(x,y,t)&=&P_B\,(y)+\tilde{p}\,(x,y,t), \quad \label{e19}\\
	h(x,t)&=&1+\tilde{h}\,(x,t), \label{e20}
	\end{eqnarray}
	\label{eq8}
\end{subequations}
\noindent where $\tilde{u}$, $\tilde{v}$, and $\tilde{p}$ are the perturbed quantities of horizontal velocity, vertical velocity and pressure, respectively. Further, the perturbation about the mean free surface is denoted by $\tilde{h}$. Eqs.~\eqref{e17}--\eqref{e20} are substituted into the dimensionless governing equations \eqref{e8}--\eqref{e10} along with the boundary conditions \eqref{e11}--\eqref{e14}.  The resultant equations are linearized by removing the higher order perturbation terms due to the smallness as compared to that of base flow, which is given by,
\begin{subequations}
	\begin{eqnarray}
	\frac{\partial \tilde{u}}{\partial x}+\frac{\partial \tilde{v}}{\partial y}&=&0,\label{e21}\\
	Re\left[\frac{\partial \tilde{u}}{\partial t}+U_B\,\frac{\partial \tilde{u}}{\partial x}+\tilde{v}\,\frac{d U_B}{d y}\right]&=&-Re\,\frac{\partial \tilde{p}}{\partial x}+\frac{\partial^2 \tilde{u}}{\partial x^2}+\frac{\partial^2 \tilde{u}}{\partial y^2},\label{e22}\\
	Re\left[\frac{\partial \tilde{v}}{\partial t}+U_B\,\frac{\partial \tilde{v}}{\partial x}\right]&=&-Re\,\frac{\partial \tilde{p}}{\partial y}+\frac{\partial^2 \tilde{v}}{\partial x^2}+\frac{\partial^2 \tilde{v}}{\partial y^2}, \label{e23}\\
	\tilde{u}= 0~~&\mbox{and}&~~\tilde{v}= 0~~\mbox{at}~~y=0,\label{e24}\\
	\tilde{v}&=&\frac{\partial \tilde{h}}{\partial t}+\tilde{u}\,\frac{\partial \tilde{h}}{\partial x}~~\text{at}~~y = 1,\label{e25}\\
	\tilde{h}\,\frac{d^2 U_B}{d y^2}&=&-\frac{\partial \tilde{u}}{\partial y}-\frac{\partial \tilde{v}}{\partial x} \quad \text{at} \quad y=1,\label{e26}\\
	Re\tilde{p} + Re\,\tilde{h}\,\frac{d U_B}{d y}
	= 2\,\frac{\partial \tilde{v}}{\partial y}
	+ \alpha\,\frac{\partial^4 \tilde{h}}{\partial x^4}&+&\beta\,\frac{\partial^2 \tilde{h}}{\partial x^2}
	- \gamma\,Re\,\frac{\partial^2 \tilde{h}}{\partial t^2}  ~~~\text{at}~~~ y=1.
	\label{e27}
	\end{eqnarray}\label{eq9}
\end{subequations}
\noindent Substituting Eq. \eqref{e21} and \eqref{e22} in Eq. \eqref{e26} and \eqref{e27} respectively yields,
\begin{subequations}
	\begin{eqnarray}
	\frac{\partial \tilde{h}}{\partial x}\,\frac{d^2 U_B}{d y^2}&=&\frac{\partial^2 \tilde{v}}{\partial y^2}-\frac{\partial^2 \tilde{v}}{\partial x^2} \quad \text{at} \quad y=1, \label{e28}\\
	2\cot\theta\,\frac{\partial^2 \tilde{h}}{\partial x^2} + \alpha\,\frac{\partial^6 \tilde{h}}{\partial x^6} +\beta\,\frac{\partial^4 \tilde{h}}{\partial x^4} &-& \gamma\,Re\,\left(\frac{\partial^3 \tilde{v}}{\partial t \partial x^2}- U_B \frac{\partial^4 \tilde{h}}{\partial t \partial x^3}\right)\,  +  3 \frac{\partial^3 \tilde{v}}{\partial y \partial x^2}+\frac{\partial^3 \tilde{v}}{\partial y^3} \nonumber\\
	\hspace{0cm}-Re\,\frac{\partial^2 \tilde{v}}{\partial y \partial t} -Re U_B\,\frac{\partial^2 \tilde{v}}{\partial x \partial y} &=& 0 ~~~ \text{at} ~~~ y=1. \label{e29}
	\end{eqnarray}\label{eq10}
\end{subequations}
\noindent The above set of equations are the perturbation equations for the considered flow system. Let $\tilde{\psi}$ be the stream function perturbation of the two-dimensional flow.  Thus,  velocity perturbations along $x$ and $y$-axis can be expressed as $\displaystyle \tilde{u}\,=\,\partial \tilde{\psi}/\partial y$ and $\displaystyle \tilde{v}\,=-\,\partial \tilde{\psi}/\partial x$. We are interested in wave-like solution of the disturbed flow, which suggest to consider the standard normal mode technique. The wave-amplitude form of such solution read,
\begin{subequations}
	\begin{eqnarray}
	\tilde{\psi}\,(x,y,t)&=&\phi(y)\,\mathrm{e}^{\mathrm{i}k(x-ct)}\label{e30}
	\\
	\tilde{h}\,(x,t)&=&\eta\,\mathrm{e}^{\mathrm{i}k(x-ct)},\label{e31}
	\end{eqnarray} 
	\label{eq11}
\end{subequations}
\noindent where $k$ and $c$ represent the wavenumber and complex wave speed, respectively. Further, using the normal mode solutions in the perturbation equations, the following Orr-Sommerfeld system is obtained,
\begin{subequations}
	\begin{eqnarray}
	(D_y^{2} - k^{2})^{2} \phi - Re[\mathrm{i} k (U_B - c )(D_y^{2}-k^{2})\phi &-& \mathrm{i} k\, D_{y}^2 U_B\, \phi] = 0,
	\label{e32} \\
	D_y \phi = 0 \quad \text{and} \quad \phi &=& 0 \quad \text{at} \quad y = 0, \label{e33} \\
	\phi +{ \eta} (U_B - c) &=& 0 \quad \text{at} \quad y = 1,
	\label{e34}\\
	D_{y}^2U_B\, {\eta}  + \left(D_y^{2} + k^{2}\right) \phi &=& 0 \quad \text{at} \quad y = 1, \label{e35} \\
	-\mathrm{i}k D_y^{3}\phi +[Re k^{2} \left(c - U_B \right) +3\mathrm{i} k^{3}] D_y\phi&-&\gamma\,Re\,k^{4}c\phi \nonumber
	\\
	-[\left( 2k^{2}\cot\theta -k^{4}\beta+\alpha\,k^6\, \right)  
	-\gamma\,Re\,ck^{4}U_B]\eta &=&0 \quad\text{at} \quad y=1. \label{e36}
	\end{eqnarray}\label{eq12}
\end{subequations}
\noindent Here, $D_y$ represents derivative with respect to $y$. Taking the limit as $\alpha\to 0$, $\gamma \to 0$ and $\beta = -\beta_0\big(Re^2 \sin\theta\big)^{-1/3}$ (where the parameter $\beta_0=\big(\sigma/\rho\big)\big(2/g\nu^4\big)^{1/3}$), the above system of equations becomes the Orr-Sommerfeld system for the fluid flow down an inclined plane in the absence of elastic plate (\cite{chin1986gravity}). For a given $k$, one need to solve the Orr-Sommerfeld system to obtain the eigenvalue $c = c_r + \mathrm{i}c_i$ as a function of the other flow parameters. 
The temporal growth rate of the perturbation waves is given by $\omega_i=k\,c_i$. The system is said to be temporally unstable if $\omega_i >0$.
\section{Long-wave analysis}
\label{LWA}
The long wave instability is analysed  by approximating the Orr-Sommerfeld system (\eqref{e32}--\eqref{e36}), letting the wavenumber $k \to 0$.  Extending  the  analysis  of  \cite{benjamin1957wave} and \cite{yih1963stability}, quantities $\phi$, $c$ and $\eta$ are expressed as a series expansion in powers of $k$,
\begin{subequations}
	\begin{eqnarray}
	\phi&=&\phi_0+k\,\phi_1+O(k^2),\label{e37}\\
	c&=&c_0+k\,c_1+O(k^2),\label{e38}\\
	\eta&=&\eta_0+k\,\eta_1+O(k^2).\label{e39}
	\end{eqnarray}
	\label{eq37_eq39}
\end{subequations}
\noindent Without loss of generality, the mean position of the floating plate is considered as $\eta_0=1$. Substituting the above expansion \eqref{e37}--\eqref{e39} in the Orr-Sommerfeld system of equations \eqref{e32}--\eqref{e36} and ignoring the higher order terms (i.e. $O(k^2)$, $O(k^3)$,.....) yields the following zeroth order approximation, 
\begin{subequations}
	\begin{eqnarray}
	D_y^4 \phi_0&=&0,\label{e40}\\
	D_y\phi_0=0\quad\text{and}\quad\phi_0&=&0\quad\text{at} \quad y=0,\label{e41}\\
	\eta_0-\frac{\phi_0}{(c_0-U_B)}&=&0\quad\text{at}\quad y=1,\label{e42}\\
	D_y^2 \phi_0-2\,\eta_0&=&0\quad\text{at}\quad y=1,\label{e43}\\
	D_y^3 \phi_0&=&0\quad\text{at}\quad y=1,\label{e44}
	\end{eqnarray}
	\label{eq14}
\end{subequations}
\noindent and first order approximation,
\begin{subequations}
	\begin{eqnarray}
	D_y^4 \phi_1+\mathrm{i}Re\big[(c_0-U_B)\,D_y^2 \phi_0
	-2\,\phi_0\big]&=&0,\label{e45}\\
	D_y\phi_1=0\quad\text{and}\quad\phi_1&=&0 \quad\text{at} \quad y=0,\label{e46}\\
	\phi_1+c_1\,\eta_0+(c_0-U_B)\,\eta_1&=&0\quad\text{at}\quad y=1,\label{e47}\\
	D_y^2 \phi_1-2\,\eta_1&=&0\quad\text{at}\quad y=1,\label{e48}\\
	Re\,(c_0-U_B)\,D_y\phi_0-2\,\cot\theta\,\eta_0-\mathrm{i}\,D_y^3\phi_1&=&0\quad\text{at}\quad y=1,\label{e49}
	\end{eqnarray}
	\label{eq15}
\end{subequations}
\noindent 
Note that $y = 0$ and $y = 1$ are respectively correspond the position of the wall and the floating plate after non-dimensionalization.
Now, solving the zeroth order equation \eqref{e40} along with boundary conditions \eqref{e41}--\eqref{e44} gives the solution as follows,
\begin{eqnarray}
\phi_0=y^2\quad \text{and}\quad c_0=2 \label{e50}. 
\end{eqnarray}
On the other hand, the solution to the first order approximation is given as,
\begin{eqnarray}
\phi_1&=&4\mathrm{i}\,Re\,\bigg(\frac{y^5}{5!}-\frac{y^4}{4!}\bigg)+2\mathrm{i}\cot\theta\,\frac{y^3}{3!}+   \bigg(2\eta_1-2\mathrm{i}\cot\theta+\frac{8\mathrm{i}Re}{3!}\bigg)\,\frac{y^2}{2!},\label{e51}\\
c_1&=&\mathrm{i}\,\bigg(-\frac{8}{15}\,Re+\frac{2}{3}\cot\theta\bigg).\label{e52}
\end{eqnarray} 
The expression of $\phi$ and  $c$ in the context of long-wave approximation be written as,
\begin{eqnarray}
c&=&2+\mathrm{i}\,k\,\bigg(-\frac{8}{15}\,Re+\frac{2}{3}\cot\theta\bigg),\label{e53}\\
\phi&=&y^2+k\,\bigg[4\mathrm{i}\,Re\,\bigg(\frac{y^5}{5!}-\frac{y^4}{4!}\bigg)+2\mathrm{i}\cot\theta\,\frac{y^3}{3!}+    
\bigg(2\eta_1-2\mathrm{i}\cot\theta+\frac{8\mathrm{i}Re}{3!}\bigg)\,\frac{y^2}{2!}\bigg].\label{e54}
\end{eqnarray}
\noindent Since the real part of $c$ represent the phase velocity of the long waves (small $k$ limit), which is approximately two, while the surface velocity of the unperturbed flow is $\left.U_B\right|_{y = 1} = 1$, the  long waves travel faster than the surface velocity of the unperturbed flow. Further, the growth rate of the long-wave disturbance is nothing but the value of $Im(c) \,(= Im(c_1))$. It is observed that the value of $c_1$ does not depend on the structural parameters of floating elastic plate (i.e., $\alpha$, $\beta$ and $\gamma$) and is identical to the case without a plate, which ascertains independency of the long waves on structural parameters. The following more sophisticated analysis will capture the influence of the plate parameters on the instability of the considered flow.
\vspace{-0.52cm}

\section{Analysis with small film aspect ratio}
\label{SAR}
Let us now consider the configuration when the film aspect ratio is small enough.  Under this approximation, the fluid film thickness is small compared to the wavelength of the surface elevation. To carry out the analysis, we introduce different length scales for  each axis.
The non-dimensional form of surface elevation $h$, time $t$, pressure $p$, horizontal component of velocity $u$, vertical component of velocity $v$, structural rigidity $EI$, compressive force $N$, and uniform mass per unit length $m_p$ are given as
\begin{align}
&\bar{x}=\frac{x}{l}, \quad \bar{y}=\frac{y}{H}, \quad \bar{h}=\frac{h}{H}, \quad \bar{t}=\frac{tV}{l}, \quad \bar{p}=\frac{p}{\rho V^{2}}, \quad \bar{u}=\frac{u}{V}, \quad \bar{v}=\frac{v\,l}{VH},\nonumber\\
&\alpha=\frac{EI H}{\rho V^2l^4}, \quad \beta=\frac{NH}{\rho V^2l^2},\quad \gamma=\frac{m_pH}{\rho l^2}, \label{e55}
\end{align}
where $l$ and $H$ are the horizontal and vertical reference length scales, which correspond to the wavelength of surface elevation and length of mean free surface. On applying Eq.~\eqref{e55}, the set of dimensionless governing equations \eqref{e1}--\eqref{e3} along with the boundary conditions \eqref{e4}--\eqref{e7} can be re-written as,
\begin{subequations}
	\begin{eqnarray}
	\frac{\partial u}{\partial x}+\frac{\partial v}{\partial y}&=&0,\label{e56}\\
	\epsilon\,Re\left[\frac{\partial u}{\partial t}+u\,\frac{\partial u}{\partial x}+v\,\frac{\partial u}{\partial y}\right]&=&-\epsilon\,Re\frac{\partial p}{\partial x}+2+\epsilon^2\,\frac{\partial^2 u}{\partial x^2}+\frac{\partial^2 u}{\partial y^2},\label{e57}\\
	\epsilon^2\,Re\left[\frac{\partial v}{\partial t}+u\,\frac{\partial v}{\partial x}+v\,\frac{\partial v}{\partial y}\right]&=&-Re\,\frac{\partial p}{\partial y}-2\cot{\theta}+\epsilon\,\left[\epsilon^2\,\frac{\partial^2 v}{\partial x^2}+\frac{\partial^2 v}{\partial y^2}\right],\label{e58}\\
	v&=&\frac{\partial h}{\partial t}+u\,\frac{\partial h}{\partial x} \quad \text{at} \quad y=h(x,t),\label{e59}
	\end{eqnarray}\vspace{0.00cm}\\
	\begin{eqnarray}
	4\,\epsilon^2\,&&\frac{\partial u}{\partial x}\,\frac{\partial h}{\partial x}-\left[1-\epsilon^2\,\frac{\partial^2 h}{\partial x^2}\right]\left[\frac{\partial u}{\partial y}+\epsilon^2\,\frac{\partial v}{\partial x}\right]=0 \quad \text{at} \quad y=h(x,t),\label{e60}
	\\
	p=&&\frac{2\,\epsilon}{Re\,\left[1+\,\epsilon^2\,\displaystyle(\partial h/\partial x)^2\right]}\,\left[\epsilon^2\,\bigg(\frac{\partial u}{\partial x}\bigg)\bigg(\frac{\partial h}{\partial x}\bigg)^2-\bigg(\frac{\partial u}{\partial y}+\epsilon^2\,\frac{\partial v}{\partial x}\bigg)\frac{\partial h}{\partial x}+\frac{\partial v}{\partial y}\right]-\gamma\,\frac{\partial^2 h}{\partial t^2}\nonumber\\
	+&&\alpha\,\frac{\partial^2}{\partial x^2}\left[\frac{\displaystyle\partial^2h/\partial x^2}{(1+\epsilon^2\{\partial h/\partial x\}^{2})^{3/2}}\right]+\beta\,\frac{\partial^2 h/\partial x^2}{(1+\epsilon^2\{\partial h/\partial x\}^{2})^{3/2}}
	\quad \text{at} \,\, y = h(x,t),\label{e61}
	\end{eqnarray}\vspace{-0.5cm}
	\begin{eqnarray}
	u=0~~\mbox{and}~~v=0 \quad \text{at} \quad y=0\label{e62},
	\end{eqnarray}
	\label{eq22}
\end{subequations}
\noindent where $\epsilon=H/l<<1$ is the aspect ratio of the fluid film. 
The dynamical quantities such as $u$, $v$ and $p$ are expanded in the powers of the aspect ratio $\epsilon$, which is given as,
\begin{eqnarray}
u=u_0+\epsilon u_1+O(\epsilon^2), \,\, \,\,
v=v_0+\epsilon v_1+O(\epsilon^2) \,\,\,\,\mbox{and}\,\,\, 
p=p_0+\epsilon p_1+O(\epsilon^2).  \,\,\,\, \label{e63}
\end{eqnarray}
We note that our aim is to find out a general equation for the free surface evolution using the approximate solution of velocity and pressure.     It is noted that the power expansion of these dynamical quantities is restricted to $O(\epsilon)$. The analytic form of $u$, $v$ and $p$ is obtained by solving the above system of equations \eqref{e56}--\eqref{e62} by using the asymptotic expansion as in Eq.~\eqref{e63}. The leading order equations associated with the above system are obtained by substituting Eq.~\eqref{e63}, which are given by,
\begin{subequations}
	\begin{eqnarray}
	\frac{\partial u_0}{\partial x}+\frac{\partial v_0}{\partial y}&=&0,\label{e64}\\
	Re\,\frac{\partial p_0}{\partial y}&=&-2\cot{\theta},\label{e65}\\
	\frac{\partial^2 u_0}{\partial y^2}&=&-2.\label{e66}
	\end{eqnarray}
	\label{eq24}
\end{subequations}
\noindent Further, the above equations satisfy the following boundary conditions,
\begin{subequations}
	\begin{eqnarray}
	v_0&=&\frac{\partial h}{\partial t}+u_0 \frac{\partial h}{\partial x} \quad \text{at} \quad y=h\,\big(x,t\big),\label{e67}\\
	\frac{\partial u_0}{\partial y}&=&0\quad \text{at} \quad y=h\,\big(x,t\big),\label{e68}\\    
	p_0&=&\alpha\,\frac{\partial^4 h}{\partial x^4} 
	+\beta\,\frac{\partial^2 h}{\partial x^2}-\gamma 
	\frac{\partial^2 h}{\partial t^2}\quad \text{at} \quad y=h\,\big(x,t\big),\label{e69}\\
	u_0&=&0,\quad v_0=0 \quad \text{at} \quad y=0.\label{e70}
	\end{eqnarray}
	\label{eq25}
\end{subequations}
\noindent The leading order horizontal and vertical components of the velocity along with the pressure are obtained as,
\begin{subequations}
	\begin{eqnarray}
	u_0&=&-y^2+2hy,\label{e71}\\
	v_0&=&-y^2\frac{\partial h}{\partial x},\label{e72}\\
	p_0&=&\frac{2\cot{\theta}}{Re}(h-y)+\alpha\,\frac{\partial^4 h}{\partial x^4} +\beta\,\frac{\partial^2 h}{\partial x^2}-\gamma    \frac{\partial^2 h}{\partial t^2}.\label{e73}
	\end{eqnarray}
	\label{eq26}
\end{subequations}
\noindent The above solutions give the base state of the velocity and pressure, which are same as those given by Eq.~\eqref{eq7}. Similarly, the first order governing equations are given as,
\begin{subequations}
	\begin{eqnarray}
	\frac{\partial u_1}{\partial x}+\frac{\partial v_1}{\partial y}&=&0,\label{e74}\\
	Re\,\left[\frac{\partial u_0}{\partial t}+u_0\,\frac{\partial u}{\partial x}+v_0\,\frac{\partial u_0}{\partial y}\right]&=&-Re\,\frac{\partial p_0}{\partial x}+\frac{\partial^2 u_1}{\partial y^2},\label{e75}\\
	Re\,\frac{\partial p_1}{\partial y}-\frac{\partial^2 v_0}{\partial y^2}&=&0.\label{e76}
	\end{eqnarray}
	\label{eq27}
\end{subequations}
The above system is subject to the following boundary conditions,
\begin{subequations}
	\begin{eqnarray}
	v_1&=&u_1\,\frac{\partial h}{\partial x}\quad \text{at} \quad y=h(x,t),\label{e77}\\
	\frac{\partial u_1}{\partial y}&=&0\quad \text{at} \quad y=h(x,t),\label{e78}\\
	Re p_1&=&2\,\left[\frac{\partial v_0}{\partial y}-\frac{\partial u_0}{\partial y}\,\frac{\partial h}{\partial x}\right]\quad \text{at} \quad y=h(x,t),\label{e79}\\
	u_1&=&0\quad\text{and}\quad v_1=0\quad \text{at} \quad y=0. \label{e80}
	\end{eqnarray}
	\label{eq28}
\end{subequations}
\noindent The first order horizontal component of velocity is obtained by solving \eqref{e75} along with the boundary conditions \eqref{e78} and \eqref{e80} to give,
\begin{eqnarray}
u_1=Re\,\bigg[4\,h^2\,\frac{\partial h}{\partial x}\bigg(\frac{h^2 y}{2}-\frac{y^3}{6}\bigg)+\frac{2}{3}\,h\,\frac{\partial h}{\partial x}\bigg(\frac{y^4}{4}-h^3y\bigg)+\frac{2\cot \theta}{Re}\frac{\partial h}{\partial x}\bigg(\frac{y^2}{2}-hy\bigg)-\alpha\,\frac{\partial^5 h}{\partial x^5}\nonumber\\
\bigg(hy-\frac{y^2}{2}\bigg)-\beta\,\frac{\partial^3 h}{\partial x^3}\left(hy-\frac{y^2}{2}\right)+\gamma\,\bigg(48\,h^2\,\bigg\{\frac{\partial h}{\partial x}\bigg\}^3+48\,h^3\frac{\partial h}{\partial x}\frac{\partial^2 h}{\partial x^2}+4h^4\frac{\partial^3 h}{\partial x^3}\bigg)\nonumber\\
\left(hy-\frac{y^2}{2}\right)\bigg].~~~~~~\label{e81}
\end{eqnarray}
\noindent Now, by using the first order continuity equation \eqref{e74} along with Eq. \eqref{e80} , the vertical component of the velocity is given as,
\begin{eqnarray}
v_1=Re\,\bigg[8h\bigg(\frac{\partial h}{\partial x}\bigg)^2\bigg(\frac{y^4}{24}-\frac{h^2y^2}{4}\bigg)+4h^2\frac{\partial^2 h}{\partial x^2}\bigg(\frac{y^4}{24}-\frac{h^2y^2}{4}\bigg)-h^3\bigg(\frac{\partial h}{\partial x}\bigg)^2y^2+\frac{2}{3}\bigg(\frac{\partial h}{\partial x}\bigg)^2\nonumber
\\\bigg(\frac{h^3y^2}{2}-\frac{y^5}{20}\bigg)+\frac{2}{3}\frac{\partial^2 h}{\partial x^2}\bigg(\frac{h^4 y^2}{2}-\frac{hy^5}{20}\bigg)+\frac{2\cot \theta}{Re}\frac{\partial^2 h}{\partial x^2}\bigg(\frac{hy^2}{2}-\frac{y^3}{6}\bigg)+\frac{\cot\theta}{Re}\bigg(\frac{\partial h}{\partial x}\bigg)^2\nonumber\\
y^2-\alpha\bigg(\frac{\partial^6 h}{\partial x^6}\bigg\{\frac{y^3}{6}-\frac{hy^2}{2}\bigg\}-\frac{\partial^5 h}{\partial x^5}\,\frac{\partial h}{\partial x}\frac{y^2}{2}\bigg)-\beta\bigg(\frac{\partial^4 h}{\partial x^4}\bigg\{\frac{y^3}{6}-\frac{hy^2}{2}\bigg\}-\frac{\partial^3 h}{\partial x^3}\frac{\partial h}{\partial x}\frac{y^2}{2}\bigg)\nonumber\\
+\gamma\bigg(96h\bigg\{\frac{\partial h}{\partial x}\bigg\}^4+288h^2\bigg\{\frac{\partial h}{\partial x}\bigg\}^2\,\bigg\{\frac{\partial^2 h}{\partial x^2}\bigg\}+48h^3\bigg\{\frac{\partial^2 h}{\partial x^2}\bigg\}^2+64h^3\bigg\{\frac{\partial h}{\partial x}\bigg\}\,\bigg\{\frac{\partial^3 h}{\partial x^3}\bigg\}\nonumber\\
+4h^4 \frac{\partial^4 h}{\partial x^4}\,\bigg)\,\bigg(\frac{y^3}{6}-\frac{hy^2}{2}\bigg)^3-\gamma\,\bigg(48h^2\bigg\{\frac{\partial h}{\partial x}\bigg\}+48h^3\,\bigg\{\frac{\partial h}{\partial x}\bigg\}\,\bigg\{\frac{\partial^2 h}{\partial x^2}\bigg\}+4h^4\,\bigg\{\frac{\partial^3 h}{\partial x^3}\bigg\}\nonumber\\
\bigg\{\frac{\partial h}{\partial x}\bigg\}\bigg)\frac{y^2}{2}\,\bigg].\hspace{1cm}\label{e82}
\end{eqnarray}
\noindent Similarly, the first order pressure is obtained from Eq. \eqref{e76} along with the boundary condition \eqref{e79}, which is given by,
\begin{equation}
p_1=-\frac{2}{Re}\big(y+h\big)\,\frac{\partial h}{\partial x}.\label{e83}    
\end{equation}
\noindent Substituting Eqs.~\eqref{e71}--\eqref{e72} and \eqref{e81}--\eqref{e82} in the kinematic boundary condition \eqref{e4} results in the non-linear first-order BE describing the surface elevation of floating elastic plate, which is given as,
\begin{eqnarray}
\frac{\partial h}{\partial t}+A(h)\,\frac{\partial h}{\partial x}+ \epsilon\,\frac{\partial}{\partial x}\bigg[B(h)\,\frac{\partial h}{\partial x}+\big(C(h)+D(h)\big)\,\frac{\partial^2 h}{\partial x^2}+E(h)\,\frac{\partial^5 h}{\partial x^5}+\nonumber\\
F(h)\,\frac{\partial h}{\partial x}\,\frac{\partial^2 h}{\partial x^2}+K(h)\bigg\{\frac{\partial h}{\partial x}\bigg\}^3\bigg]=0.\label{e84}
\end{eqnarray}
where
\begin{eqnarray}
&\displaystyle A(h)=2h^2,\quad B(h)=\left(\frac{8}{15}\,Re h^6-\frac{2}{3}\,\cot\theta\,h^3\right),\quad C(h)=\frac{4\gamma}{3}\,Re\,h^7,\quad D(h)=-\frac{\beta}{3}\,Re\,h^3,\nonumber\\
&\hspace{-3.8cm}\displaystyle E(h)=-\frac{\alpha}{3}\,Re\,h^3,\quad F(h)=16\gamma\,Re\,h^6 \quad \text{and}\quad K(h)=16\gamma\,Re\,h^5. \nonumber
\end{eqnarray}
\noindent Further, the normal mode analysis is carried out by assuming the surface elevation of the floating elastic plate as,
\begin{equation}
h(x,t)=1+\tilde{h}(x,t) \quad \text{such that} \quad \tilde{h}(x,t)=\eta\,e^{\mathrm{i}(kx-\omega_r t)+\omega_i\,t},  \label{e85} 
\end{equation}
where $\omega_r=kc_r$ and $\omega_i=kc_i$ are respectively the real and imaginary part of the temporal frequency $\omega$,
and $\tilde{h}$ being the deflection of the plate with respect to the mean free surface. By substituting Eq. \eqref{e85} in the non-linear equation \eqref{e84} and expanding using Taylor series expansion, the unsteady equation obtained as,
\begin{equation}
\frac{\partial \tilde{h}}{\partial t}+\mathcal{A}_1\,\frac{\partial \tilde{h}}{\partial x}+\epsilon\,\mathcal{B}_1\,\frac{\partial^2 \tilde{h}}{\partial x^2}+\epsilon\,(\mathcal{C}_1+\mathcal{D}_1)\,\frac{\partial^4 \tilde{h}}{\partial x^4}+\epsilon\,\mathcal{E}_1\,\frac{\partial^6 \tilde{h}}{\partial x^6}=\mathcal{N},\label{e86}
\end{equation}
where the group of non-linear terms, 
\begin{eqnarray}
&\displaystyle \mathcal{N}=-\bigg[\tilde{h}\,\mathcal{A}_1'+\frac{\tilde{h}^2}{2}\mathcal{A}_1''\bigg]\,\frac{\partial \tilde{h}}{\partial x}-\epsilon\,\bigg[\tilde{h}\,\mathcal{B}_1'+\frac{\tilde{h}^2}{2}\mathcal{B}_1''\bigg]\,\frac{\partial^2 \tilde{h}}{\partial x^2}-\epsilon\,\bigg[\tilde{h}\,\mathcal{C}_1'+\frac{\tilde{h}^2}{2}\mathcal{C}_1''+\tilde{h}\,\mathcal{D}_1'+\frac{\tilde{h}^2}{2}\mathcal{D}_1''\bigg]\,\frac{\partial^4 \tilde{h}}{\partial x^4}\nonumber\\
&\hspace*{0.5cm} \displaystyle-\epsilon\,\bigg[\tilde{h}\,\mathcal{E}_1'+\frac{\tilde{h}^2}{2}\mathcal{E}_1''\bigg]\,\frac{\partial^6 \tilde{h}}{\partial x^6}-\epsilon\,\bigg[\tilde{h}\,\mathcal{F}_1'+\frac{\tilde{h}^2}{2}\mathcal{F}_1''\bigg]\,\left[\bigg(\frac{\partial^2 \tilde{h}}{\partial x^2}\bigg)^2+\frac{\partial \tilde{h}}{\partial x}\,\frac{\partial^3 \tilde{h}}{\partial x^3}\right]-3\,\epsilon\,\mathcal{K}_1\,\bigg(\frac{\partial \tilde{h}}{\partial x}\bigg)^2\,\frac{\partial^2 \tilde{h}}{\partial x^2}\nonumber\\
&\hspace*{0.8cm}\displaystyle-\epsilon\,\big[B_1'+\tilde{h}\,B_1''\big]\,\bigg(\frac{\partial \tilde{h}}{\partial x}\bigg)^2-\epsilon\,\big[\mathcal{C}_1'+\tilde{h}\,\mathcal{C}_1''+\mathcal{D}_1'+\tilde{h}\,\mathcal{D}_1''\big]\,\frac{\partial \tilde{h}}{\partial x}\,\frac{\partial^2 \tilde{h}}{\partial x^2}-\epsilon\,\big[\mathcal{E}_1'+\tilde{h}\,\mathcal{E}_1''\big]\,\frac{\partial \tilde{h}}{\partial x}\,\frac{\partial^5 \tilde{h}}{\partial x^5}\nonumber\\
&\hspace*{9.3cm}\displaystyle-\epsilon\,\big[\mathcal{F}_1'+\tilde{h}\mathcal{F}_1''\big]\,\bigg(\frac{\partial \tilde{h}}{\partial x}\bigg)^2\,\frac{\partial^2 \tilde{h}}{\partial x^2}.\nonumber
\end{eqnarray}
such that
\begin{eqnarray}
&\mathcal{A}_1 = \left. A \right|_{h=1},\quad \mathcal{A}_1'=\left. A' \right|_{h=1}, \quad \mathcal{A}_1''=\left. A'' \right|_{h=1},\nonumber\\
&\mathcal{B}_1 = \left. B \right|_{h=1},\quad \mathcal{B}_1'=\left. B' \right|_{h=1}, \quad \mathcal{B}_1''=\left. B'' \right|_{h=1},\nonumber\\
&\mathcal{C}_1 = \left. C \right|_{h=1},\quad \mathcal{C}_1'=\left. C' \right|_{h=1}, \quad \mathcal{C}_1''=\left. C'' \right|_{h=1},\nonumber\\
&\mathcal{D}_1 = \left. D \right|_{h=1},\quad \mathcal{D}_1'=\left. D' \right|_{h=1}, \quad \mathcal{D}_1''=\left. D'' \right|_{h=1},\nonumber\\
&\mathcal{E}_1 = \left. E \right|_{h=1},\quad \mathcal{E}_1'=\left. E' \right|_{h=1}, \quad \mathcal{E}_1''=\left. E'' \right|_{h=1},\nonumber\\
&\mathcal{F}_1 = \left. F \right|_{h=1},\quad \mathcal{F}_1'=\left. F' \right|_{h=1}, \quad \mathcal{F}_1''=\left. F'' \right|_{h=1},\nonumber\\
&\mathcal{K}_1 = \left. K \right|_{h=1},\quad \mathcal{K}_1'=\left. K' \right|_{h=1}, \quad \mathcal{K}_1''=\left. K'' \right|_{h=1}.\label{4.15}
\end{eqnarray}
By suppressing the non-linear terms (i.e. $\mathcal{N}=0$), the analytical expression for linear phase speed $\omega_r$ and growth rate $\omega_i$ are given by 
\begin{eqnarray}
&\displaystyle \omega_r=2k\quad \text{and}\quad\omega_i=\epsilon Re\left[\bigg(\frac{8}{15}-\frac{2}{3}\frac{\cot\theta}{Re}\bigg)\,k^2+\frac{(\beta-4\,\gamma)\,k^4-\alpha\,k^6}{3}\right],\label{e87}  
\end{eqnarray}
which ensures the dependency of disturbance growth rate on the plate parameters. Now letting $\omega_i\to 0$, the value of critical wavenumber is derived as,
\begin{eqnarray}
&\hspace*{-1cm}\displaystyle k_c=\sqrt{\frac{\displaystyle\bigg(\frac{4}{3}\gamma\,Re-\frac{1}{3}\,\beta\,Re\bigg)+\sqrt{\displaystyle\bigg(\frac{1}{3}\,\beta\,Re-\frac{4}{3}\gamma\,Re\bigg)^2+\displaystyle\frac{4}{3}\alpha\,Re\,\bigg(\frac{8}{15}\,Re-\frac{2}{3}\cot\theta\bigg)}}{\displaystyle\frac{2}{3}\alpha\,Re}}.\label{e88}    
\end{eqnarray}
In similar way, the value of critical $Re$ from the expression of growth rate is obtained as, 
\begin{eqnarray}
&\displaystyle Re_c=\frac{\displaystyle\frac{2}{3}\cot\theta}{\displaystyle\bigg(\frac{8}{15}-\frac{\alpha k^4}{3}+\frac{\beta\,k^2}{3}-\frac{4\gamma\,k^2}{3}\bigg)}.\label{e89}
\end{eqnarray}
However, the value of $Re_c$ along the long-wave limit by letting $k \to 0$, is given by
\begin{equation}
Re_c = \frac{5}{4}\cot\theta.\label{e90}    
\end{equation}
This is the cut-off value of Reynolds number after which the flow becomes linearly unstable. It may be noted that the value of critical Reynolds number depends only on the inclination angle $\theta$. Further, the value of $Re_c$ obtained in the case of floating elastic plate is same as the thin Newtonian film flow down an inclined plane  (\cite{chin1986gravity}).

\section{Weakly nonlinear stability analysis} \label{WN}

The weakly nonlinear stability analysis becomes essential for the small but finite amplitude disturbances, where the nonlinear effects turn out prominent (i.e. $\mathcal{N}\neq 0$ in Eq.~\eqref{e86}) and the linear theory fails. A weakly nonlinear analysis is performed for the considered flow system following the works of \cite{oron2004subcritical, krishna1977nonlinear} and \cite{sadiq2008thin}. After applying  the slight perturbation to free surface using Eq. \eqref{e85}, the non-linear evolution equation \eqref{e86} is obtained. The solution to Eq. \eqref{e86} for the long-wave of arbitrary amplitudes is obtained through this technique. Around the neighborhood of criticality measured from distance
$\delta$, the slow independent variables are given by,
\begin{subequations}
	\begin{eqnarray}
	&\displaystyle X=\delta x, \quad T_1=\delta t \quad \text{and}\quad  T_2=\delta^2 t,\label{e91}\\
	&\displaystyle \frac{\partial}{\partial t}=\frac{\partial }{\partial t}+\delta\frac{\partial}{\partial T_1}+\delta^2\frac{\partial}{\partial T_2} \quad \text{and}\quad \frac{\partial}{\partial x}=\frac{\partial }{\partial x}+\delta\frac{\partial}{\partial X}, \label{e92}
	\end{eqnarray}    
	\label{eq39}
\end{subequations}
\noindent Using the method of multiple scales, the nonlinear equation \eqref{e86} can be written in the form of 
\begin{equation}
\big[\mathcal{L}_0+\delta\,\mathcal{L}_1+\delta^2\,\mathcal{L}_2\big]\,\tilde{h}=-\delta^2\,\mathcal{N}_1-\delta^3\,\mathcal{N}_2,\label{e93}
\end{equation}
where $\tilde{h}$ can be written in the series form as $\tilde{h}=\delta h_0 + \delta^2 h_1+ \delta^3 h_3$. Further, the operators $\mathcal{L}_0$, $\mathcal{L}_1$ and $\mathcal{L}_2$ along with the non-linear terms $\mathcal{N}_1$ and $\mathcal{N}_2$ are given by
\begin{subequations}
	\begin{eqnarray}
	\displaystyle \mathcal{L}_0&=&\frac{\partial}{\partial t}+\mathcal{A}_1\,\frac{\partial }{\partial x}+\epsilon\, \mathcal{B}_1\,\frac{\partial^2}{\partial x^2}+\epsilon\,(\mathcal{C}_1+\mathcal{D}_1)\,\frac{\partial^4}{\partial x^4}+\epsilon\,\mathcal{E}_1\,\frac{\partial^6}{\partial x^6},\label{e94}\\
	\displaystyle \mathcal{L}_1&=&\frac{\partial}{\partial T_1}+\mathcal{A}_1\,\frac{\partial}{\partial X}+2\epsilon\,\mathcal{B}_1\,\frac{\partial^2}{\partial x \partial X}+4\epsilon(\mathcal{C}_1+\mathcal{D}_1)\frac{\partial^4}{\partial x^3\partial X}+6\epsilon\,\mathcal{E}_1\,\frac{\partial^6}{\partial x^5 \partial X},\label{e95}\\
	\displaystyle \mathcal{L}_2&=&\frac{\partial}{\partial T_2}+\epsilon\,\mathcal{B}_1\,\frac{\partial^2}{\partial X^2}+6\epsilon(\mathcal{C}_1+\mathcal{D}_1)\frac{\partial^4}{\partial x^2\partial X^2}+15\epsilon\,\mathcal{E}_1\,\frac{\partial^6}{\partial x^4 \partial X^2},\label{e96}\\
	\displaystyle \mathcal{N}_1&=&\frac{\mathcal{A}_1''h_1^2}{2}\frac{\partial h_1}{\partial x}+\epsilon \left[\frac{\mathcal{B}_1''\,h_1^2}{2}\,\frac{\partial^2 h_1}{\partial x^2}+\frac{\mathcal{C}_1''h_1^2}{2}\frac{\partial^4 h_1}{\partial x^4}+\frac{\mathcal{D}_1''h_1^2}{2}\frac{\partial^4 h_1}{\partial x^4}+\frac{\mathcal{E}_1''h_1^2}{2}\frac{\partial^6 h_1}{\partial x^6}\right],\label{e97}\\
	\displaystyle \mathcal{N}_2&=&\mathcal{A}_1'\bigg[h_1\,\bigg(\frac{\partial h_2}{\partial x}+\frac{\partial h_1}{\partial X}\bigg)+h_2\,\frac{\partial h_1}{\partial x}\bigg]+\frac{\mathcal{A}_1''h_1^2}{2}\,\frac{\partial h_1}{\partial x}+\epsilon\,\mathcal{B}_1'\,\bigg[h_1\bigg(2\,\frac{\partial^2 h_1}{\partial x \partial X}+\frac{\partial^2 h_2}{\partial x^2}\bigg)\nonumber\\
	\displaystyle &&+h_2\frac{\partial h_1}{\partial^2 x^2}\bigg]+\frac{\epsilon\,h_1^2\,\mathcal{B}_1''}{2}\,\frac{\partial^2 h_1}{\partial x^2}+\epsilon\,\big(\mathcal{C}_1'+\mathcal{D}_1'\big)\,\bigg[h_1\bigg(4\frac{\partial^4 h_1}{\partial^3 x\partial X}+\frac{\partial^4 h_2}{\partial x^4}\bigg)+h_2\nonumber\\
	\displaystyle \,&&\frac{\partial^4 h_1}{\partial x^4}\bigg] +\frac{\epsilon\,h_1^2\,(\mathcal{C}_1''+\mathcal{D}'')}{2}\,\frac{\partial^4 h_1}{\partial x^4}+\epsilon\,\mathcal{E}_1'\bigg[h_1\bigg(6\frac{\partial^6 h_1}{\partial x^5\partial X}+\frac{\partial^6 h_2}{\partial x^6}\bigg)+h_2\,\frac{\partial^6 h_1}{\partial x^6}\bigg]\nonumber
	\end{eqnarray}
	\begin{eqnarray}	
	\displaystyle &&+\frac{\epsilon\,\mathcal{E}_1''\,h_1^2}{2}\,\frac{\partial^6 h_1}{\partial x^6}+\epsilon\,\mathcal{F}_1\,\bigg[2\frac{\partial^2 h_1}{\partial x^2}\bigg(2\,\frac{\partial^2 h_1}{\partial x \partial X}+\frac{\partial^2 h_2}{\partial x^2}\bigg)\bigg]+\epsilon\,\mathcal{F}_1'h_1\bigg(\frac{\partial^2 h_1}{\partial x^2}\bigg)^2+\epsilon\,\,\nonumber \\
	\displaystyle &&\big(\mathcal{F}_1+\mathcal{C}_1'+\mathcal{D}_1'\big)\,\bigg[\frac{\partial h_1}{\partial x}\bigg(3\frac{\partial^3 h_1}{\partial x^2 \partial X}+\frac{\partial^3 h_2}{\partial x^3}\bigg)+\frac{\partial^3 h_1}{\partial x^3}\,\bigg(\frac{\partial h_2}{\partial x}+\frac{\partial h_1}{\partial X}\bigg)\bigg]+\epsilon\,\big(\mathcal{F}_1'\nonumber\\
	\displaystyle &&+\mathcal{C}_1''+\mathcal{D}_1''\big)\,h_1\,\frac{\partial h_1}{\partial x}\frac{\partial^3 h_1}{\partial x^3}+\epsilon\,\big(3\mathcal{G}_1+\mathcal{F}_1'\big)\,\bigg(\frac{\partial h_1}{\partial x}\bigg)^2\,\frac{\partial^2 h_1}{\partial x^2}+2\,\epsilon\,\mathcal{B}_1'\,\frac{\partial h_1}{\partial x}\bigg(\frac{\partial h_2}{\partial x}\nonumber\\
	\displaystyle&&+\frac{\partial h_1}{\partial X}\bigg)+\epsilon\,\mathcal{B}_1''\,h_1\,\bigg(\frac{\partial h_1}{\partial x}\bigg)^2+\epsilon\,\mathcal{E}_1'\bigg[\frac{\partial h_1}{\partial x}\bigg(5\,\frac{\partial^6 h_1}{\partial x^4 \partial X}+\frac{\partial^5 h_2}{\partial x^5}\bigg)+\frac{\partial^5 h_1}{\partial x^5}\bigg(\frac{\partial h_2}{\partial x}\nonumber\\
	&&\displaystyle +\frac{\partial h_1}{\partial X}\bigg)\bigg] +\epsilon\,\mathcal{E}_1''\,h_1\frac{\partial h_1}{\partial x}\,\frac{\partial^5 h_1}{\partial x^5}.\label{e98}
	\end{eqnarray}\label{eq41}
\end{subequations}
\noindent where $\mathcal{A}_1$, $\mathcal{B}_1$, $\mathcal{C}_1$, $\mathcal{D}_1$, $\mathcal{E}_1$, $\mathcal{F}_1$ and $\mathcal{G}_1$ are the constants as given in Eq. \eqref{4.15}. The first order solution $O(\delta)$ of equation $\mathcal{L}_0h_1=0$ is given by
\begin{equation}
h_1=\eta e^{\mathsf{i}(kx-\omega_rt)}+\bar{\eta}\,e^{-\mathsf{i}(kx-\omega_rt)},\label{e99}
\end{equation}
\noindent where $\eta(X, T_1, T_2)$ be the nonlinear amplitude function and its complex conjugate is denoted by $\bar{\eta}(X, T_1, T_2)$. Similarly, in th case of second order equation $O(\delta^2)$, the solution is obtained as follows,
\begin{eqnarray}
\mathcal{L}_0h_2+\mathcal{L}_1h_1&=&-\mathcal{N}_2,\label{e100}\\
\mathcal{L}_0h_2&=&-\bigg[\frac{\partial}{\partial T_1}+\mathcal{S}\,\frac{\partial}{\partial X}\bigg]\,\eta\,e^{\mathsf{i}(kx-\omega_r t)}+\mathcal{H}\,\eta^2\,\,e^{\mathsf{i}(kx-\omega_r t)}+c.c,\label{e101}
\end{eqnarray}
\noindent where $c.c$ denotes the complex conjugate of terms in right hand side of Eq. \eqref{e101}. Further, the coefficients $\mathcal{S}$ and $\mathcal{H}$ is given as,
\begin{eqnarray}
\mathcal{S}&=&\mathcal{S}_r+\mathsf{i}\,\mathcal{S}_i,\nonumber~~~\mathcal{S}_r=\mathcal{A}_1,\nonumber\\
\mathcal{S}_i&=&2\,\epsilon k \mathcal{B}_1-4\epsilon\,k^3(\mathcal{C}_1+\mathcal{D}_1)-6\,\epsilon\,k^5\,\mathcal{E}_1,\nonumber\\
\mathcal{H}&=&\mathcal{H}_r+\mathsf{i}\,\mathcal{H}_i,\nonumber\\
\mathcal{H}_r&=&2\,\epsilon\,k^6\mathcal{E}_1'-2\,\epsilon\,k^4\,(\mathcal{C}_1'+\mathcal{D}_1'+\mathcal{F}_1)+\epsilon\,k^2\,\mathcal{B}_1',\nonumber\\
\mathcal{H}_i&=&-\mathcal{A}_1'\,k.\nonumber
\end{eqnarray}
\noindent In order to get solution, the first term in the right hand side of Eq. \eqref{e101} known as secular terms, is equated to zero. Hence, the form of solution at $O(\delta^2)$ is given as
\begin{equation}
\displaystyle h_2=m\eta^2\,e^{2\mathsf{i}(kx-\omega_rt)}+\bar{m}\,\bar{\eta}^2\,e^{-\mathsf{i}(kx-\omega_rt)},\label{e102}
\end{equation}
where $m=m_r+\mathsf{i}\,m_i$. Further, the equation of perturbed amplitude $\eta(X,T_1,T_2)$ is obtained by using $h_1$ and $h_2$ in equation \eqref{e93} at $O(\delta^3)$, which is given by
\begin{equation}
\frac{\partial \eta}{\partial T_2}+\epsilon\,\big(\mathcal{B}_1-6\,k^2(\mathcal{C}_1+\mathcal{D}_1)+15\,k^4\,\mathcal{E}_1\big)\,\frac{\partial^2 \eta}{\partial X^2}-\delta^{-2}\,\omega_i\,\eta+\big(J_r+\mathsf{i}\,J_i\big)\,\eta^2\bar{\eta}=0,\label{e103}
\end{equation}  
where
\begin{subequations}
	\begin{eqnarray}
	J_r&=&-\mathcal{A}_1'm_ik-\epsilon\,k^2\,m_r\,\mathcal{B}_1'-\frac{\epsilon\,k^2}{2}\,\mathcal{B}_1''+7\,\mathcal{C}_1'\epsilon\,k^4\,m_r+\frac{\epsilon\,k^4}{2}\,\mathcal{C}_1''+7\,\mathcal{D}_1'\epsilon\,k^4\,m_r+\frac{\epsilon\,k^4}{2}\,\mathcal{D}_1''\nonumber\\
	&&-31\epsilon\,k^6\,m_r\,\mathcal{E}_1'-\frac{\epsilon\,k^6}{2}\,\mathcal{E}_1''+2\,k^3\,\epsilon\,\big[4m_i-5km_r\big]\,\mathcal{F}_1+\epsilon\,k^4\mathcal{F}_1',\label{e104}\\
	J_i&=&\mathcal{A}_1'm_rk+\frac{\mathcal{A}_1''}{2}\,k-\epsilon\,k^2\,m_i\,\mathcal{B}_1'+7\,\epsilon\,m_i\,\mathcal{C}_1'+7\,\epsilon\,m_i\,\mathcal{D}_1'-31\epsilon\,k^6\,m_i\,\mathcal{E}_1'\nonumber\\
	&&-2\,k^3\,\epsilon\,\big[4m_r+5km_i\big]\,\mathcal{F}_1,\label{e105}\\
	m_r&=&\frac{\mathcal{H}_r}{W_r},\quad m_i=\frac{\mathcal{H}_i}{W_r},\label{e106}\\
	W_r&=&16\,k^4\epsilon\,\big(\mathcal{C}_1+\mathcal{D}_1\big)-64\,k^6\,\epsilon\mathcal{E}_1-4k^2\epsilon\,\mathcal{B}_1.\label{e107}
	\end{eqnarray}
	\label{e47}
\end{subequations}
Equation \eqref{e103} describes the weekly nonlinear behavior of film flow in the presence of floating elastic plate. Further, the form  solution to Eq. \eqref{e103} is considered as $\displaystyle\eta=\chi_0\,e^{-\mathsf{i}f(T_2)\,T_2}$ where the spatial modulation is neglected. Substituting the expression for $\eta$ in Eq. \eqref{e103} yield the Ginzburg-Landau equation, which is given by
\begin{eqnarray}
&&\frac{\partial\chi_0}{\partial T_2}+\big(J_r\,\chi_0^2-\delta^{-2}\,\omega_i\big)\,\chi_0=0,\label{e108}\\
&&\frac{\partial}{\partial T_2}\, \big[T_2\,f(T_2)\big]-J_i\,\chi_0^2=0.\label{e109}
\end{eqnarray}
The linear disturbance in the flexural flow is controlled by the signs of $J_r$ and $\omega_r$ in the second and third terms, respectively, of Eq. \eqref{e108}. Further, the nonlinearity in flexural flow is induced by these two terms. The threshold amplitude $\delta\,\chi_0$ and function $f(T_2)$ obtained from Eqs. \eqref{e108} and  \eqref{e109}, respectively, is given by
\begin{equation}
\delta\,\chi_0=\bigg(\frac{\omega_i}{J_r}\bigg)^{1/2},\quad f(T_2)=\frac{\omega_i\,J_i}{\delta^2\,J_r}.     
\end{equation}
Substituting the value of $\eta$ and $T_2$ in the expression for $\tilde{h}$ in Eq. \eqref{e85} gives the nonlinear phase speed $\omega_{nr}$ as,
\begin{equation}
\omega_{nr}=\omega_r+\frac{\omega_i\,J_i}{J_r}.
\end{equation}
This analysis is carried out to show the occurrence of supercritical and sub-critical region around the neighbourhood of criticality depending on the signs of $J_r$ and $\omega_i$. Further, the growth of waves in these regions depends on the wavenumber of initial perturbation $k$.

\section{Numerical results and discussion}
\label{NRD}
Let us begin with the numerical solving procedure of the generalized eigenvalue problem governed by the Orr-Sommerfeld system (\ref{e32})$-$(\ref{e36}). The eigenvalues and eigenfunctions are obtained by the Chebyshev collocation method. The unknowns of the system are described using the Chebyshev polynomials on $[-1,1]$. Using a linear transformation of the independent variable $y$, the Orr-Sommerfeld equations and the boundary conditions are transformed to the interval $[0,1]$. The stream-function $\phi(y)$ and it's derivatives are approximated by the truncated expansions, $$\phi^r(y) = \sum_{i=0}^n a_i T_i^r(y)$$ where $T_i(y)$ are orthogonal Chebyshev polynomials and the superscript $r$ is used to represent the $r$-th derivative with respect to $y$. The unknowns $a_i$ are the discrete Chebyshev expansion coefficients, and the non-zero finite number $\nu$ denotes the truncation level. Substituting these expansions for $\phi(y)$ (and its derivatives) in the set of equations and boundary conditions (\ref{e32})$-$(\ref{e36}), and using the recurrence relations for Chebyshev polynomials, we obtain the system of linear equations 
$${\bf AX} = c{\bf BX}$$  
where ${\bf X} = (\eta, \phi_1, \phi_2,...,\phi_n)^T$ is the vector representation of the unknowns, $\bf A$ and $\bf B$ are matrices of order $n+1$. For a given wavenumber $k$, the phase speed $c$ is the eigenvalue of the above linear system.
Based on the technique described by \cite{canuto2012spectral}, a MATLAB code is developed for the numerical computations and the eigenvalues of the system of linear equations are obtained. After using the value of the eigenvalue `$c$', the numerical solution of the problem can be found at `$n$' Chebyshev collocation points, 
$$y_j = \cos{\left[ \frac{(j-1)}{(n-1)}\pi\right]}\quad j = 1,2,...,n.$$
\begin{figure}
	\begin{center}
		\subfigure[]{\label{f2a}\includegraphics*[width=6.6cm]{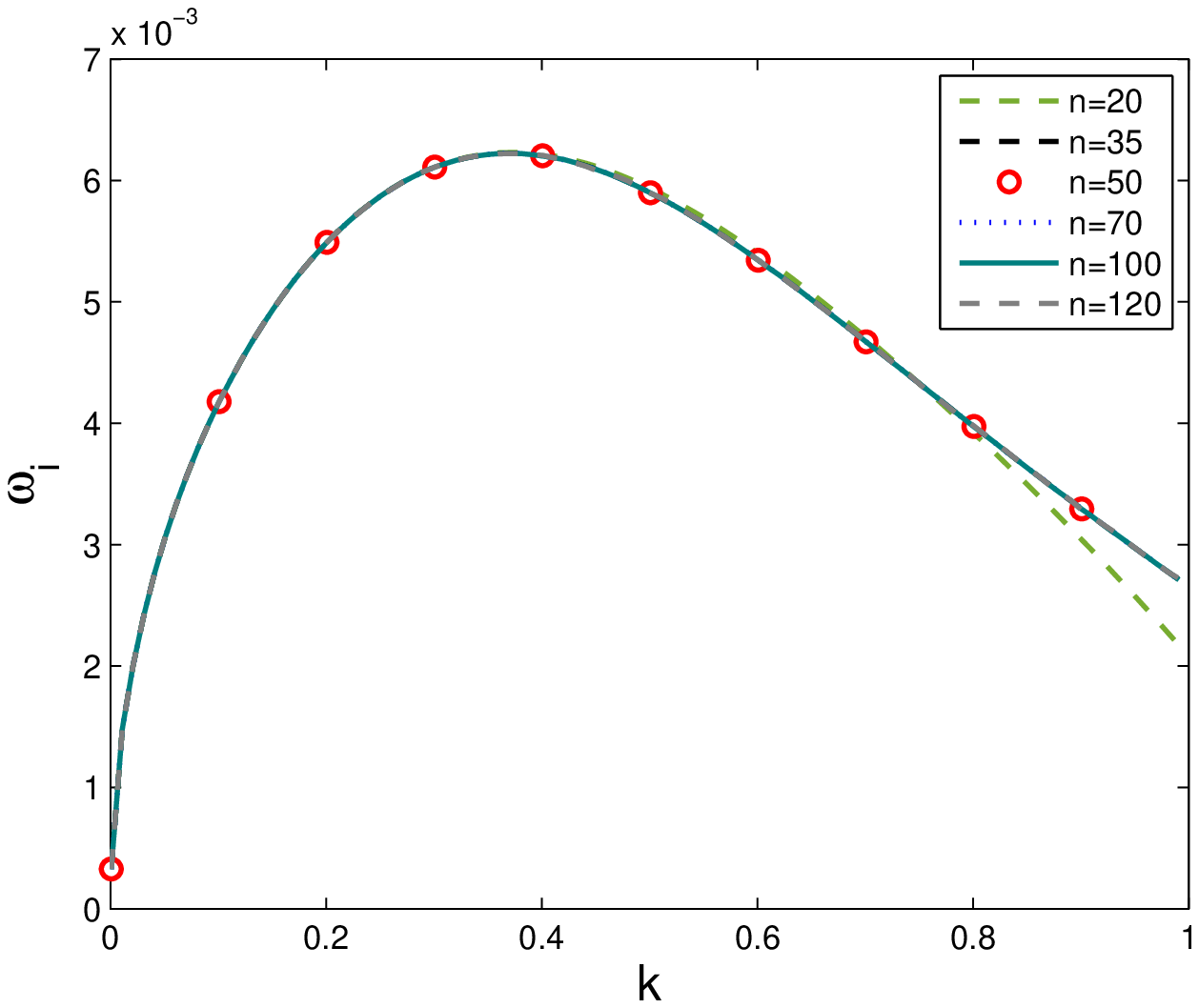}}
		\subfigure[]{\label{f2b}\includegraphics*[width=6.6cm]{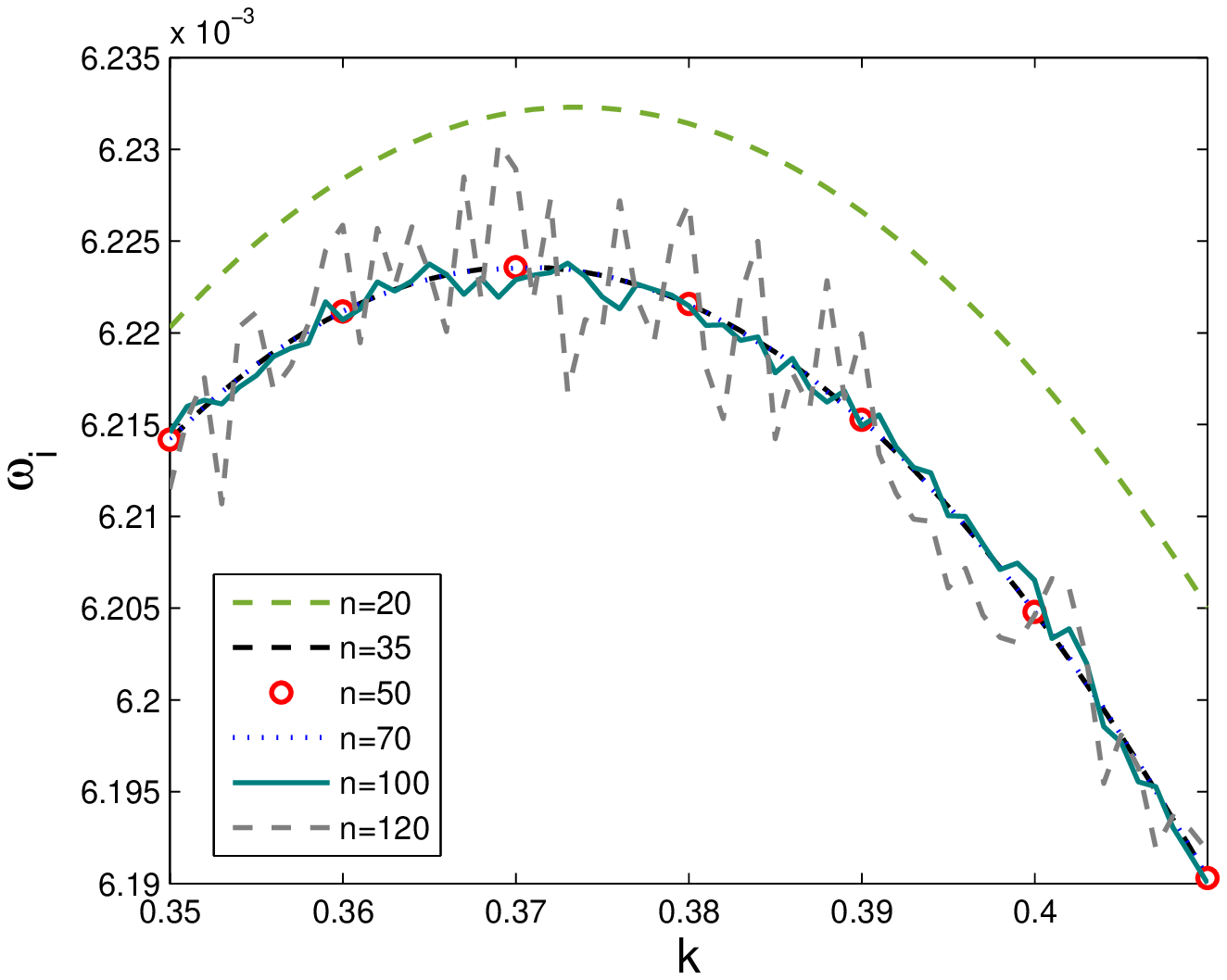}}
	\end{center}
	\caption{Grid convergence test showing the dispersion curves (growth rate $\omega_i$ as a function of wavenumber $k$) using different number of collocation points for (a) $0\leq k \leq 1$ and (b) $0.35\leq k \leq 0.41$. The fixed flow parameters are $\alpha=0.1$, $\beta=0$, $\gamma=0.03$, $Re=2500$ and $\theta=10^\circ$.}\label{f2}
\end{figure}
The complex phase velocity  $c = c_r + \mathrm{i}c_i$, obtained from the above system, are used to calculate the non-dimensional scaled temporal growth rate  $\omega_i = kc_i$ as a function of wavenumber $k$. The convergence of eigenvalues is examined by varying the number of collocation points in the computation in Fig.~\ref{f2}, where the dispersion curves ($\omega_i$ versus $k$) are plotted for a typical set of parameters. It is observed that the growth rate $\omega_i$ converges rapidly for smaller values of $k$ and more slowly for larger values of $k$. 
Clearly, the convergence is achieved with an intermediate level of grid refinement and the results of this study are generated using $n$ value between $50$ and $90$. 

\begin{figure}
	\begin{center}
		\subfigure[]{\label{f3a}\includegraphics*[width=6.6cm]{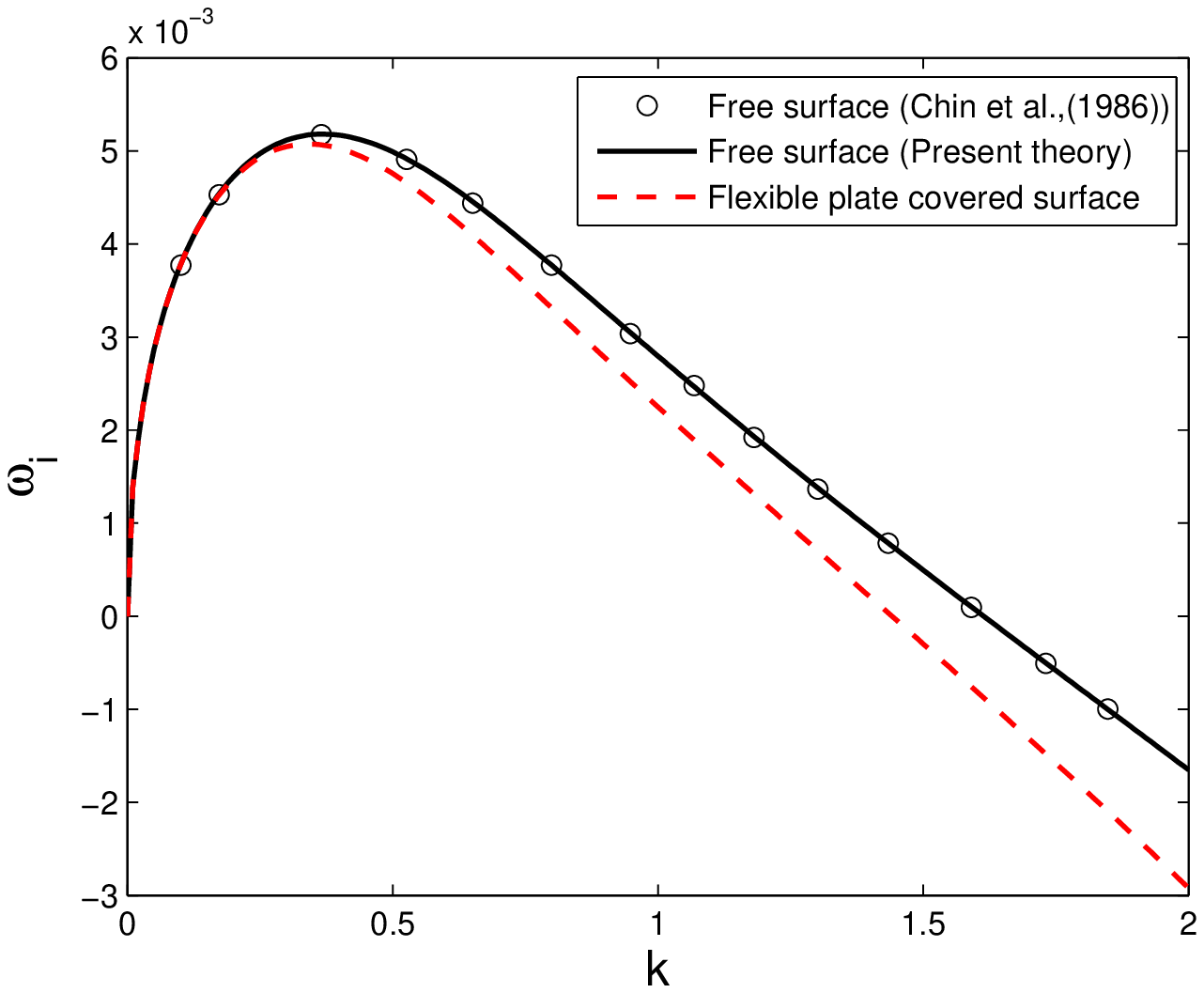}}
		\subfigure[]{\label{f3b}\includegraphics*[width=6.6cm]{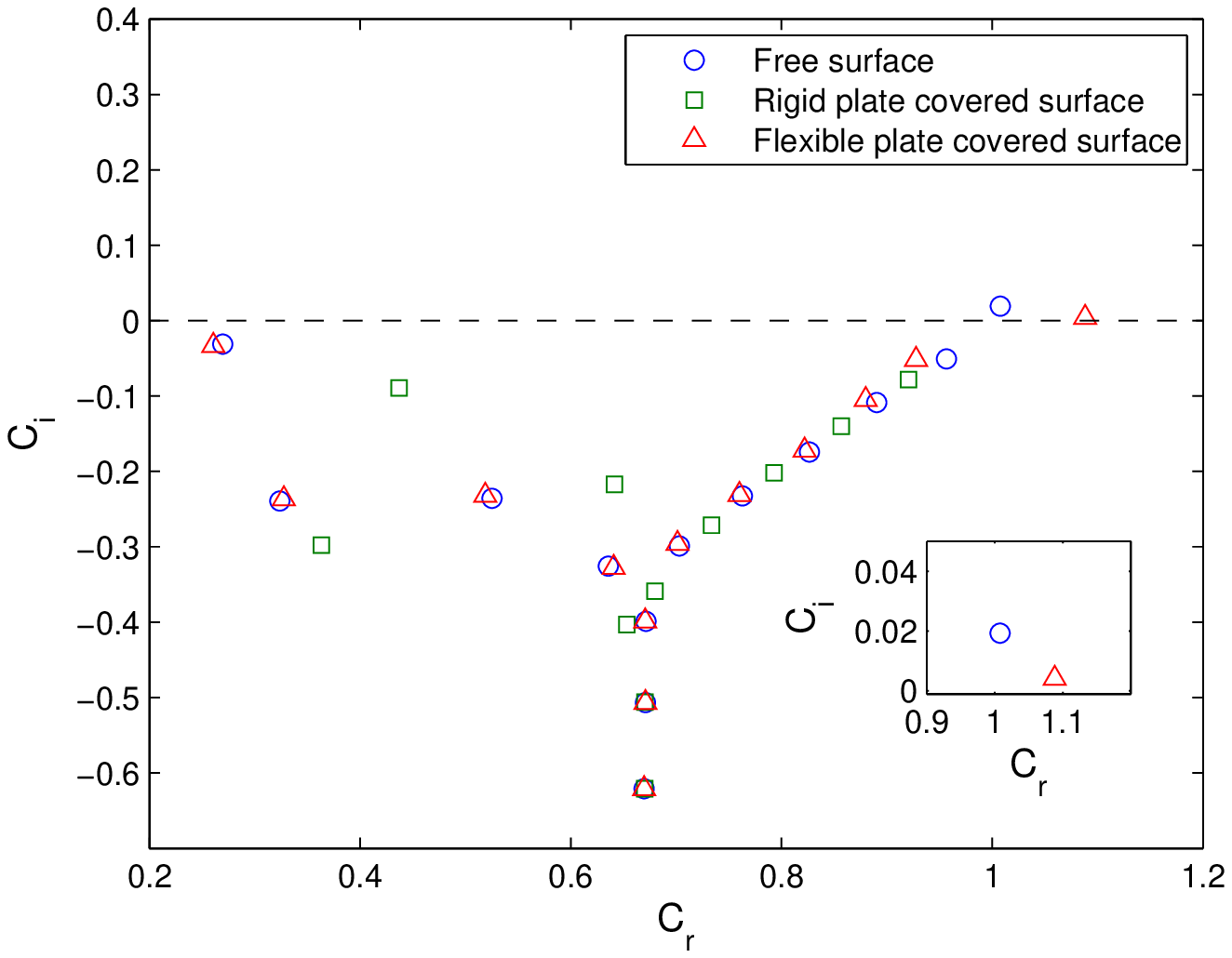}}
	\end{center}
	\caption{Comparison between classical free surface falling film and the flow bounded above by a flexible plate: (a) Growth rate of most unstable mode $\omega_i$ against the wavenumber $k$, and (b) Distribution of eigenvalues corresponding to two-dimensional disturbances. For free surface flow $\alpha=0$, $\beta=\beta_0\,\big(Re^2\,\sin\theta\big)^{-1/3}$, $\gamma=0$, flexible plate covered surface $\alpha=0.1$, $\beta=0$, $\gamma=0.03$ and rigid plate covered surface $\alpha=20$, $\beta=0$, $\gamma=10$. The other parameters are fixed as $Re=2500$, $\beta_0=4280$ and $\theta=4^\circ$. }\label{f3}
\end{figure} 
To test the reliability of our numerical code, computations were carried out for the limiting free surface flow and presented in  Fig.~\ref{f3}. Fig.~\ref{f3}(a) compares the temporal growth of the most unstable mode obtained from our computation with the prediction of \cite{chin1986gravity} in the case of free surface flow by limiting the parameters $\alpha=0$, $\beta = -\beta_0\big(Re^2 \sin\theta\big)^{-1/3}$ and $\gamma=0$. Here, the non-dimensional parameter $\beta$ corresponds to the surface-tension and $\beta_0$ be the dimensionless parameter defined by \cite{chin1986gravity} such that $\beta_0 \in (0, 5000)$. Growth rate curves obtained by the present numerical computation are in excellent agreement with the corresponding prediction of \cite{chin1986gravity} (in their figure 5). The solid dark line in Fig.~\ref{f3}(a) represents the current results for falling free surface flow, and the circles represent results of \cite{chin1986gravity}.  Result in the presence of a floating plate are drawn with a broken red line. The two-dimensional spectrum corresponding to three different flows is shown in Fig.~\ref{f3}(b). Following the observation of  \cite{chin1986gravity}, the unstable mode is the gravity mode (shown in the inset of Fig. \ref{f3}(b)) and the shear mode present in the left branch of the spectrum is the damped mode.  The positive growth rate of eigenmode reveals the occurrence of unstable two-dimensional normal modes. In the range of flow parameters used for both free surface and flexural flows, the growth rate increases initially and attains maximum value, then decreases with increasing wavenumber ($k$). The growth rate of the most unstable mode reduces for higher wavenumbers in the presence of a flexible plate compared to that for the free surface flow. However, the long-wave instability remains unaltered. Unlike the surface mode instability for free surface falling film, the existence of an unstable mode for the falling film covered with a floating plate is confirmed in Fig.~\ref{f3}(b). Such an unstable mode is absent in the case of a flow covered by a rigid surface.

\begin{figure}
	\begin{center}
		\subfigure[]{\label{f5a}\includegraphics*[width=6.6cm]{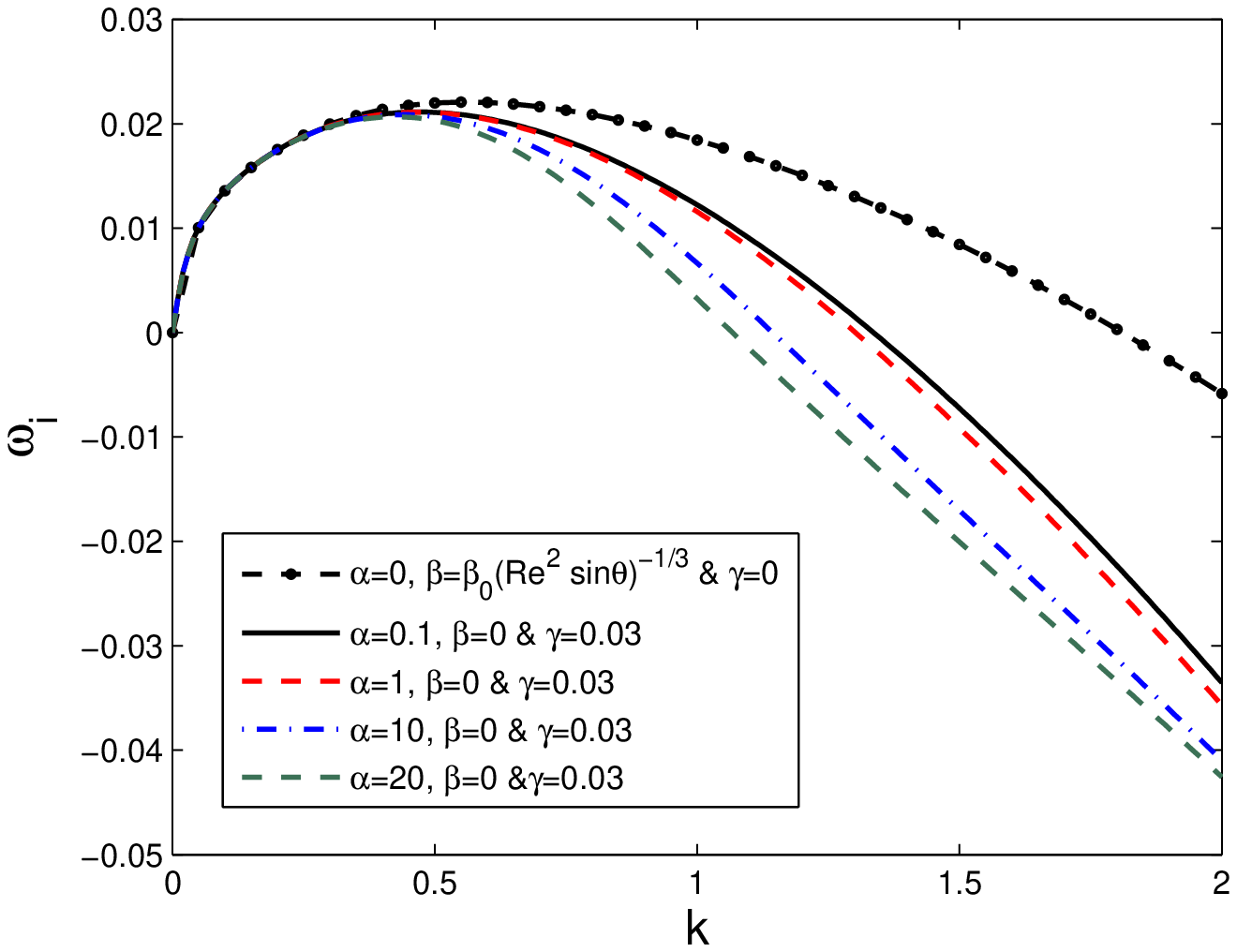}}
		\subfigure[]{\label{f5b}\includegraphics*[width=6.6cm]{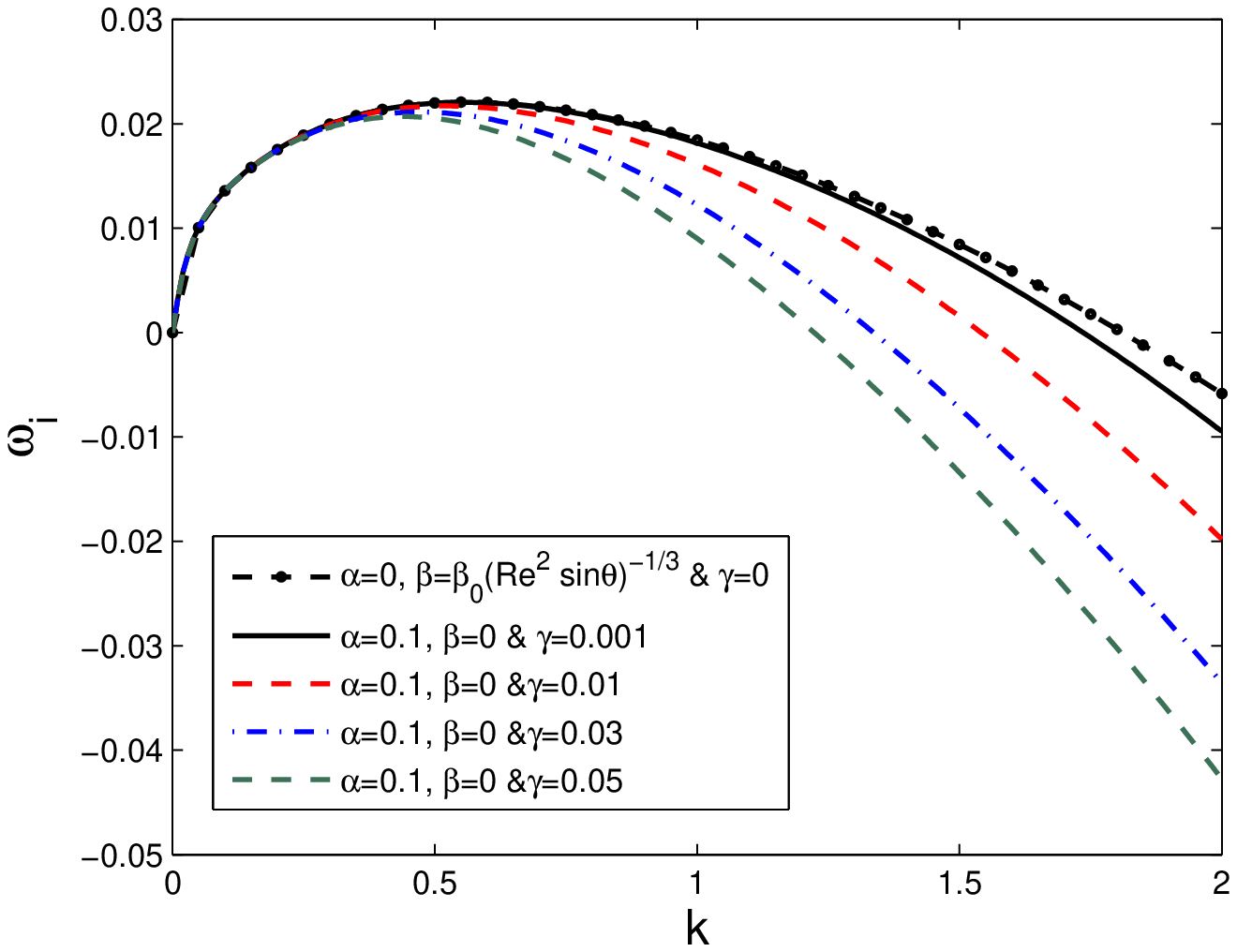}}
	\end{center}
	\caption{Growth rate of supreme unstable mode $\omega_i$ against the wavenumber $k$ showing the effect of (a) structural rigidity $\alpha$ with $\gamma=0.03$ and (b) uniform mass per unit length $\gamma$ with $\alpha=0.1$. The other parameters are $\theta=10^\circ$, $Re=200$ and $\beta_0=1$.}\label{f5}
\end{figure}

The influences of structural rigidity $\alpha$ and uniform mass per unit length $\gamma$ on the dominant growth rate $\omega_i$ (as a function of $k$) are plotted in Figs.~\ref{f5}(a) and (b), respectively. As the structural rigidity increases (Fig.~\ref{f5}(a)), the floating plate becomes more rigid and the growth rate of dominant disturbance decreases, which results in a relative stabilization. The value of $\gamma$ is directly proportional to the thickness of plate. As $\gamma$ increases the thickness of plate $d$ increases, and it behaves like a rigid wall. In consequence, the growth rate of the surface mode reduces for higher $\gamma$, which in turn suppresses the flow instability.

\begin{figure}
	\begin{center}
		\subfigure[]{\label{f4a}\includegraphics*[width=6.6cm]{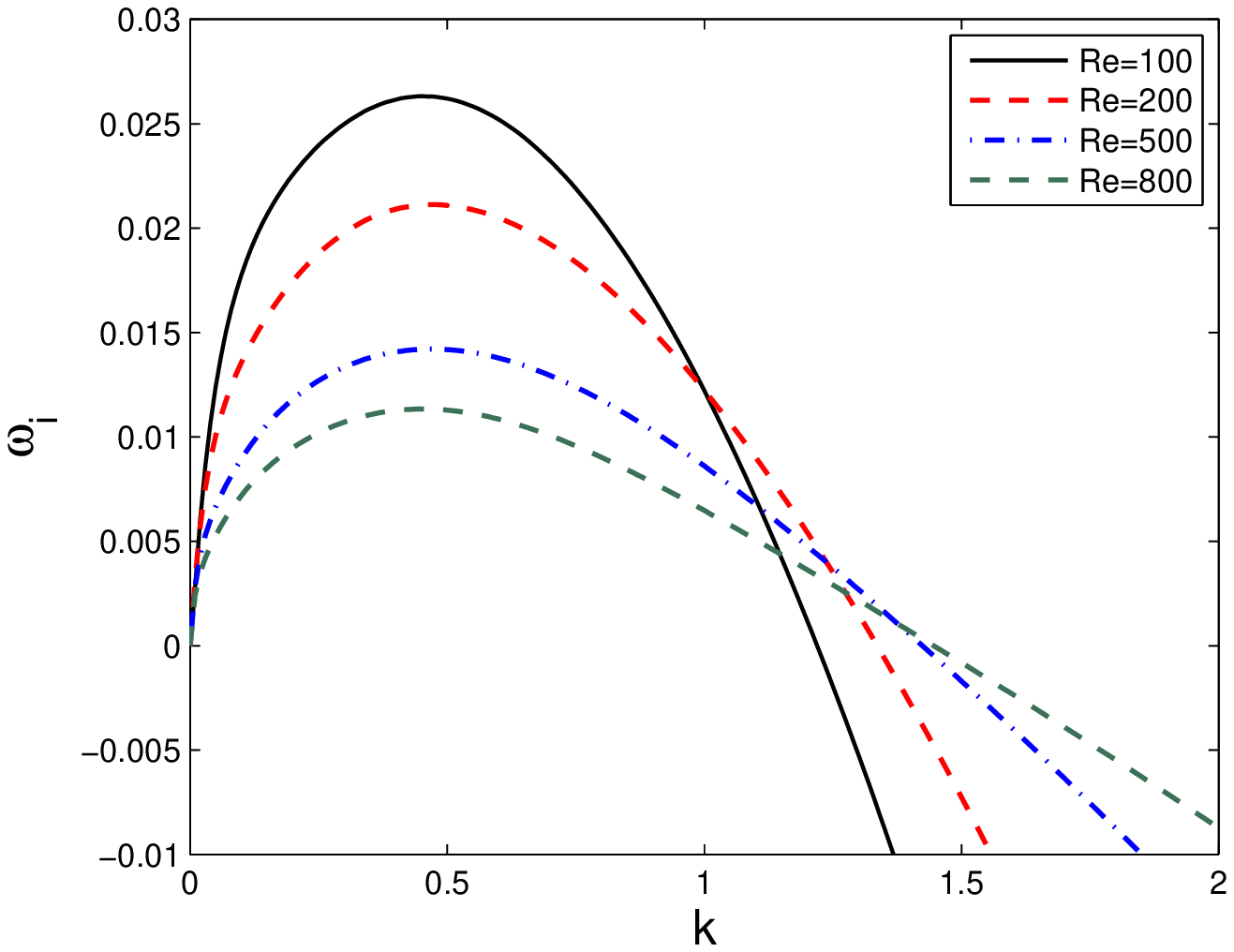}}
		\subfigure[]{\label{f4b}\includegraphics*[width=6.6cm]{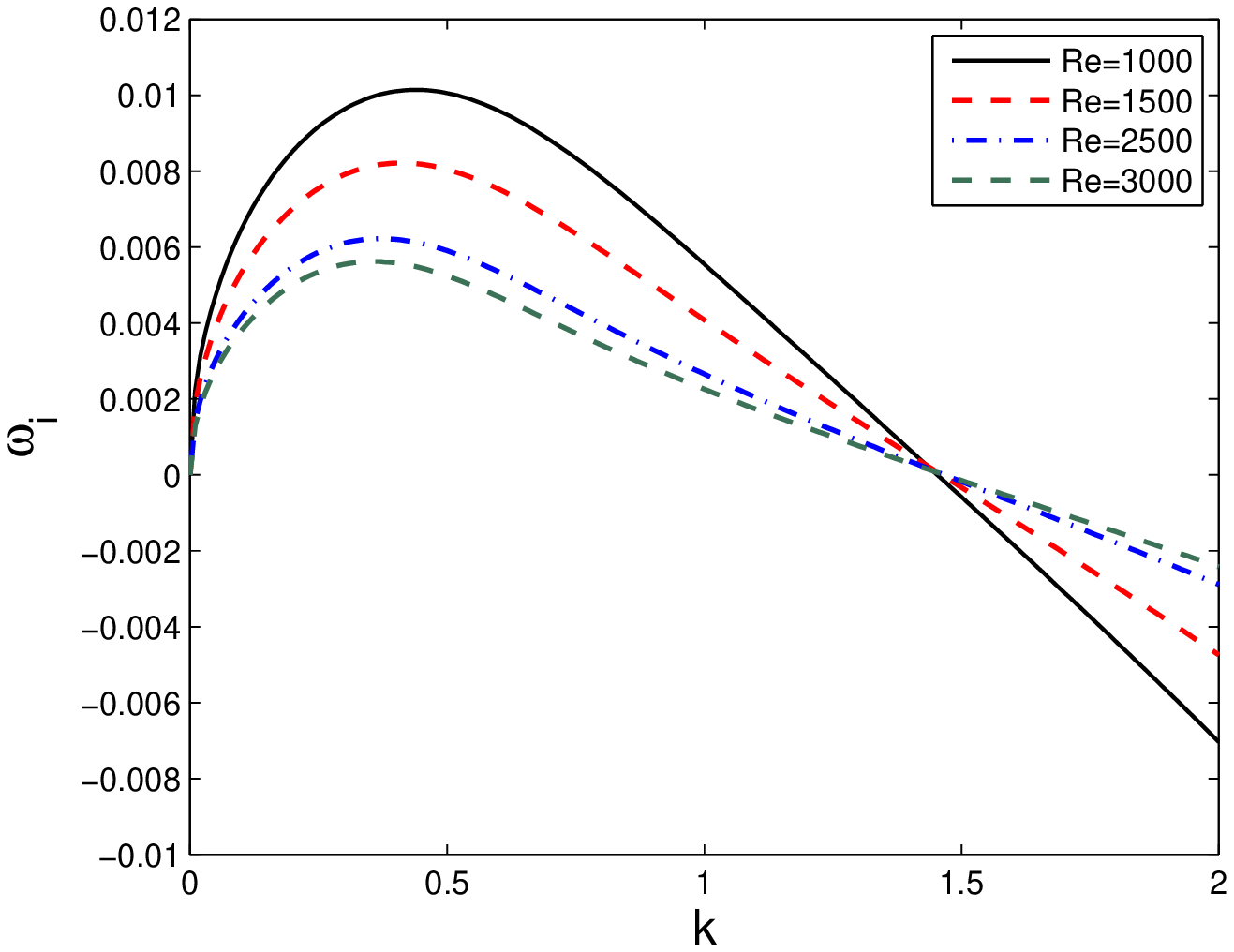}}
	\end{center}
	\caption{Growth rate of most excited mode $\omega_i$ against the wavenumber $k$ showing the effect of Reynolds number (a) $Re<1000$ and (b) $Re\geq1000$. The other parameters are $\theta=10^\circ$, $\alpha=0.1$, $\beta=0$ and $\gamma=0.03$.}\label{f4}
\end{figure}

\begin{figure}
	\begin{center}
		\subfigure[]{\label{f6a}\includegraphics*[width=6.6cm]{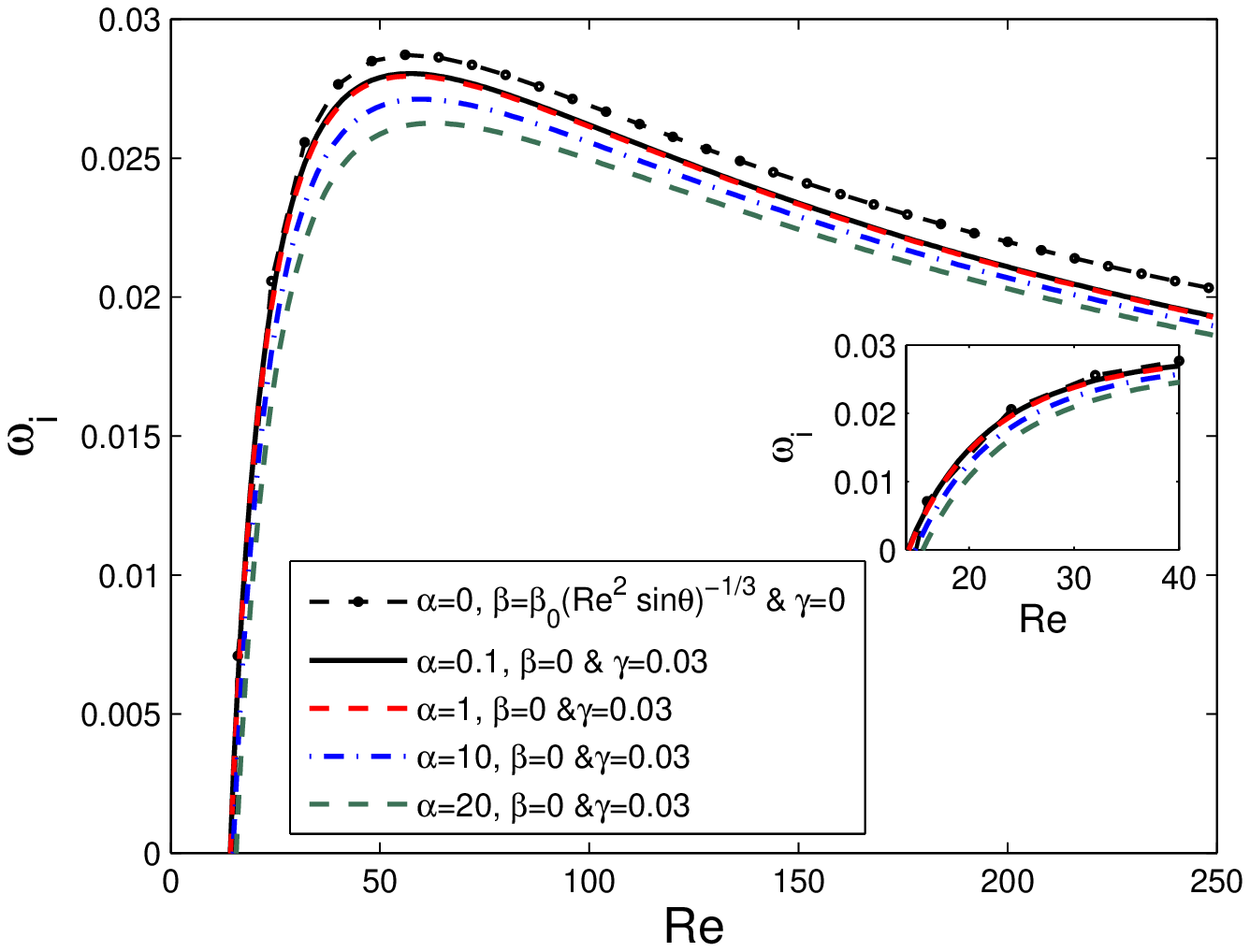}}
		\subfigure[]{\label{f6b}\includegraphics*[width=6.6cm]{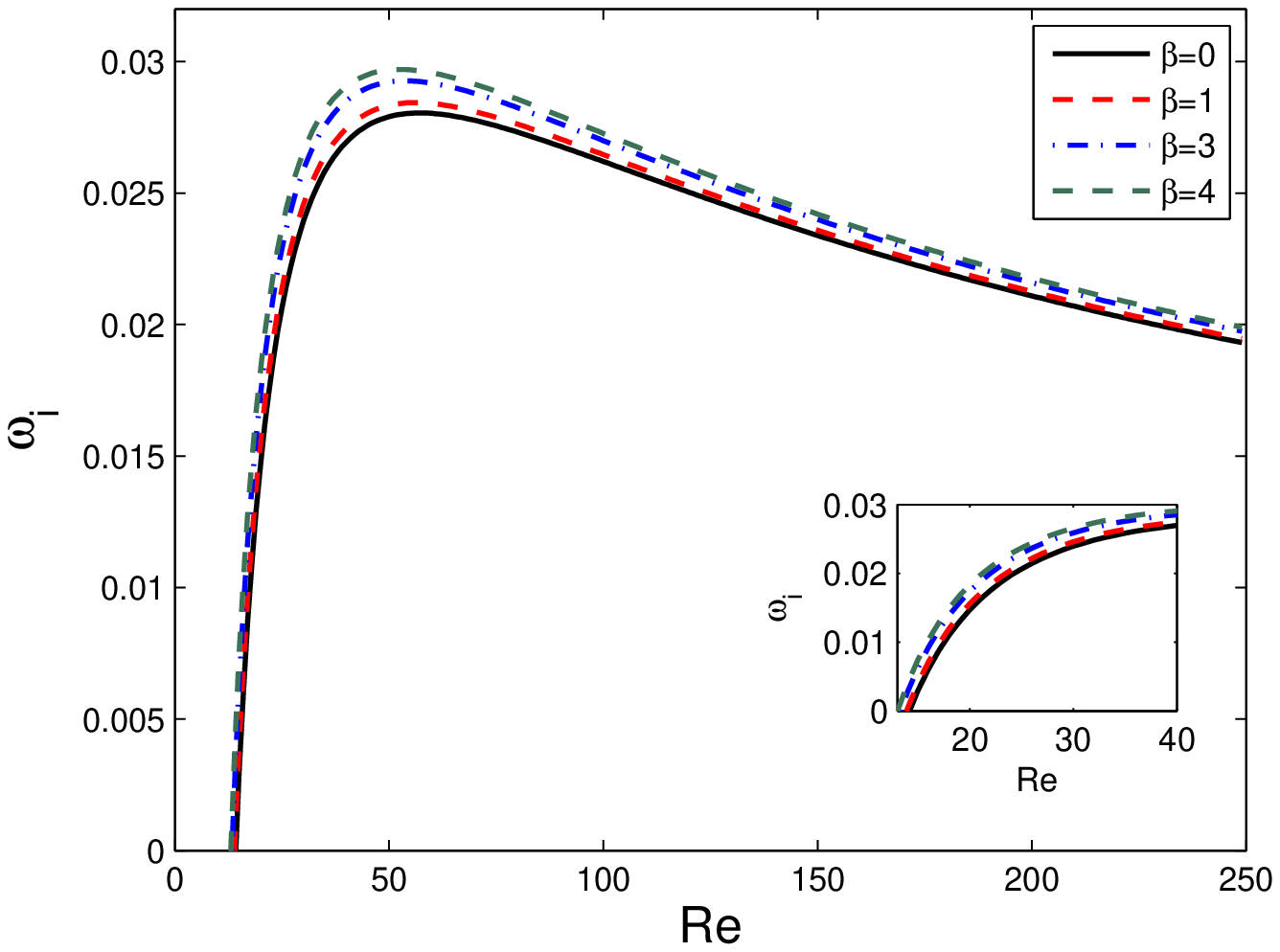}}
		\subfigure[]{\label{f6c}\includegraphics*[width=6.6cm]{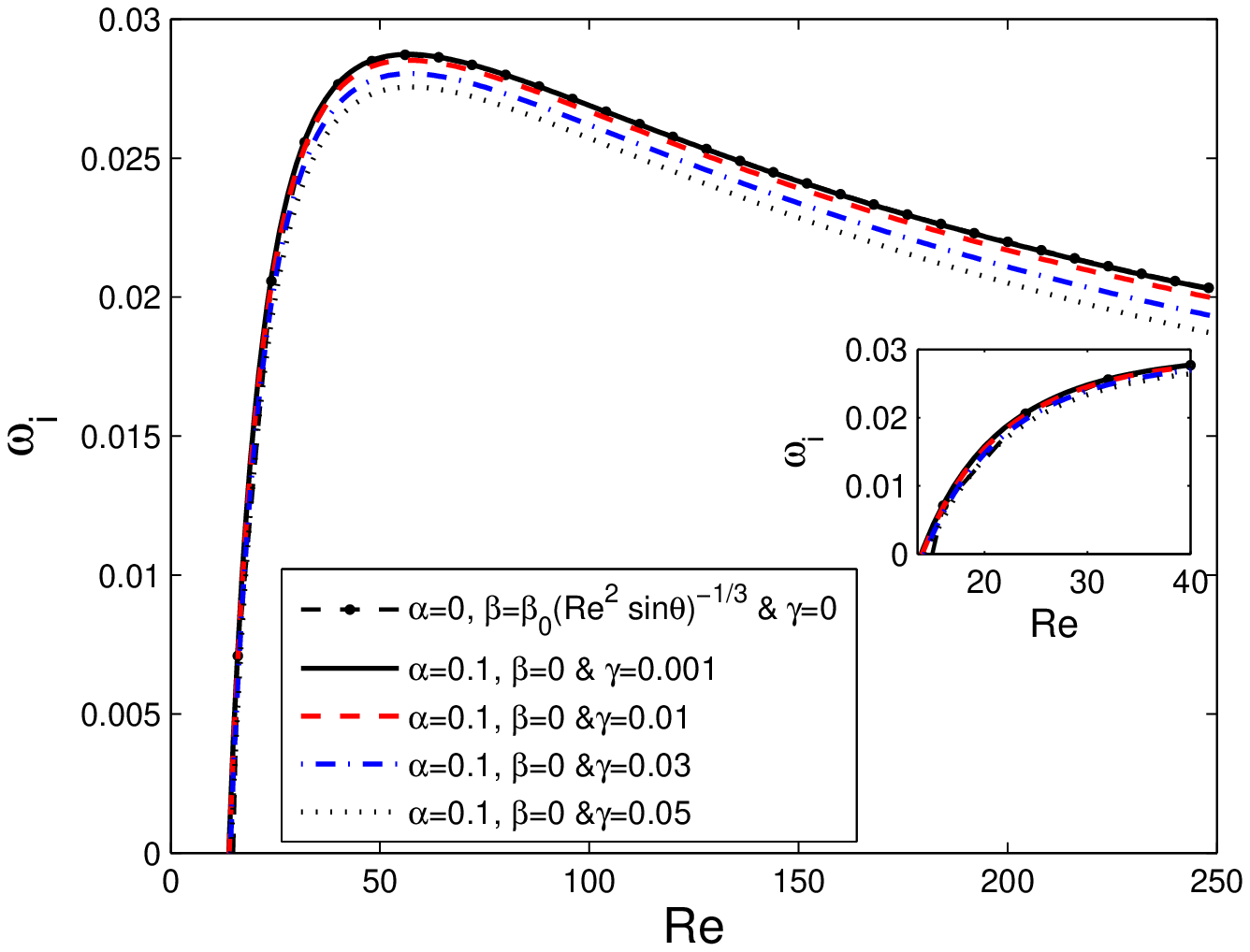}}
		\subfigure[]{\label{f6b}\includegraphics*[width=6.6cm]{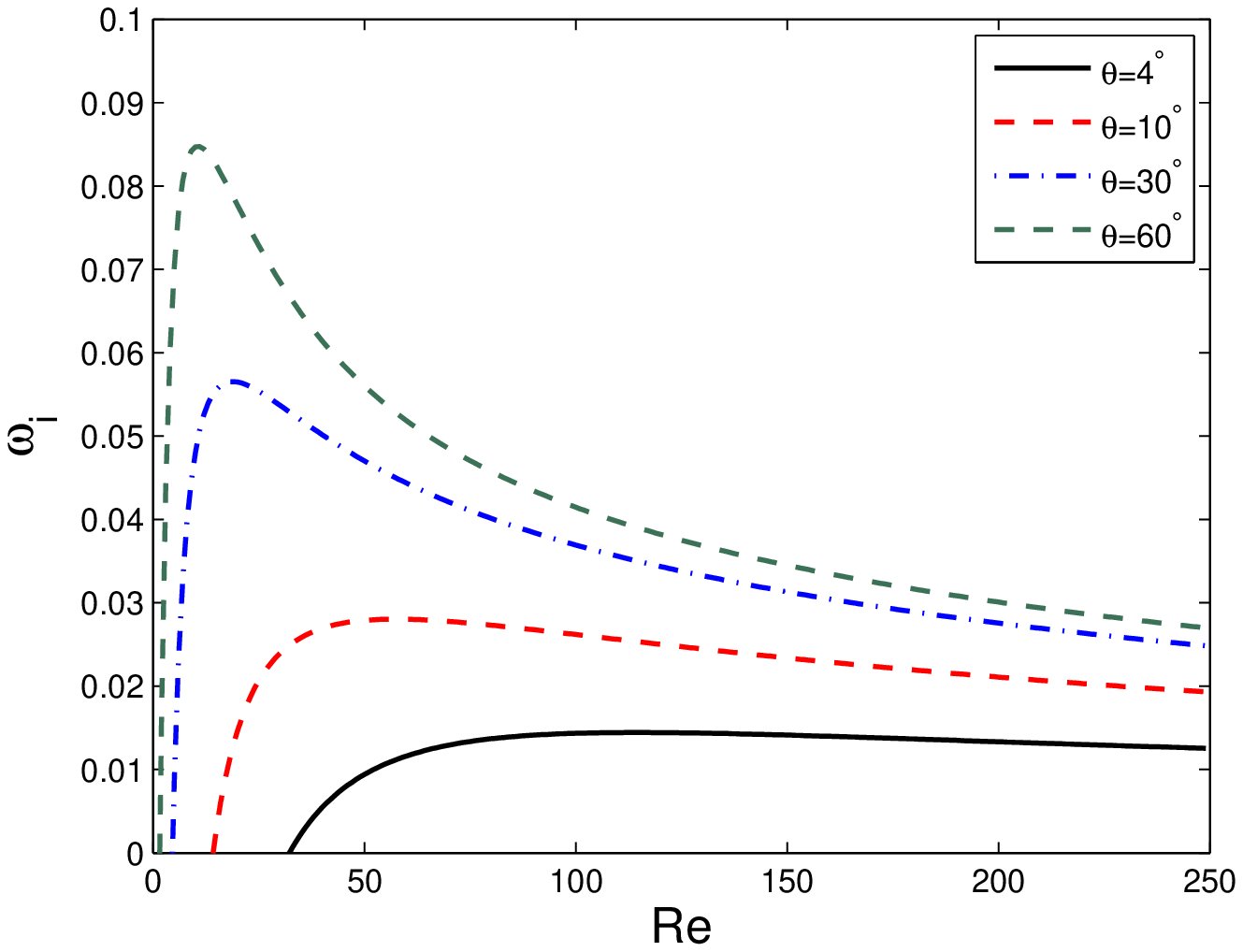}}
	\end{center}
	\caption{Inertial effects on the growth rate for different values of (a) structural rigidity $\alpha$ for $\beta=0$, $\gamma=0.03$ \& $\theta=10^\circ$ , (b) compressive force $\beta$ for $\alpha=0.1$, $\gamma=0.03$ \& $\theta=10^\circ$, (c) uniform mass per unit length $\gamma$ with $\alpha=0.1$, $\beta=0$ \& $\theta=10^\circ$ and (d) inclination angle $\theta$ with $\alpha=0.1$, $\beta=0$ \& $\gamma=0.03$. The constant parameters are $k=0.5$ and $\beta_0=1$.}\label{f6}
\end{figure}

The effect of inertia on the growth rate of the most unstable mode in the presence of floating plate is demonstrated in Figs.~\ref{f4}(a) and (b) for two different range of Reynolds number such as (a) $Re<1000$ and (b) $Re\geq1000$, respectively. The behaviour of growth rate curves in both the cases are comparable for each fixed Reynolds number and follow the conclusions of Fig.~\ref{f3}. However, as the Reynolds number is raised, inertia causes a decrease in the growth rate until a critical wavenumber, and after that, an exchange of instability occurs for shorter waves. Longwave instability is observed for each $Re$ values, and the corresponding instability is weaker for less viscous fluid (larger $Re$). However, it is not analogous at shorter-wavelengths (after a critical $k$) due to the gradual increase in the inertial effect.

The influences of viscous and inertial forces on the growth rate of the dominant mode for flexural flow with floating plate is elaborated in Fig.~\ref{f6} by fixing the wavenumber $k = 0.5$. It is clear from Fig.~\ref{f4} that the maximum growth rate of the disturbance is occurring near about $k = 0.5$. Influence of different plate parameters like structural rigidity $\alpha$, compressive force $\beta$, uniform mass per unit length $\gamma$ and inclination angle $\theta$ are also taken into account in Figs.~\ref{f6}(a), (b), (c) and (d), respectively. The results for a clean interface (without plate) are shown in Figs.~\ref{f6}(a) and (c) as a bold dashed-dotted line in order to show the stabilizing effect of parameters such as $\alpha$ and $\gamma$ for comparison, respectively. In each case, the effect of the inertia is twofold. From Figs.~\ref{f6}(a), (b), (c) and (d), it is very clear that the growth rate increases initially with respect to $Re$ and attains maximum, then decreases gradually for increasing values of $Re$.

The growth rate of the most unstable mode increases faster upto moderate range of $Re$ ($Re \leq 50$), promoting the instability. However, after the threshold $Re \approx 50$ a reverse trend is observed and at high $Re$ the variation of the growth rate is poor. Structural rigidity ($\alpha$) and uniform mass per unit length ($\gamma$) of the plate lower the growth rate of the instability significantly (Figs.~\ref{f6}(a), (c)), promoting the stability of the flow. Whereas, compressive force of the plate ($\beta$) and inclination angle ($\theta$) of the system enhance the growth rate for all Reynolds numbers (Figs.~\ref{f6}(b), (d)).

\begin{figure}
	\begin{center}
		\subfigure[]{\label{f7a}\includegraphics*[width=6.6cm]{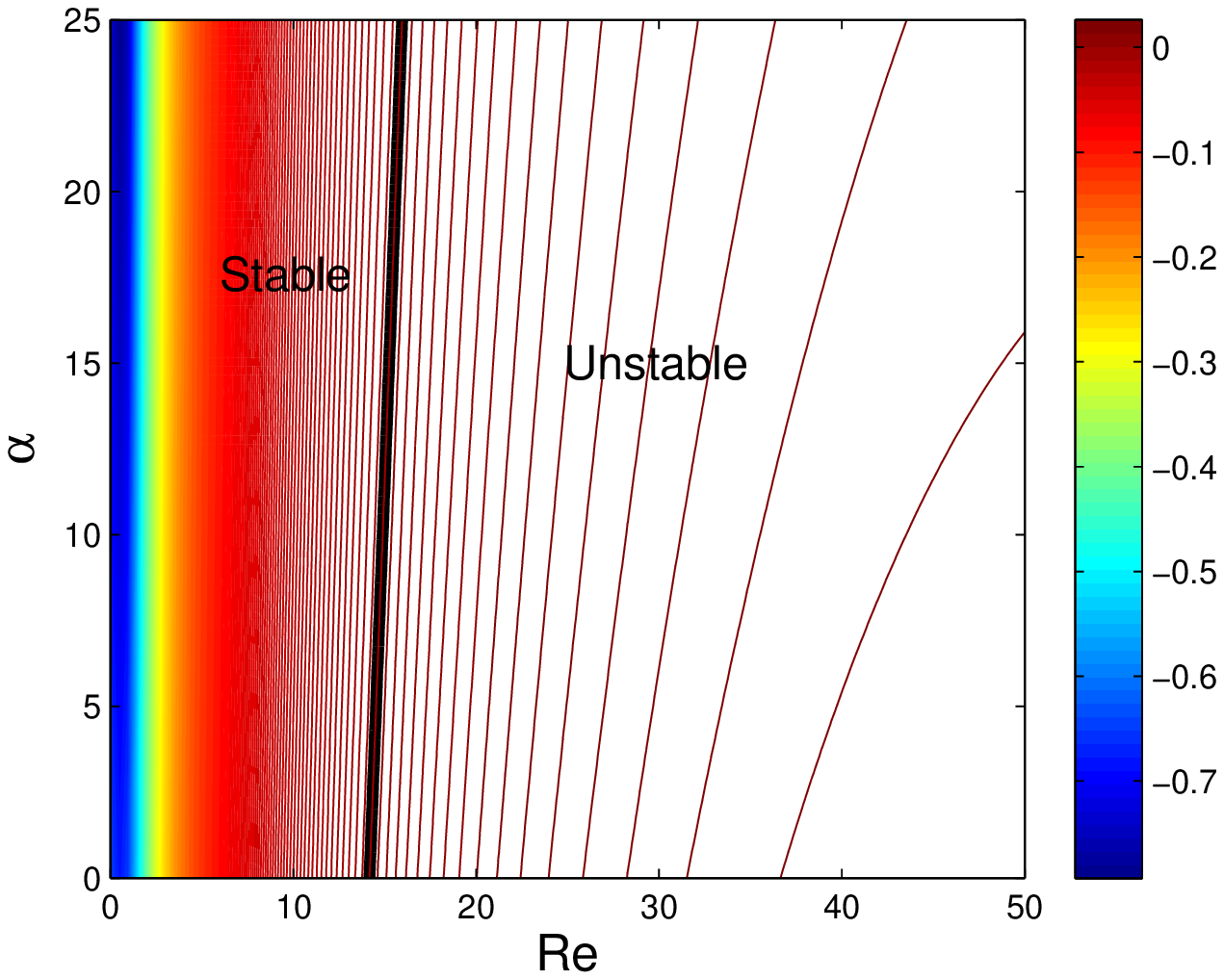}}
		\subfigure[]{\label{f7b}\includegraphics*[width=6.6cm]{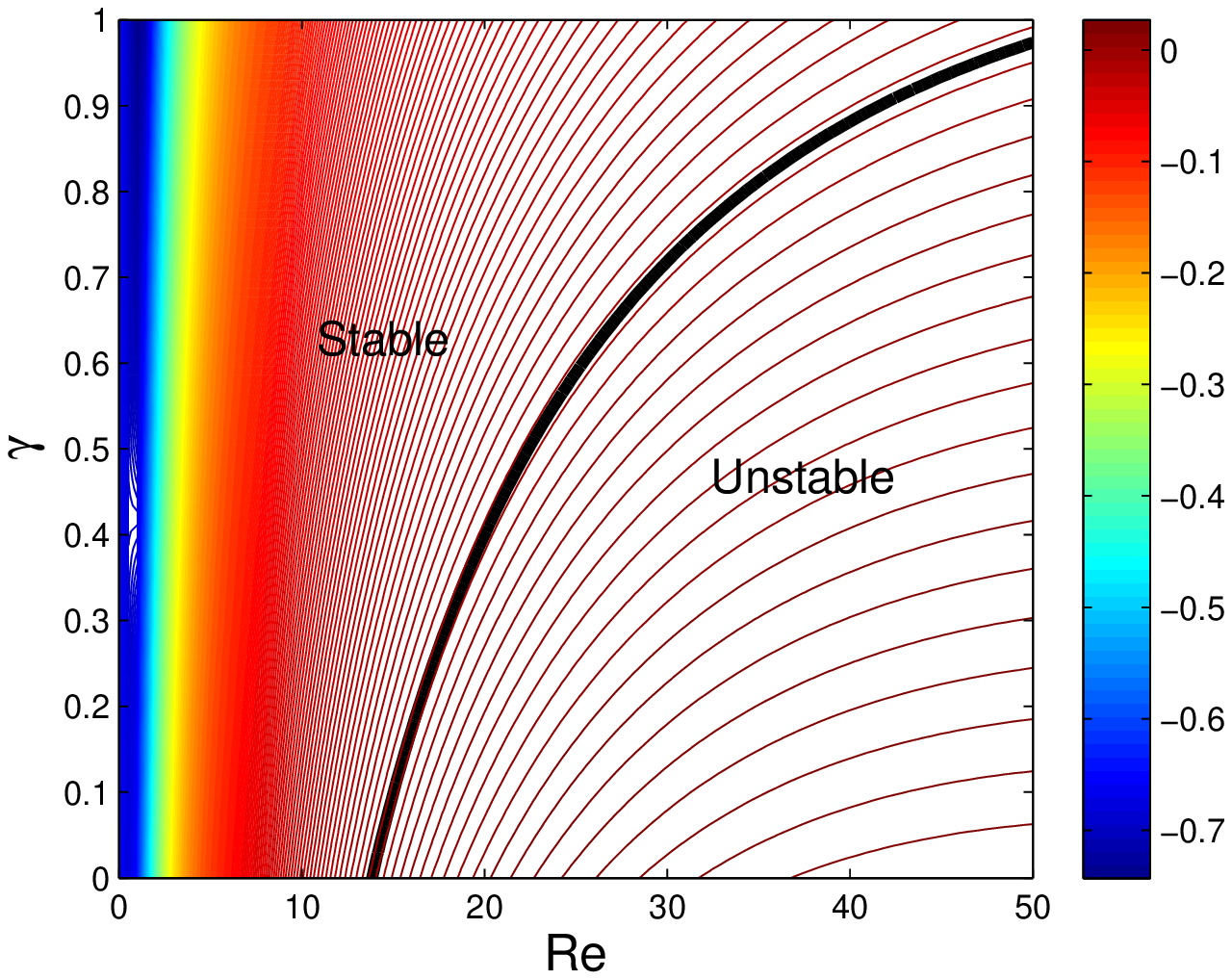}}
	\end{center}
	\caption{ Contour plot of growth rates showing the impact of (a) structural rigidity $\alpha$ with $\beta=0$ \&$\gamma=0.03$ and (b) uniform mass per unit length $\gamma$ with $\alpha=0.1$ \& $\beta=0$.     The solid lines denote the neutral stability curve ($\omega_i = 0$) and the constant parameters are $\theta=10^\circ$ \& $k=0.5$.}\label{f7}
\end{figure}
\begin{figure}
	\begin{center}
		\subfigure[]{\label{f8a}\includegraphics*[width=6.6cm]{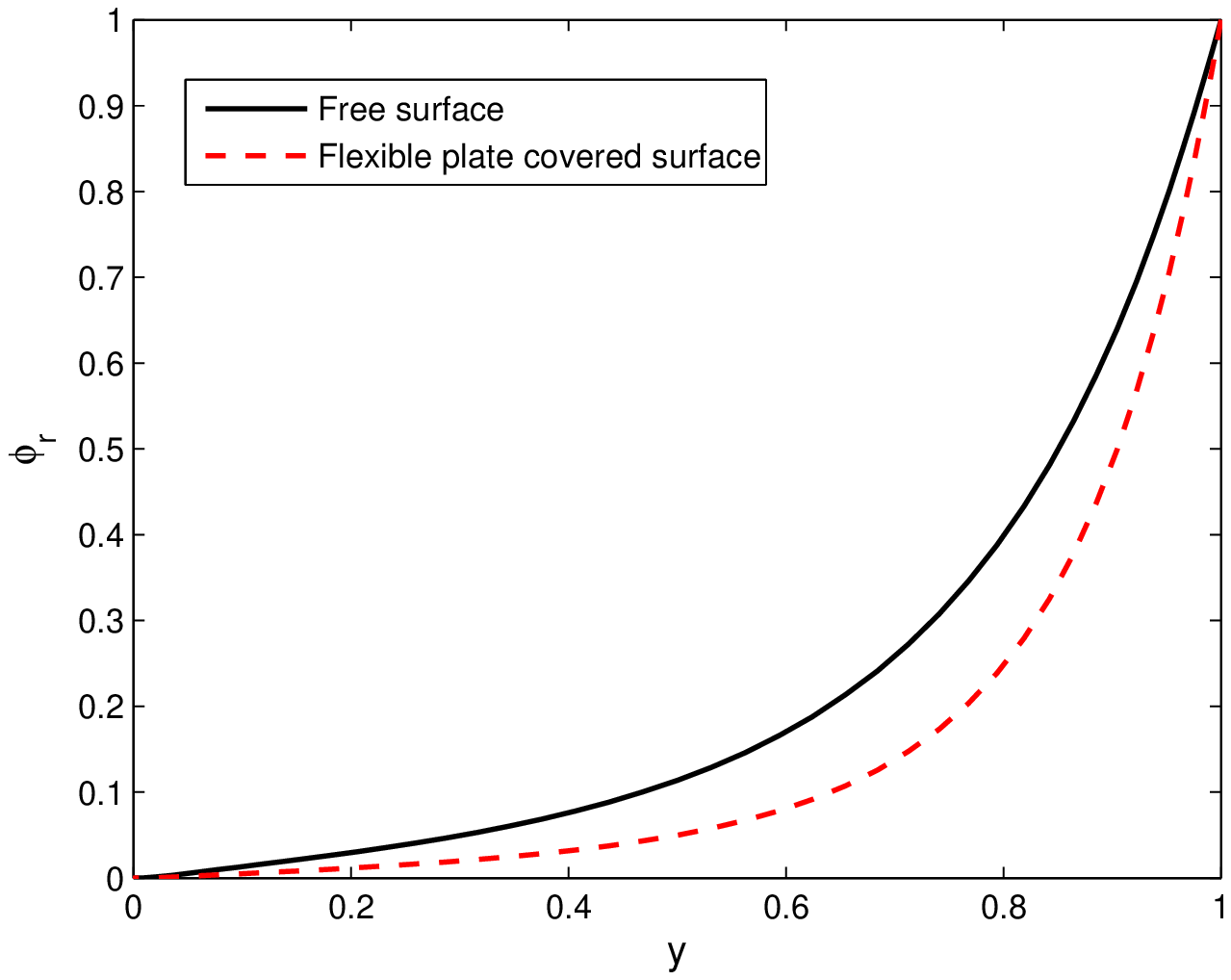}}
		\subfigure[]{\label{f8b}\includegraphics*[width=6.6cm]{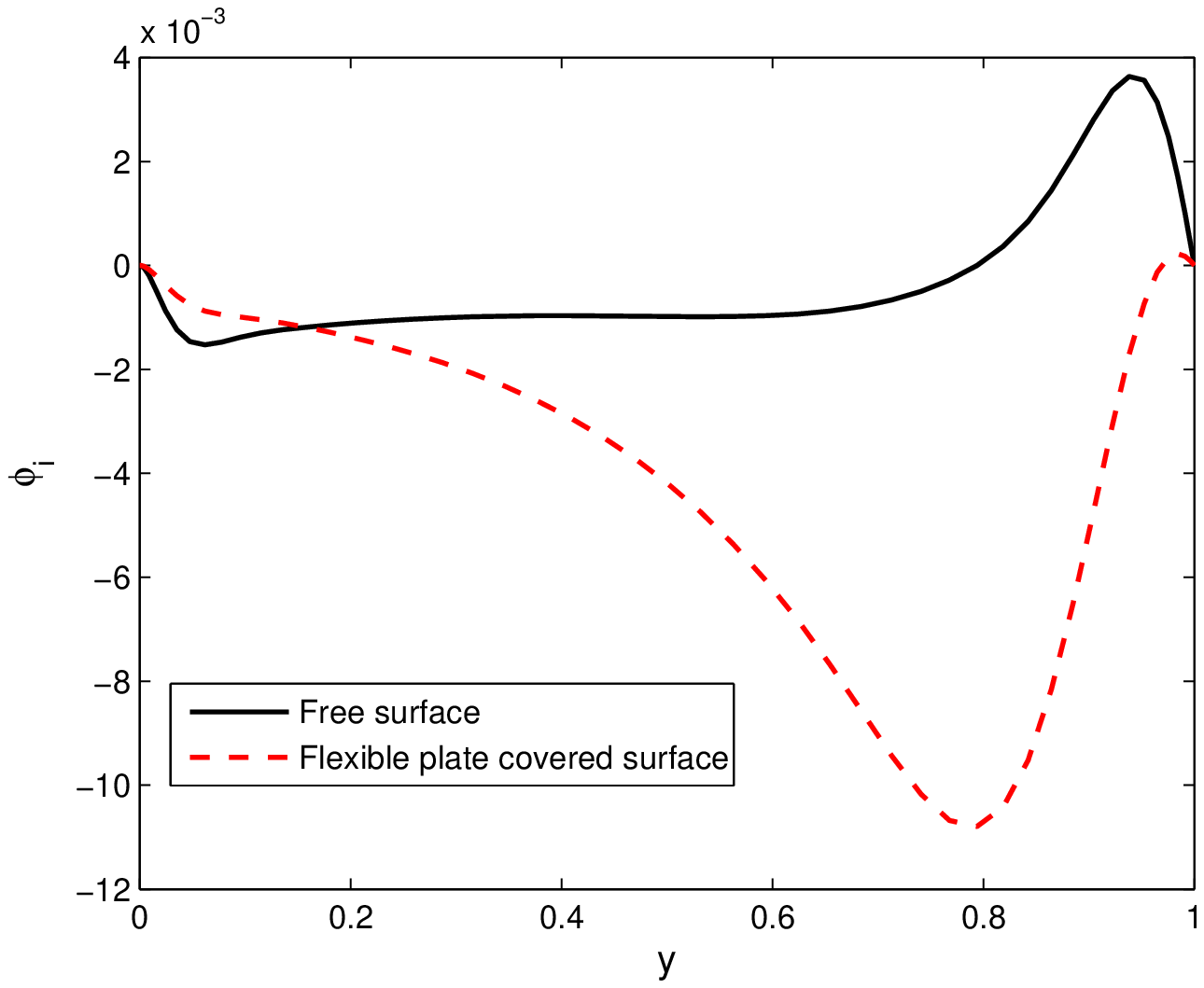}}
		\subfigure[]{\label{f8c}\includegraphics*[width=6.6cm]{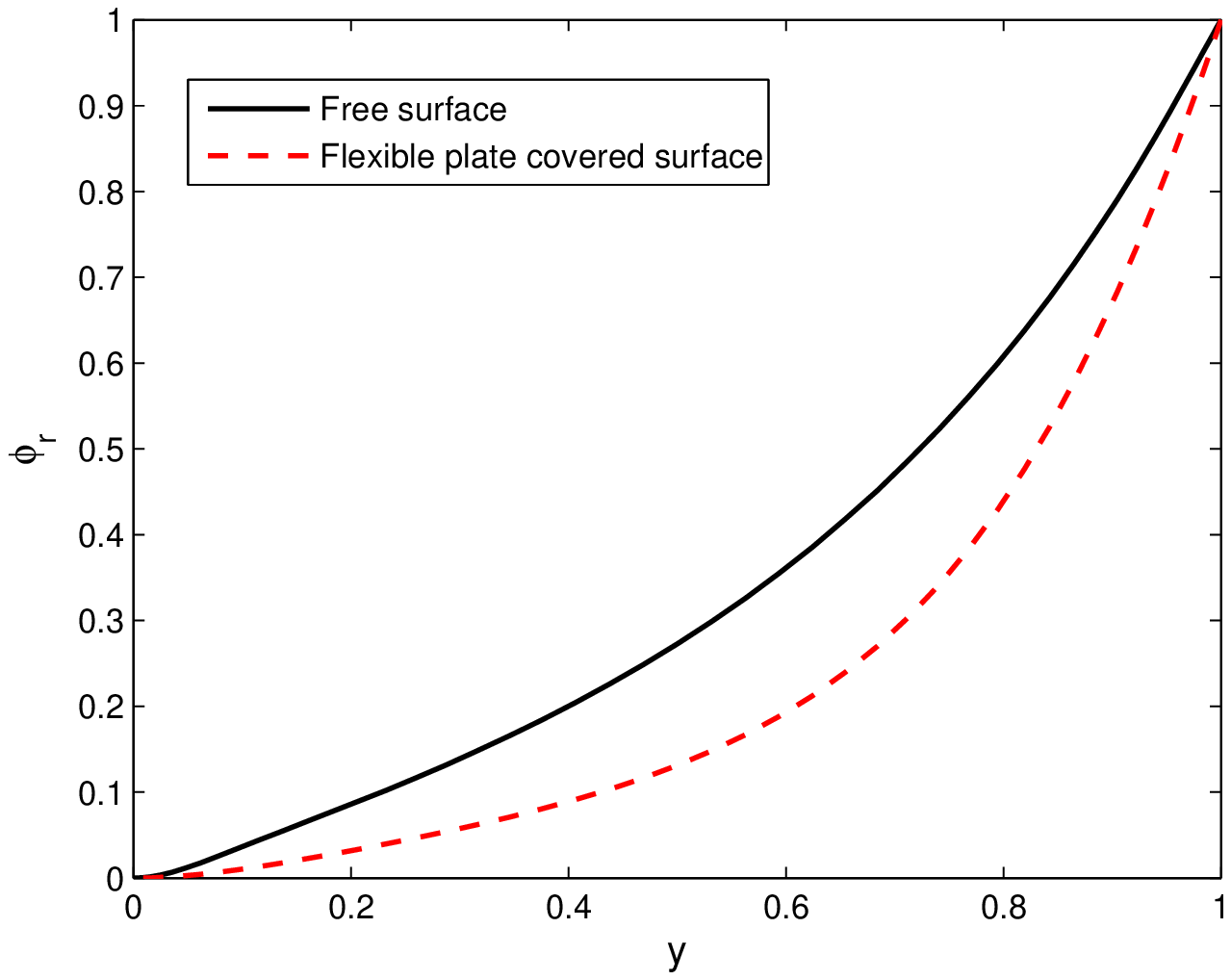}}
		\subfigure[]{\label{f8d}\includegraphics*[width=6.6cm]{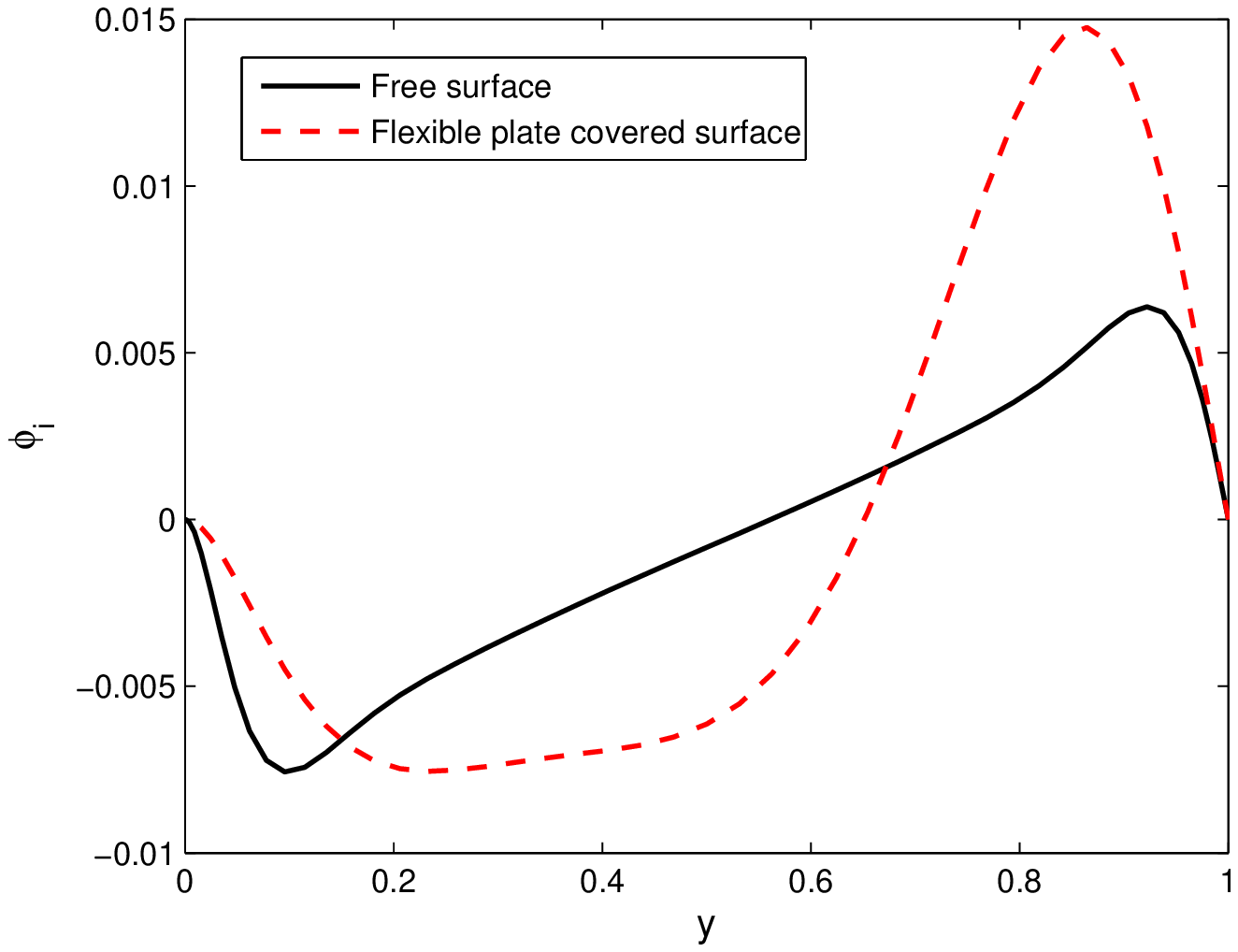}}
	\end{center}
	\caption{Real and imaginary part of eigenfunction $\phi$ associated with the most unstable mode of free surface ($\alpha=0$, $\beta=\beta_0 (Re^2\,\sin\theta)^{-1/3}$) and plate covered ($\alpha=0.1$, $\beta=0$ and $\gamma=0.03$) flow when $Re=722$ (for (a) and (b)) and $Re=50$ (for (c) and (d)). The constant parameters are $k=1.04$, $\beta_0=4280$ and $\theta=4^\circ$; The solid line in the figures (a) and (b) represents the results of \cite{chin1986gravity}.}\label{f8}
\end{figure}

In Figs.~\ref{f7}(a) and (b), the contours of the dimensionless growth rate are plotted by varying the structural rigidity $\alpha$ and uniform mass per unit length $\gamma$, respectively to get an overall knowledge of the parameter dependency for the case of zero compressive force. Here, the neutral curve ($\omega_i = 0$) is indicated by a bold solid line. The area underneath the neutral curve in Figs. \ref{f7}(a) and (b), correspond to instability. As per the results of Fig.~\ref{f7}(a), increasing $\alpha$ has weak effect on the stability of the flow. However, the stability window in the $Re - \gamma$ plane increases for larger values of $\gamma$ (Fig.~\ref{f7}(b)).

Typical structure of the physical perturbations corresponding to the most dominant mode (surface/gravity mode) are grouped in Fig.~\ref{f8} for both the free surface flow as well as the flow covered by a flexible plate. The real and imaginary parts of the eigenfunction $\phi(y)$ are plotted as a function of the normal coordinate for two different Reynolds numbers. The mode shape does not change drastically because of the presence of a floating elastic plate. However, the convexity of $\phi_r$ profile has changed, and the shape of $\phi_r, \phi_i$ is much deformed. The perturbed flow shear rate is altered by the floating plate, which changes the stability of the flow. 

\begin{figure}
	\begin{center}
		\subfigure[]{\label{f9a}\includegraphics*[width=6.6cm]{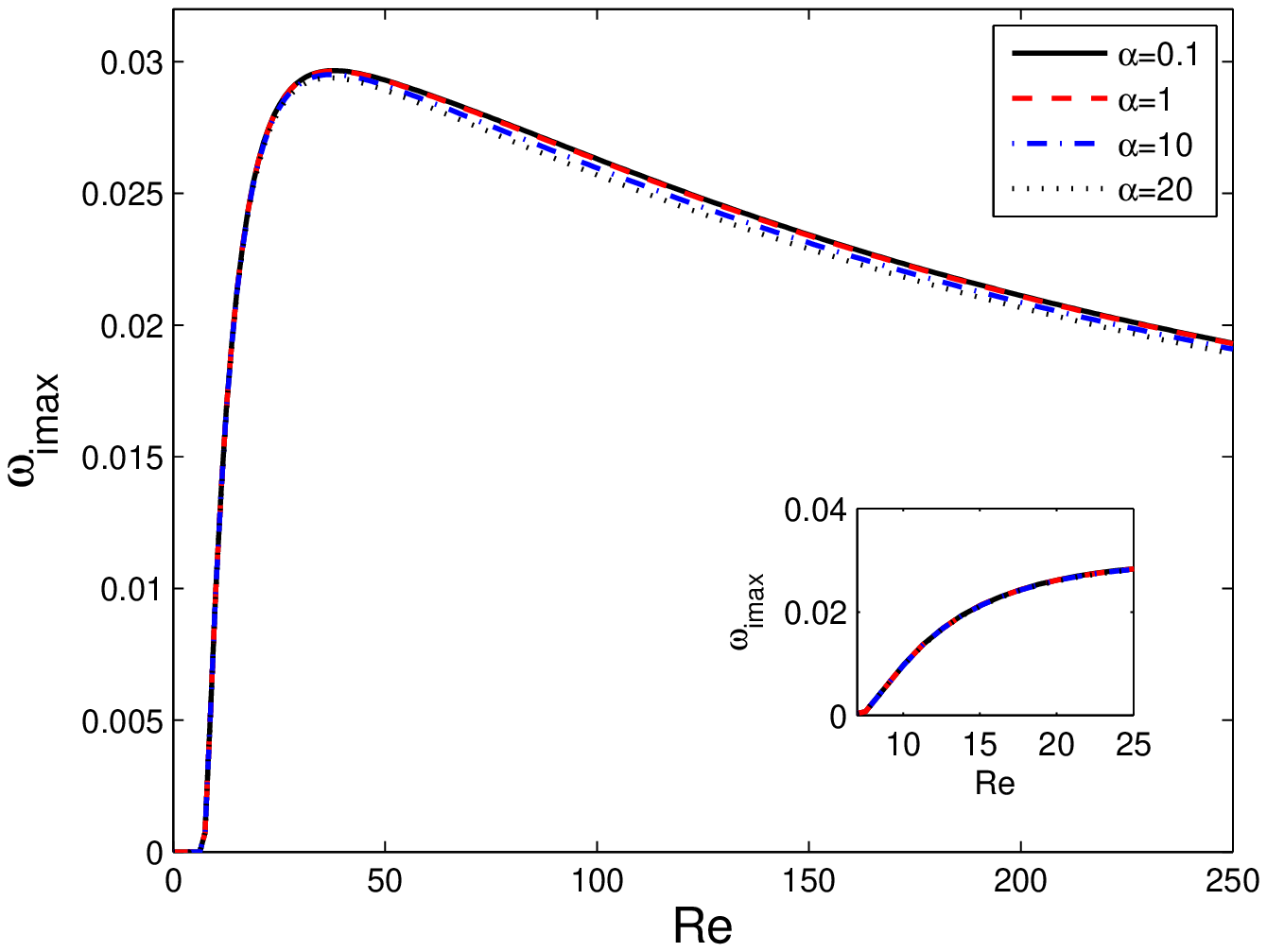}}
		\subfigure[]{\label{f9b}\includegraphics*[width=6.6cm]{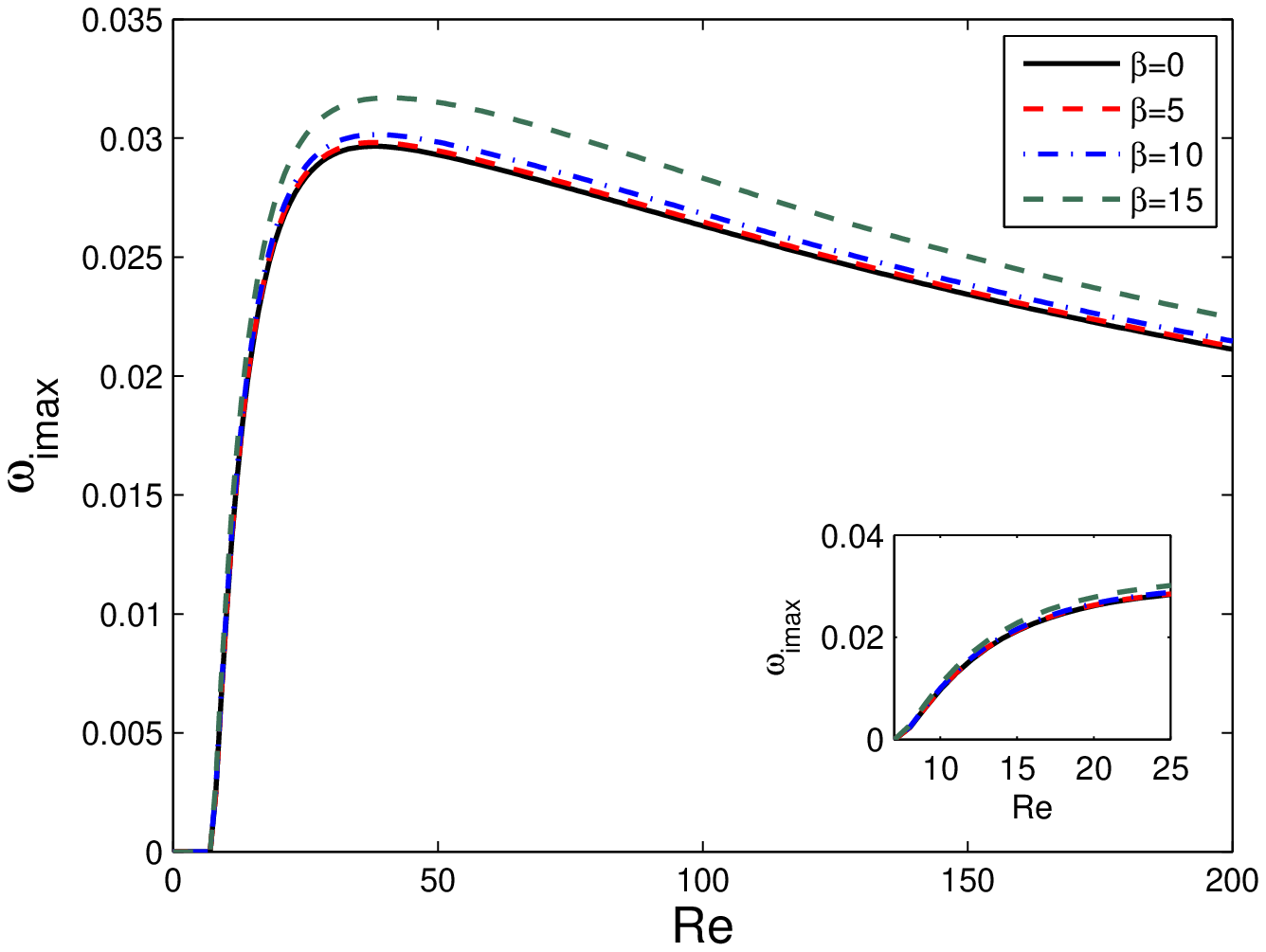}}
		\subfigure[]{\label{f9c}\includegraphics*[width=6.6cm]{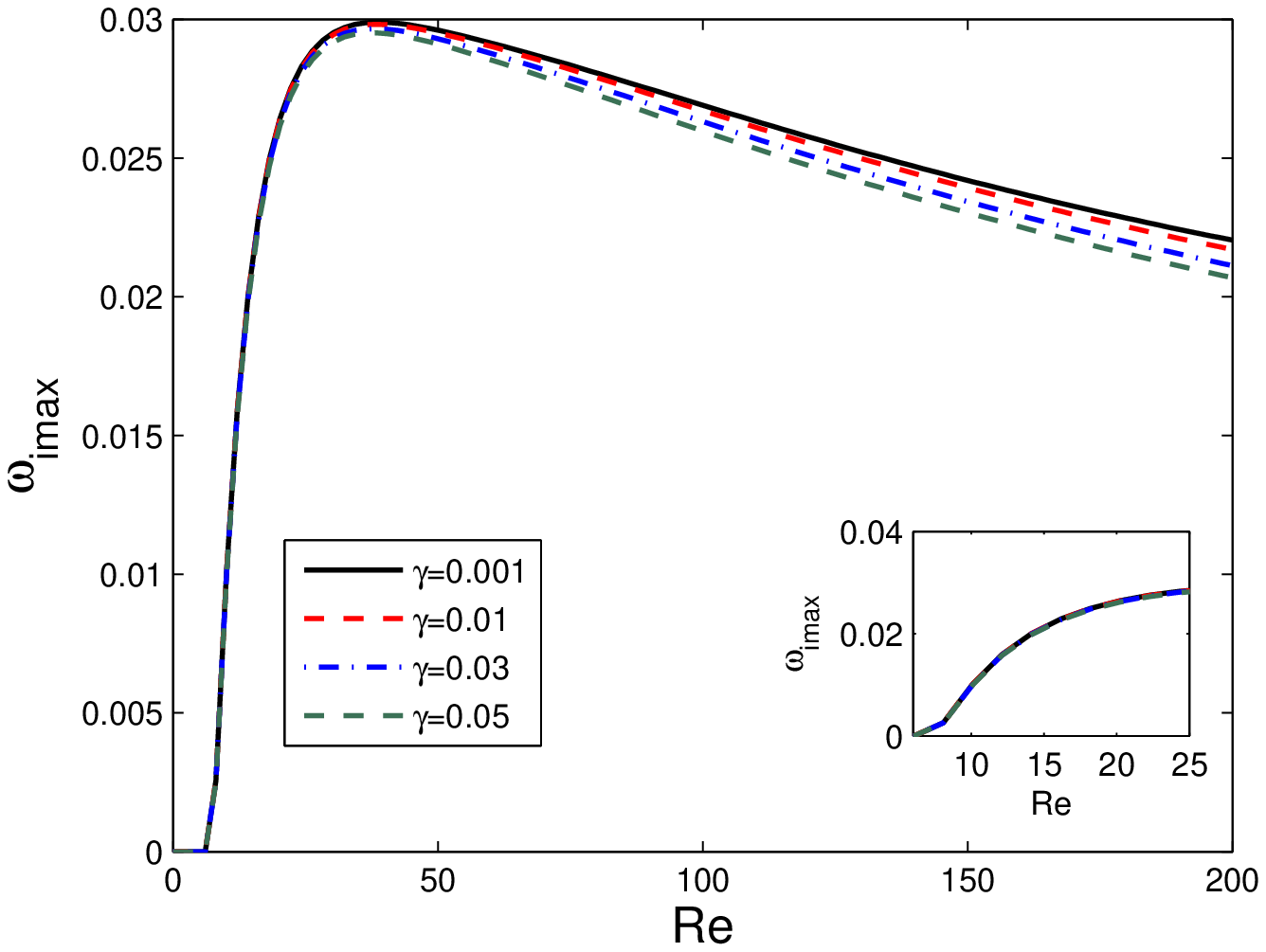}}
		\subfigure[]{\label{f9d}\includegraphics*[width=6.6cm]{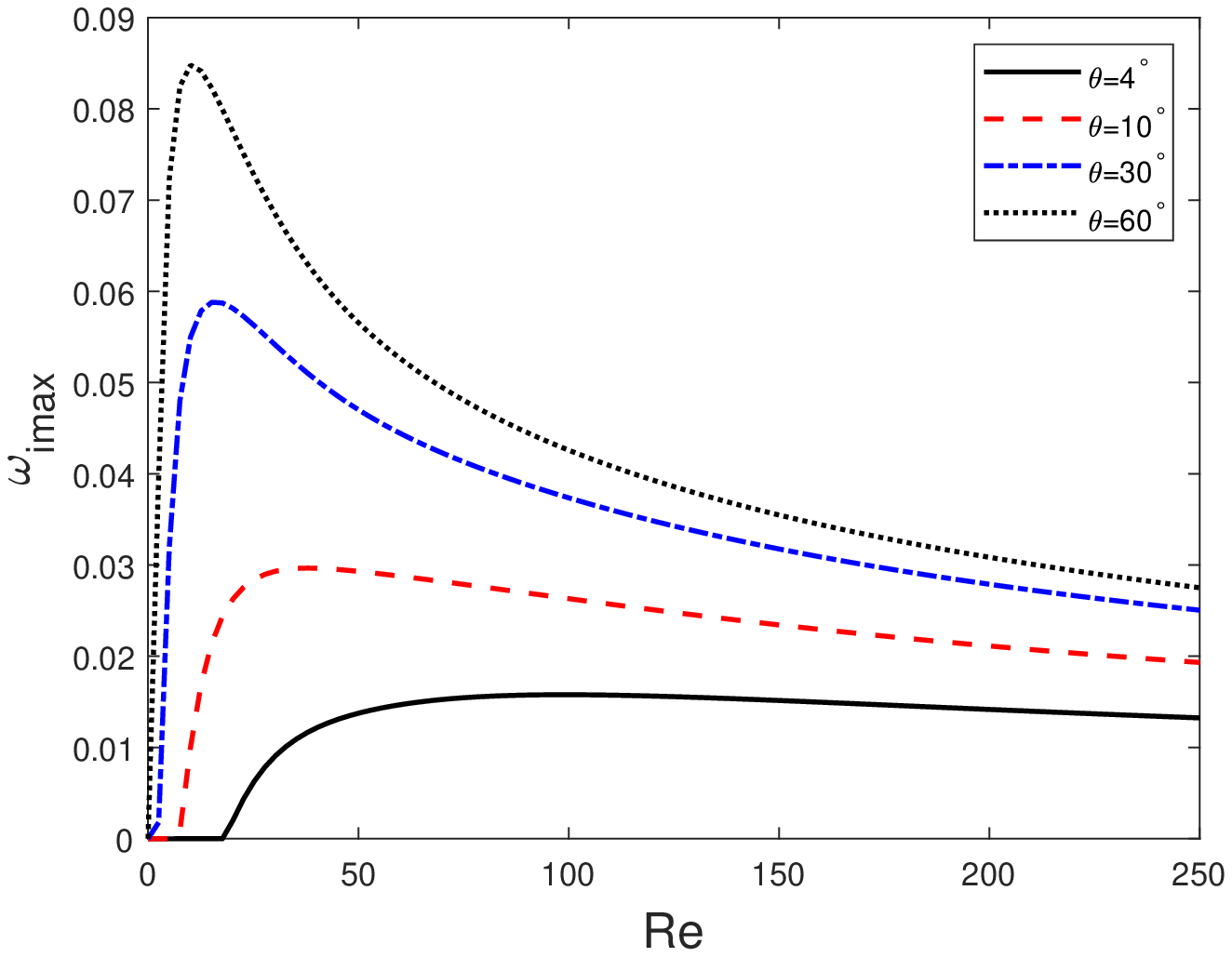}}
	\end{center}
	\caption{Maximum growth rate $\omega_{\mathsf{i}max}$ against the Reynolds number $Re$ showing the effect  of (a) structural rigidity $\alpha$ with $\beta=0$, $\gamma=0.03$ \& $\theta=10^\circ$, (b) compressive force $\beta$ with $\alpha=0.1$, $\gamma=0.03$ \& $\theta=10^\circ$, (c) uniform mass per unit length $\gamma$ with $\alpha=0.1$, $\beta=0$ \& $\theta=10^\circ$ and (d) inclination angle $\theta$ with $\alpha=0.1$, $\beta=0$ \& $\gamma=0.03$.}\label{f9}
\end{figure}
\begin{figure}
	\begin{center}
		\subfigure[]{\label{f10a}\includegraphics*[width=6.6cm]{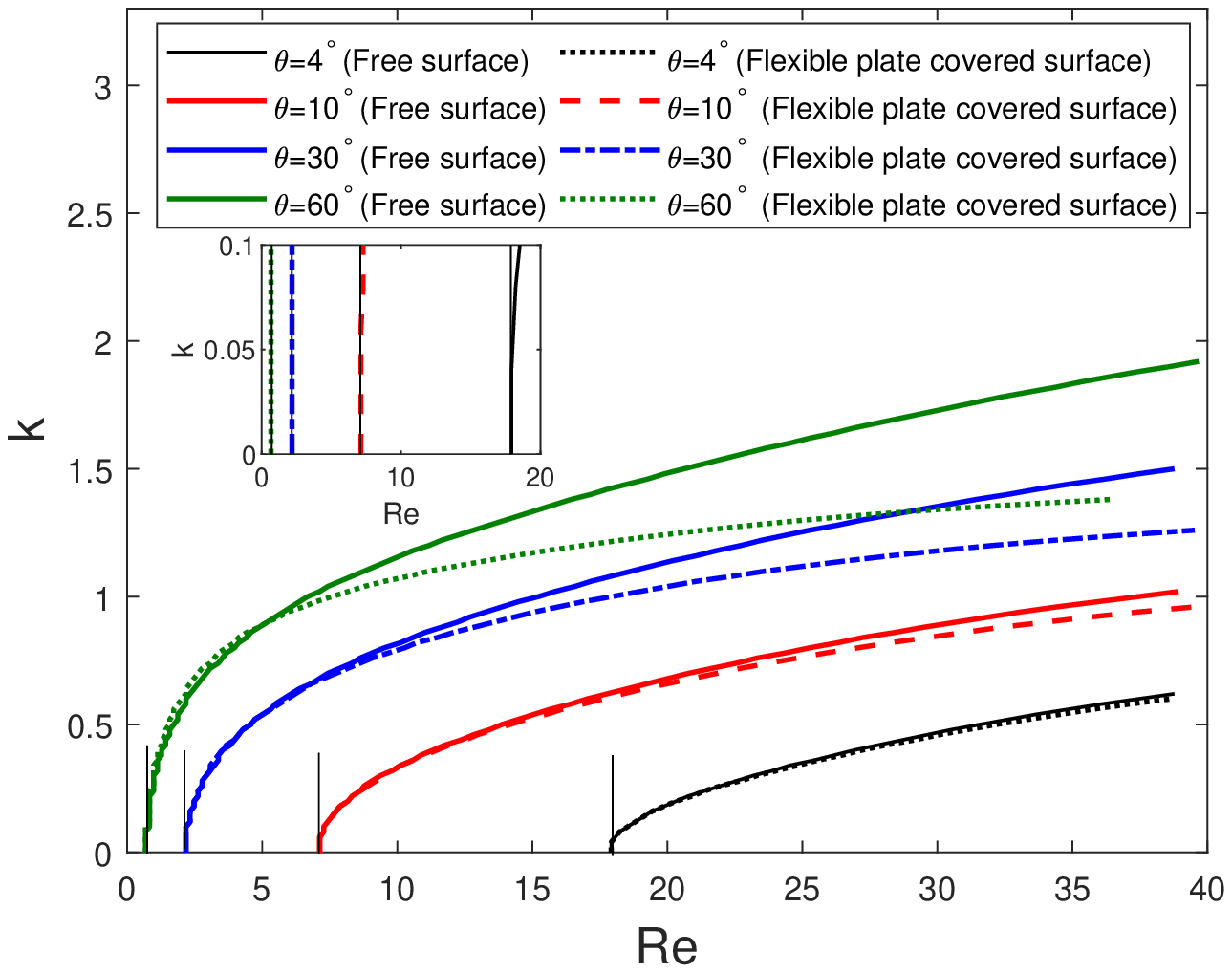}}
		\subfigure[]{\label{f10b}\includegraphics*[width=6.6cm]{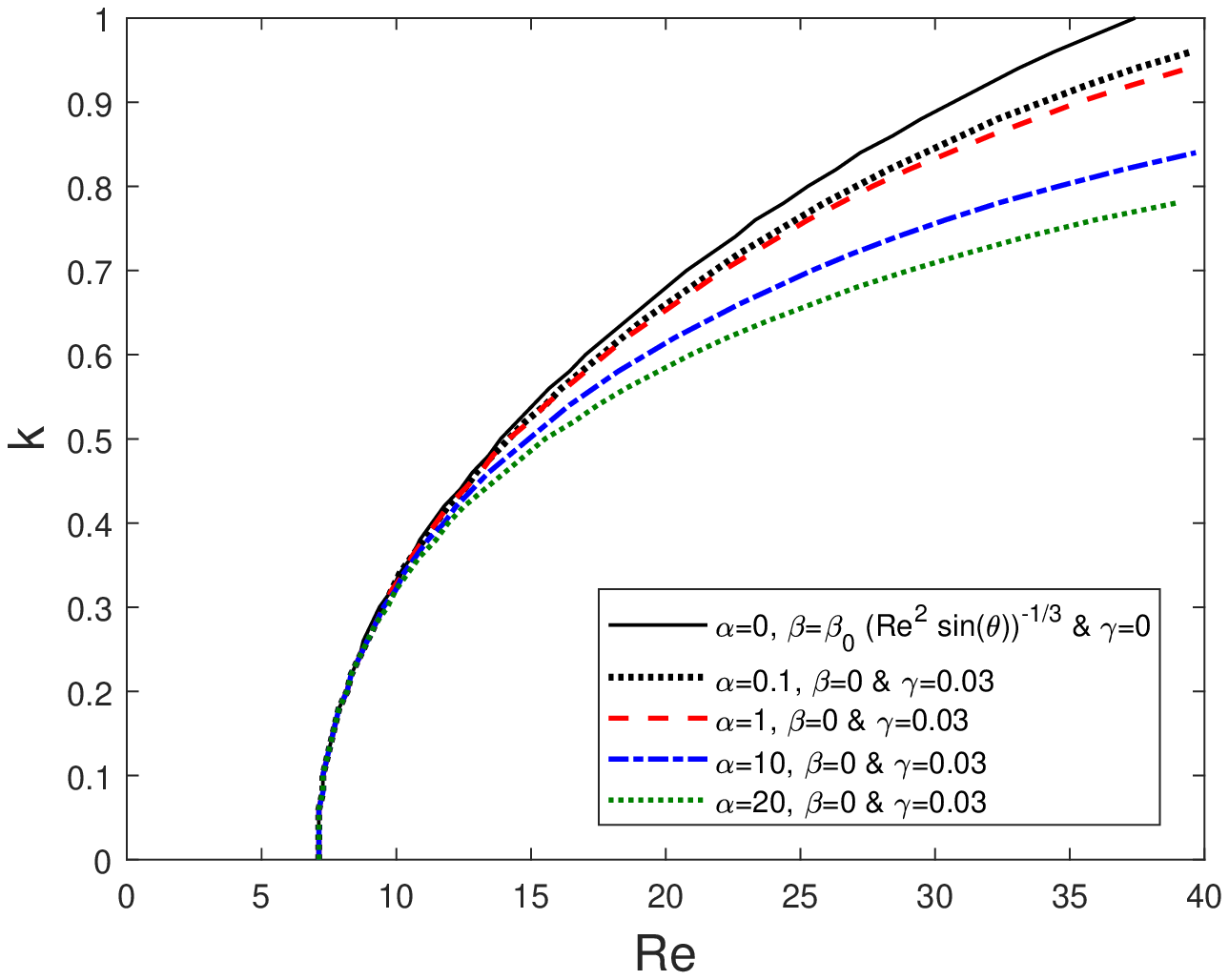}}
		\subfigure[]{\label{f10c}\includegraphics*[width=6.6cm]{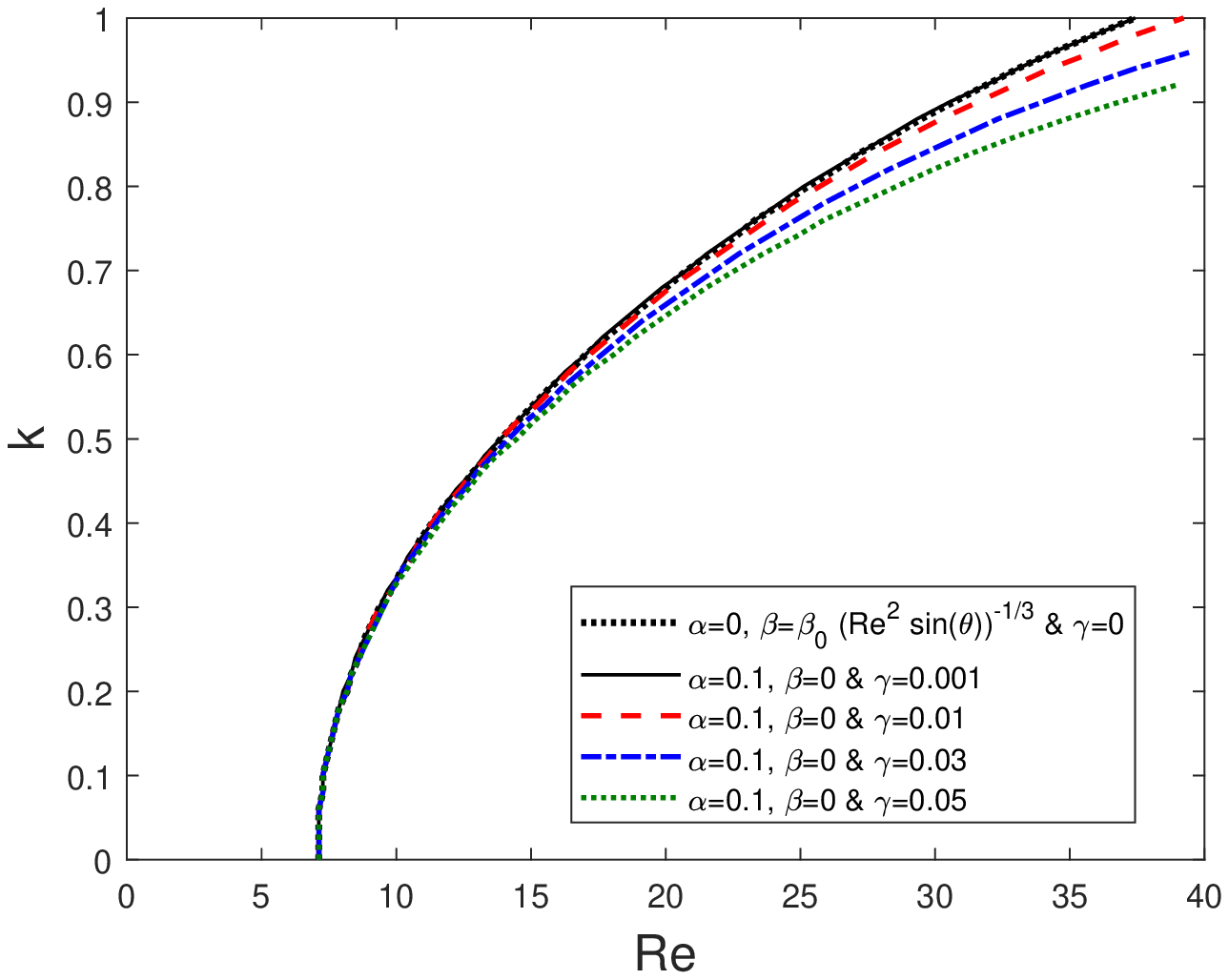}}
		\subfigure[]{\label{f10d}\includegraphics*[width=6.6cm]{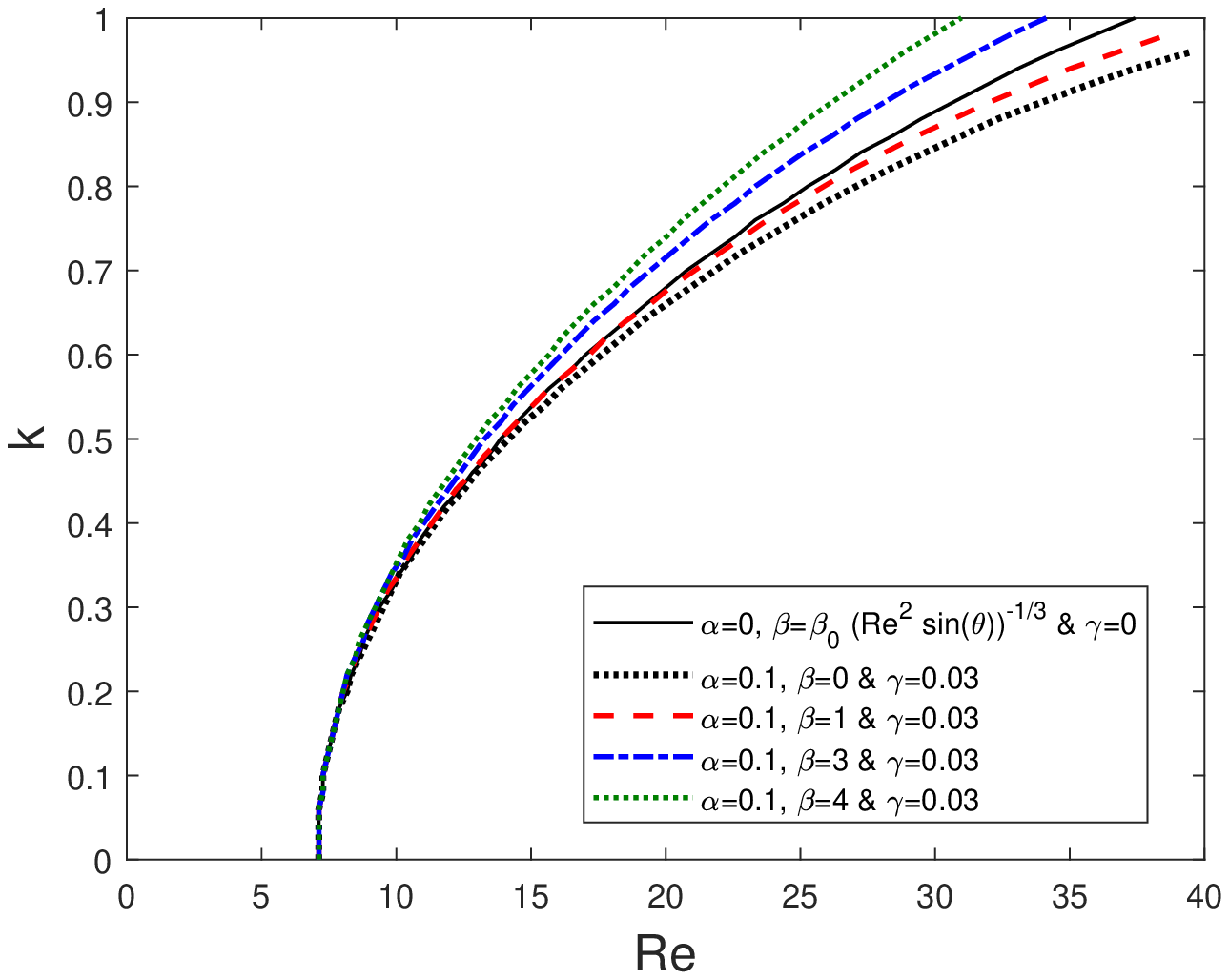}}
	\end{center}
	\caption{Neutral stability curve for the flow with floating elastic plate with different values of (a) inclination angle $\theta$ with $\alpha=0.1$, $\gamma=0.03$ \& $\beta=0$, where solid lines denote the critical Reynolds number ($Re_c$) obtained by long wave analysis, (b) structural rigidity $\alpha$ with $\theta=10^\circ$, $\gamma=0.03$ \& $\beta=0$, (c) uniform mass per unit length $\gamma$ with $\theta=10^\circ$, $\alpha=0.1$ \& $\beta=0$ and (c) compressive force $\beta$ with $\theta=10^\circ$, $\alpha=0.1$ \& $\gamma=0.03$.}
	\label{f10}
\end{figure}

The maximum growth rate of dominant disturbance over the range of unstable wavenumbers, against the Reynolds number, is plotted for different values of structural rigidity $\alpha$, uniform mass per unit length $\gamma$, compressive force $\beta$ and inclination angle $\theta$ in Figs.~\ref{f9}(a), (b), (c) and (d), respectively. In each configuration, $Re$ has a non-monotonic influence on the maximum growth rate. When the inertial force is not strong enough instability increases with $Re$. However, the scenario changes after a threshold value of $Re$ ($Re \approx 40$), where the maximum growth rate attains its highest value and then starts decreasing for larger values of $Re$. The physical parameters like structural rigidity $\alpha$ and uniform mass per unit length $\gamma$ tend to reduce the maximum growth rate  (Figs.~\ref{f9}(a) and (b)). Whereas the quantities like compressive force and inclination angle rise the maximum growth rate (Figs.~\ref{f9}(c) and (d)). In consequence, one can use the parameter range appropriately to enhance or suppress the instability in the film flow.  

In Figs.~\ref{f10}(a), (b), (c) and (d), the neutral stability curves are illustrated for different values of inclination angle $\theta$, structural rigidity $\alpha$, uniform mass per unit length $\gamma$ and compressive force $\beta$, respectively. Here, the region beneath each curve corresponds to instability. As the inclination angle is raised and the film tends to become upright, the neutral curve is shifted upwards, widening the range of unstable wavenumbers at any given Reynolds number (Fig.~\ref{f10}(a)). Furthermore, the critical Reynolds number diminished and the flow befall more unstable. The results shown with black lines near the critical $Re$ in Fig.~\ref{f10}(a) are the predictions of the long-wave approximation ($k\to 0$) expressed in Section \ref{SAR}, (Eq.~\eqref{e90}). As observed in both Figs.~\ref{f10}(b) and (c) respectively, the unstable wavenumber bandwidth diminishes for larger values of $\alpha$ and $\gamma$. On the other hand, as observed in Fig.~\ref{f10}(d), the unstable mode bandwidth increases for an increase in the value of $\beta$. However, the critical Reynolds number remains unaltered. It is also to conclude that inclination angle and compressive force influence all the unstable perturbation waves, while the structural rigidity and uniform mass per unit length can only suppress shorter wave instability.

\subsection{Results with small aspect ratio $(\epsilon \ll 1)$}

\begin{figure}
	\begin{center}
		\subfigure[]{\label{f11a}\includegraphics*[width=6.6cm]{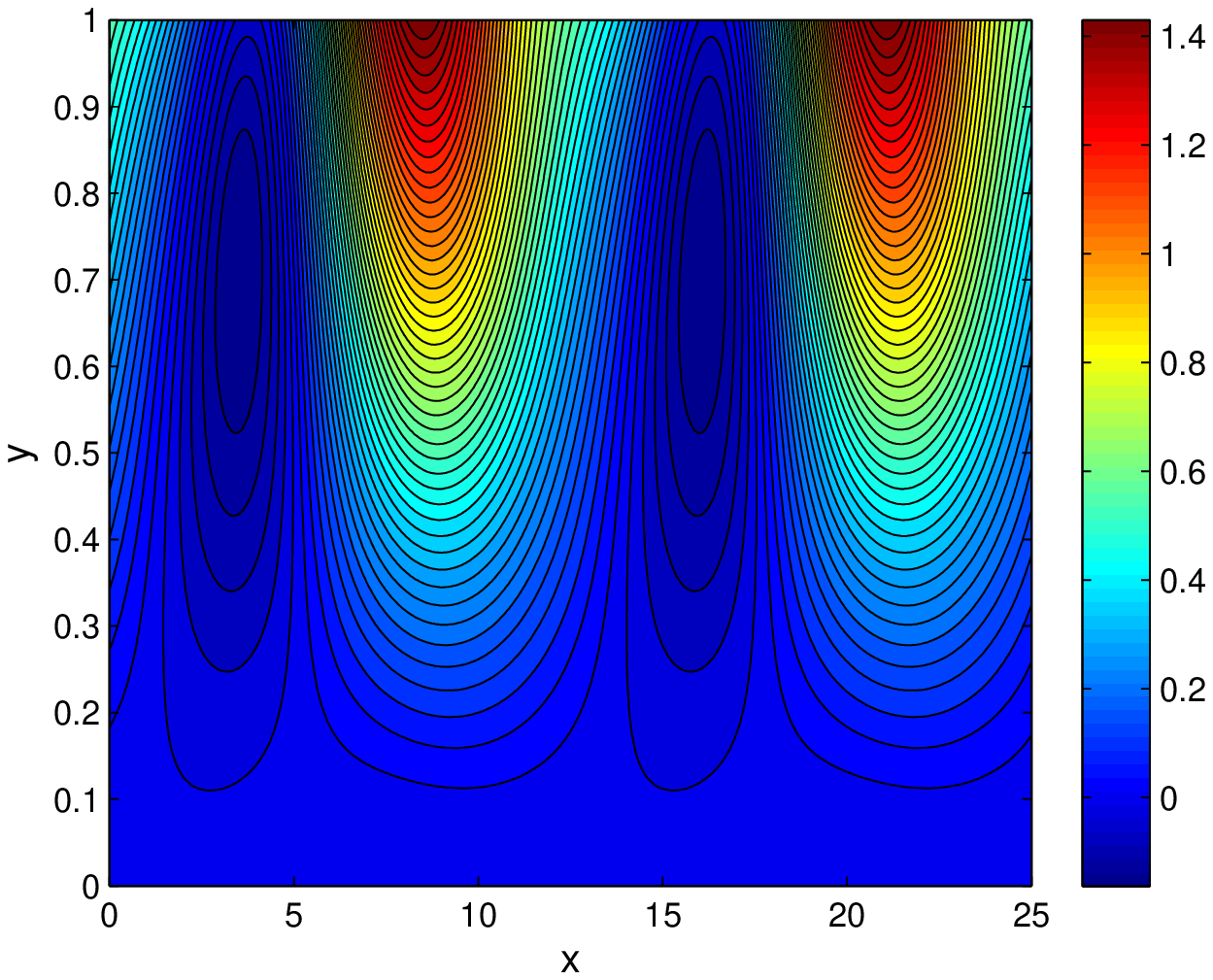}}
		\subfigure[]{\label{f11b}\includegraphics*[width=6.6cm]{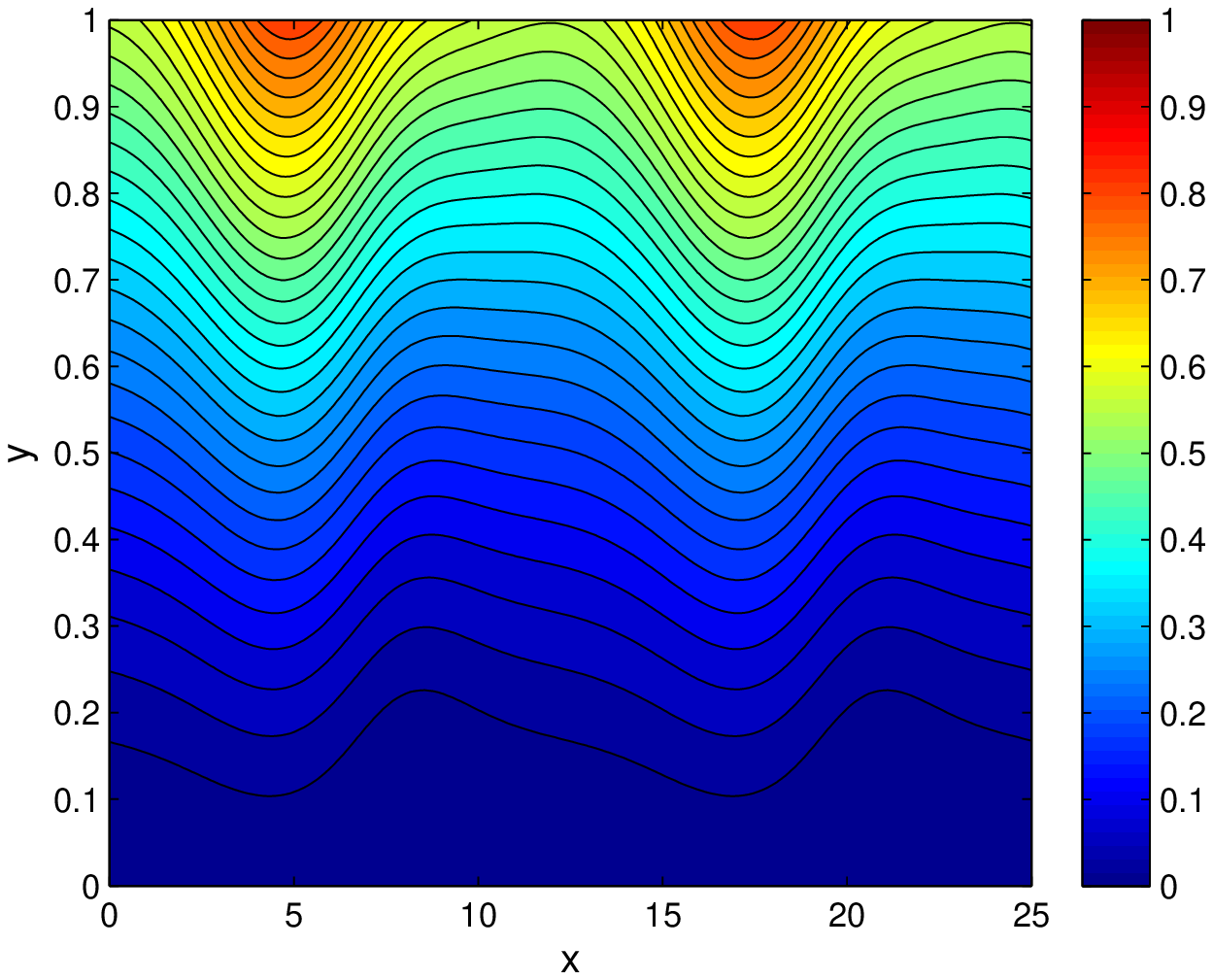}}
	\end{center}
	\caption{Average horizontal velocity perturbation for (a) free surface with $\alpha=0$, $\beta=\beta_0(Re^2\sin\theta)^{-1/3}$
		and $\gamma=0$, and (b) $\alpha=0.1$, $\beta=1$
		and $\gamma=0.03$. The constant parameters are $\beta_0=4280$, $Re=100$, $k=0.5$, $\epsilon=0.01$ and $\theta=10^\circ$.}\label{f11}
\end{figure}

In section \ref{SAR} we have focused on solving the problem analytically assuming that the ratio of the channel width to the wavelength ($\epsilon$) is sufficiently small. Using order analysis, the velocity profiles and critical parameters (like critical Reynolds number and wavenumber) for instability were calculated. Here, corresponding solution behaviours are explored, taking the value of aspect ratio as $\epsilon=0.01$. The physical quantities like velocity profiles, pressure and shearing stress are plotted and analyzed for various structural parameters and flow configurations.

Figs.~\ref{f11}(a) and (b) present the two-dimensional contour plot of horizontal velocity perturbation (averaging over the flow domain) for the classical free-surface flow as well as flexural flow with floating plate, respectively. The core of the vortices corresponds to the maximum value. For each flow system, the maximum occurs near the top surface ($y = 1$, which is respectively the free surface and the flexible plate) providing evidence in favour of a surface mode instability. Towards the rigid wall ($y= 0$), a continuous decrease in the strength confirms the stability of the shear mode. Moreover, the vortices are weak in the presence of a floating elastic plate as compared to that of free-surface flow, and the streamlines are considerably deformed. This deformation suggests a shear rate variation near the film's top surface, which is the source of the stability mechanism for the flow covered by the flexible plate. In contrast to free-surface flow, the additional core (local minimum) between two maximum core vanishes,  the perturbations are ordered and may be in consequence less excited. In view of the above figures, the free surface/flexible plate plays a leading role in the stabilization or destabilization of falling film.  

\begin{figure}
	\begin{center}
		\subfigure[]{\label{f12a}\includegraphics*[width=6.6cm]{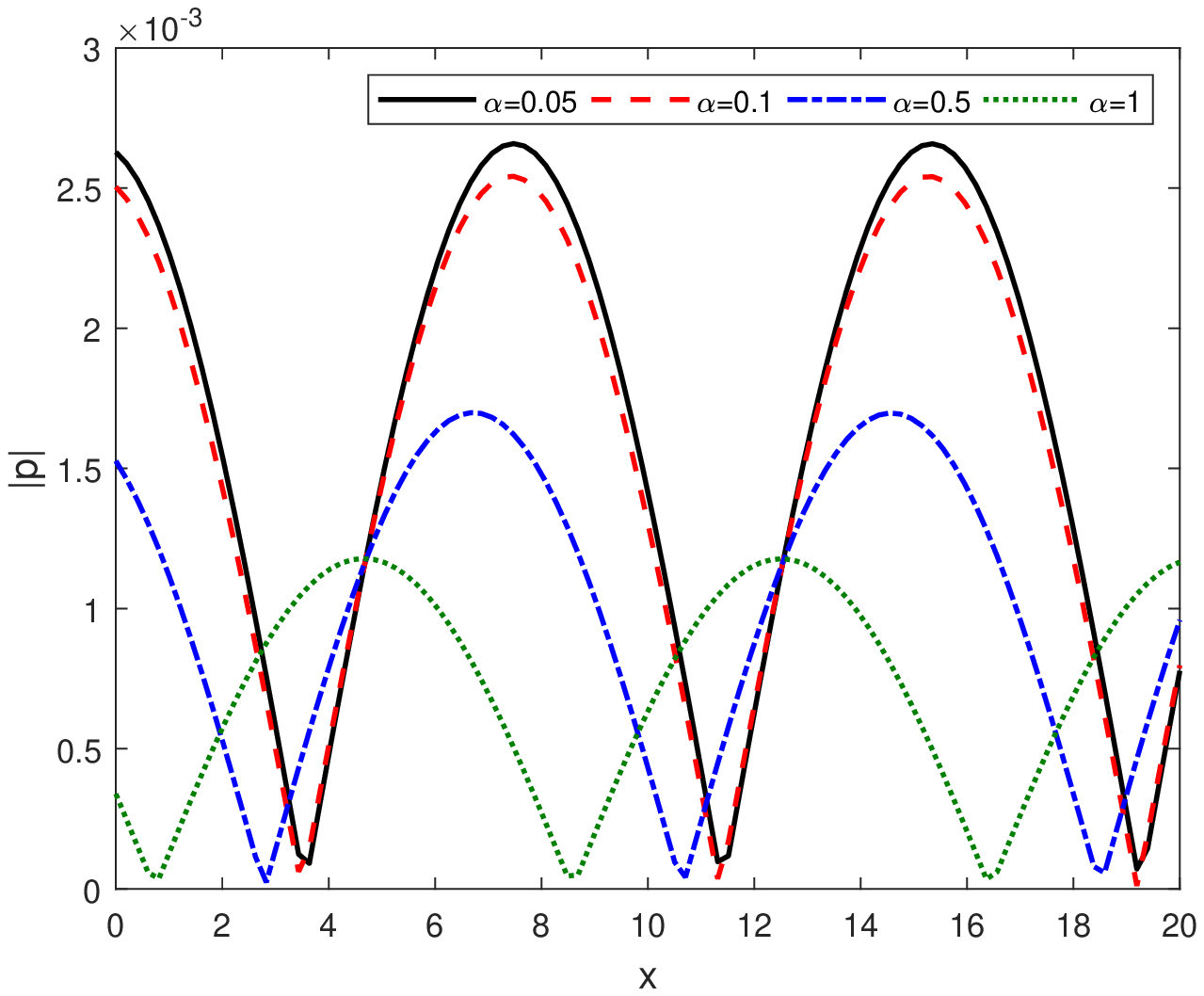}}
		\subfigure[]{\label{f12b}\includegraphics*[width=6.6cm]{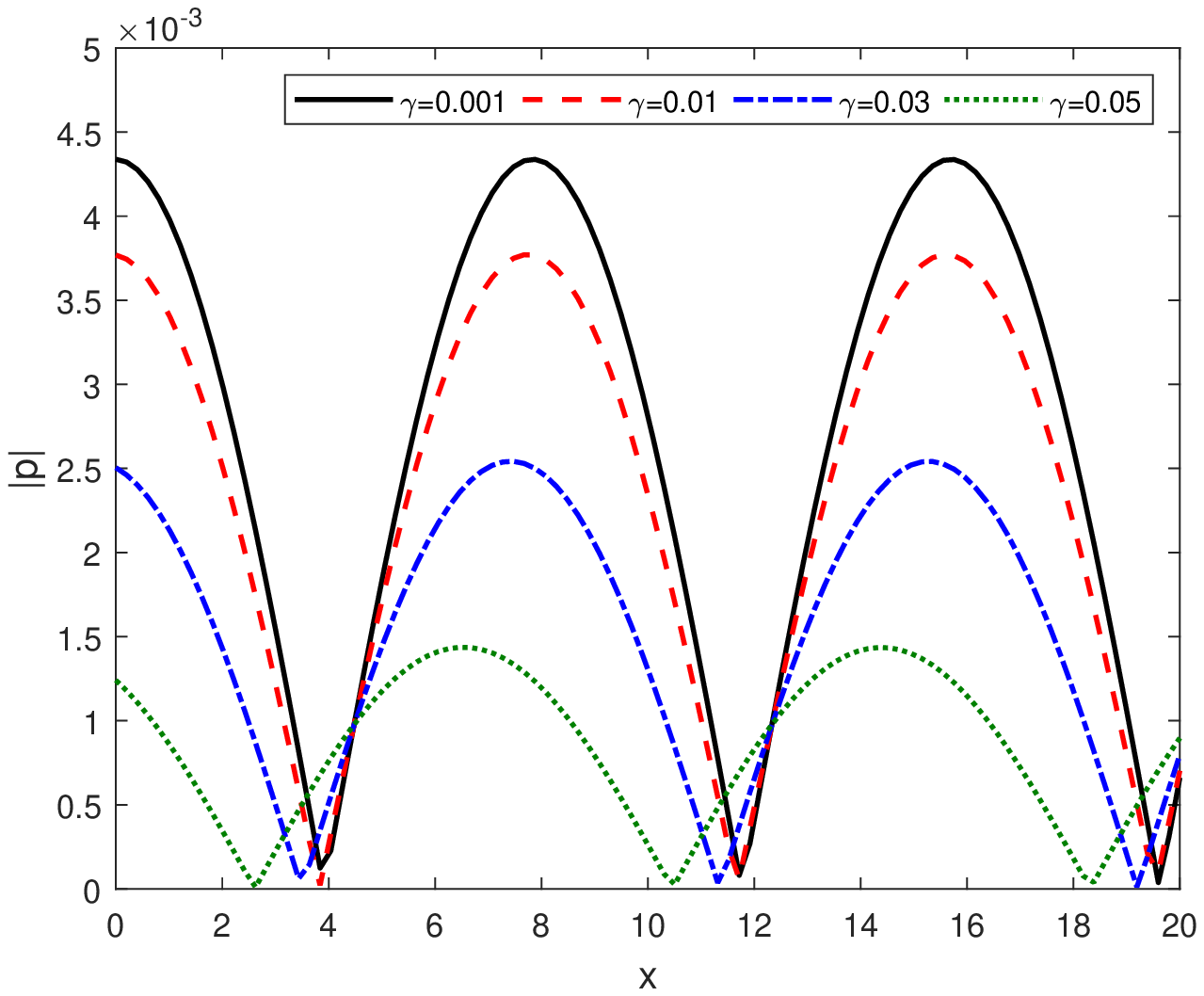}}
	\end{center}
	\caption{Absolute value of pressure acting on the floating elastic plate showing the effects of (a) structural rigidity $\alpha$ with $\gamma=0.03$ and (b) uniform mass per unit length $\gamma$ with $\alpha=0.1$. The constant parameters are $\theta=10^\circ$, $\beta=1$, $Re=100$, $k=0.4$, $\epsilon=0.01$ and $y = 1$.}\label{f12}
\end{figure}
\begin{figure}
	\begin{center}
		\subfigure[]{\label{f13a}\includegraphics*[width=6.6cm]{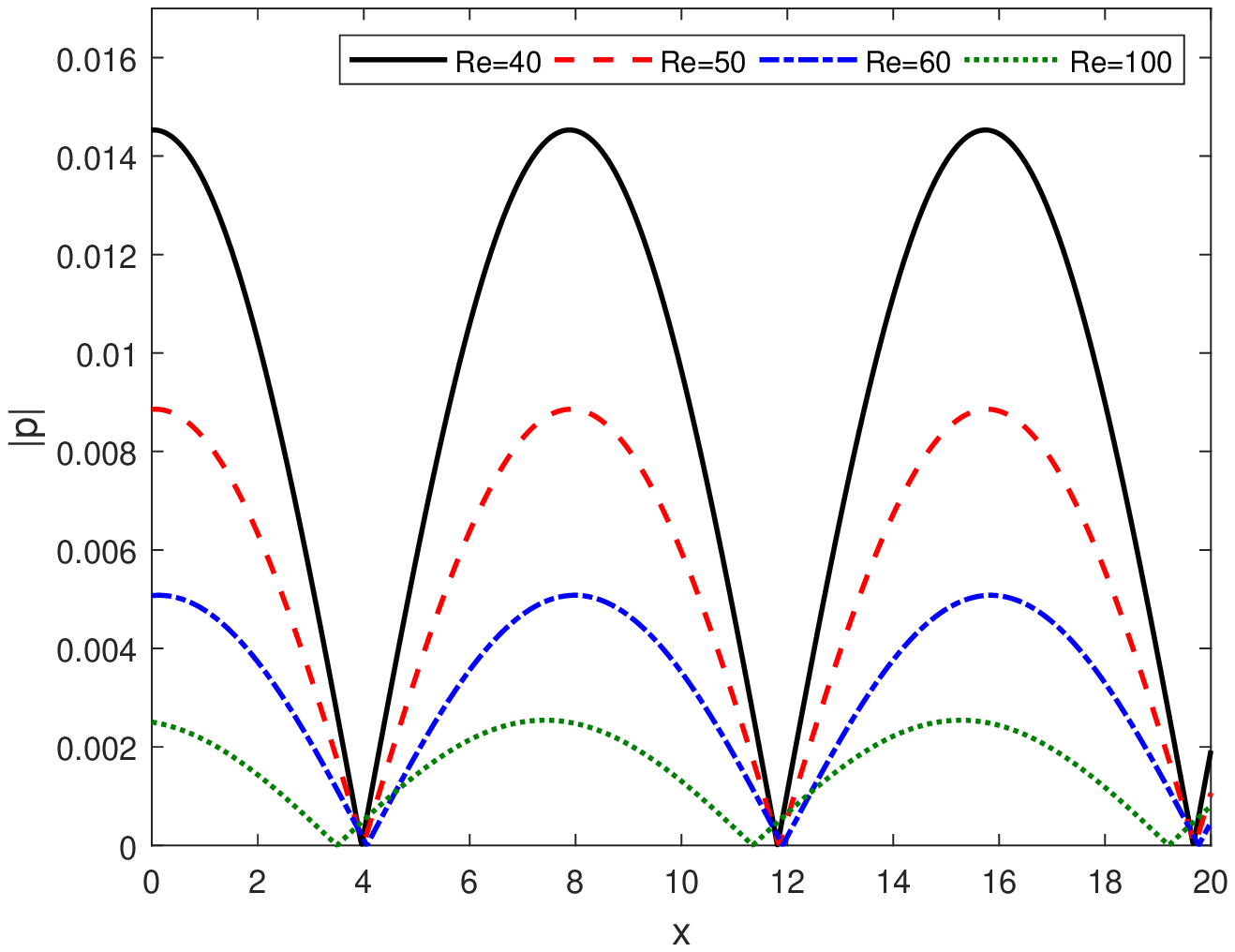}}
		\subfigure[]{\label{f13b}\includegraphics*[width=6.6cm]{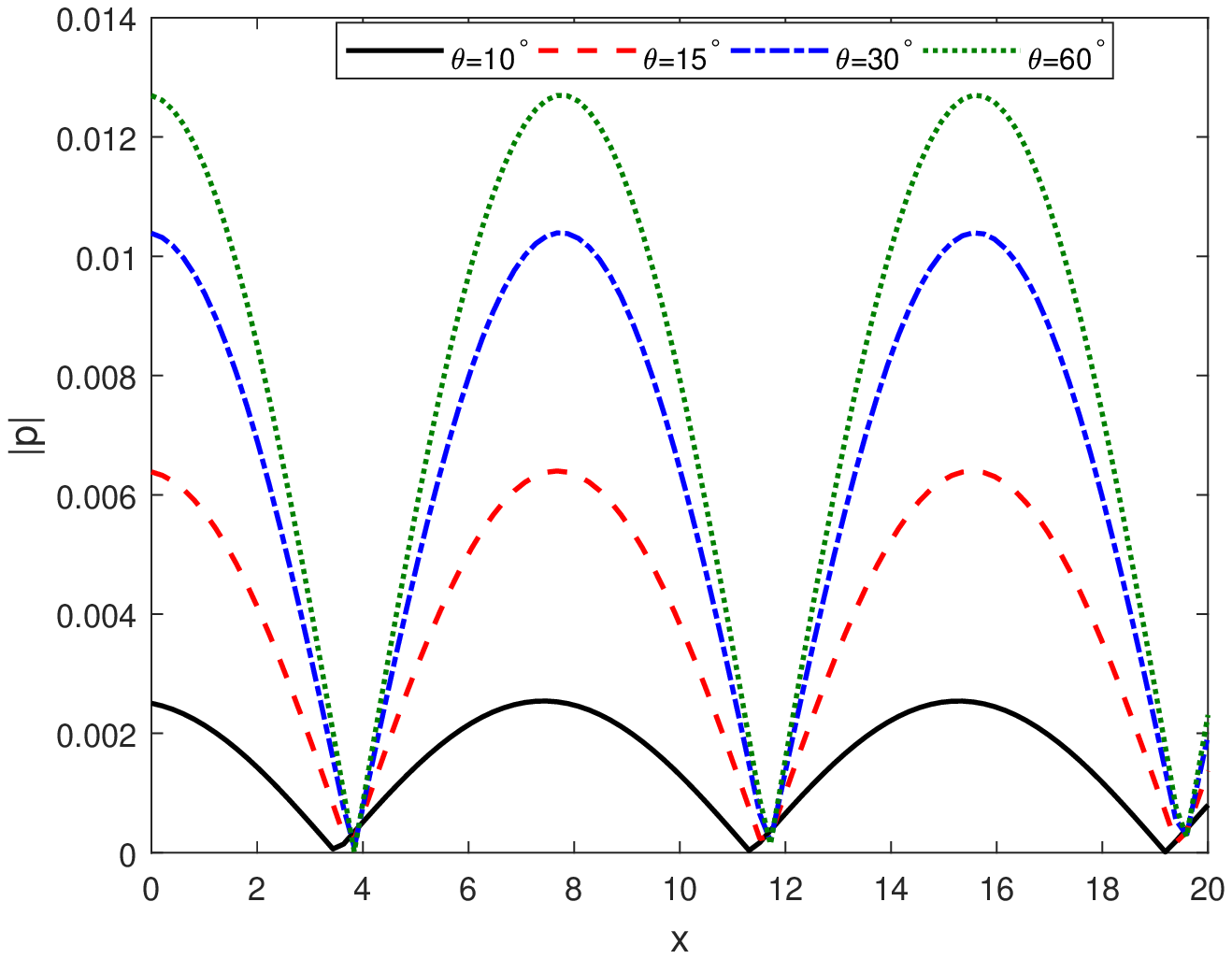}}
	\end{center}
	\caption{Absolute value of pressure acting on the floating elastic plate showing the effects of (a) Reynolds number Re with $\theta= 10^\circ$ and (b) inclination angle $\theta$ with $Re = 100$. The constant parameters are $\alpha=0.1$, $\beta=1$, $\gamma=0.03$, $k=0.4$, $\epsilon=0.01$.}\label{f13}
\end{figure}
\begin{figure}
	\begin{center}
		\subfigure[]{\label{f14}\includegraphics*[width=6.6cm]{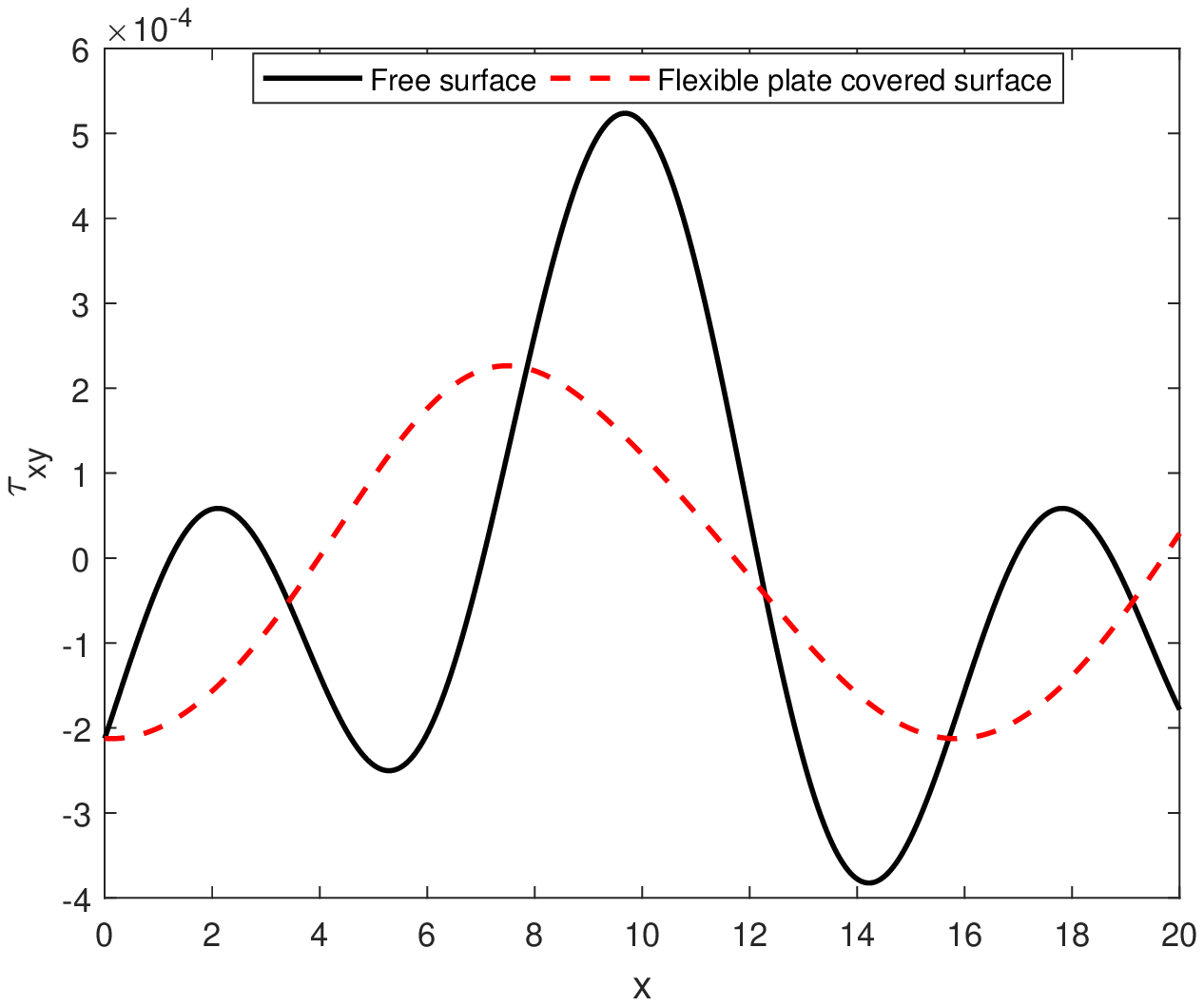}}
	\end{center}
	\caption{Comparison of shear stress at $y = 1$ between the free surface flow with $\alpha=0$, $\beta=\beta_0(Re^2\,\sin(\theta))^{-1/3}$ and $\gamma=0$ and flexible plate covered flow with $\alpha=0.1$, $\beta=1$ and $\gamma=0.03$. The constant parameters are $\theta=10^\circ$, $\beta_0=4280$, $k=0.4$, $\epsilon=0.01$.}\label{f14}
\end{figure}

In Fig.~\ref{f12}, the pressure acting on the floating elastic plate is plotted for different values of structural rigidity $\alpha$ and uniform mass per unit length $\gamma$. On increasing the values of $\alpha$ (Fig.~\ref{f12}(a)), the floating plate becomes more rigid and reduces the acting pressure. Consequently, the floating elastic plate experiences less pressure for higher values of $\gamma$ as in Fig.~\ref{f12}(b). It was found that both $\alpha$ and $\gamma$ have a stabilizing effect. 

The variation of the pressure distributions across the floating elastic plate are shown in  Figs.~\ref{f13}(a) and (b) for different values of Reynolds number ($Re$) and inclination angle ($\theta$), respectively. The pressure exerted on the floating plate decreases with larger $Re$ (Fig.~\ref{f13}(a)). However, a destabilizing effect is substantiated by a higher inclination angle, because of the enhancement of the absolute pressure acting on the floating elastic plate in Fig.~\ref{f13}(b). The shear stress distribution along the free surface and the floating plate is compared in Fig.~\ref{f14}. For the case of classical free surface flow, the shearing stress fluctuation on the fluid surface is more compared to that in the presence of the floating elastic plate. Note that, on placing the elastic plate at the top of fluid layer, the shear stress of fluid is partially balanced by the reverse stress of the floating elastic plate. 

\subsection{Results based on weakly nonlinear analysis}

\begin{figure}
	\begin{center}
		\subfigure[$\beta=3000\epsilon^2$]{\label{f15a}\includegraphics*[width=4.4cm]{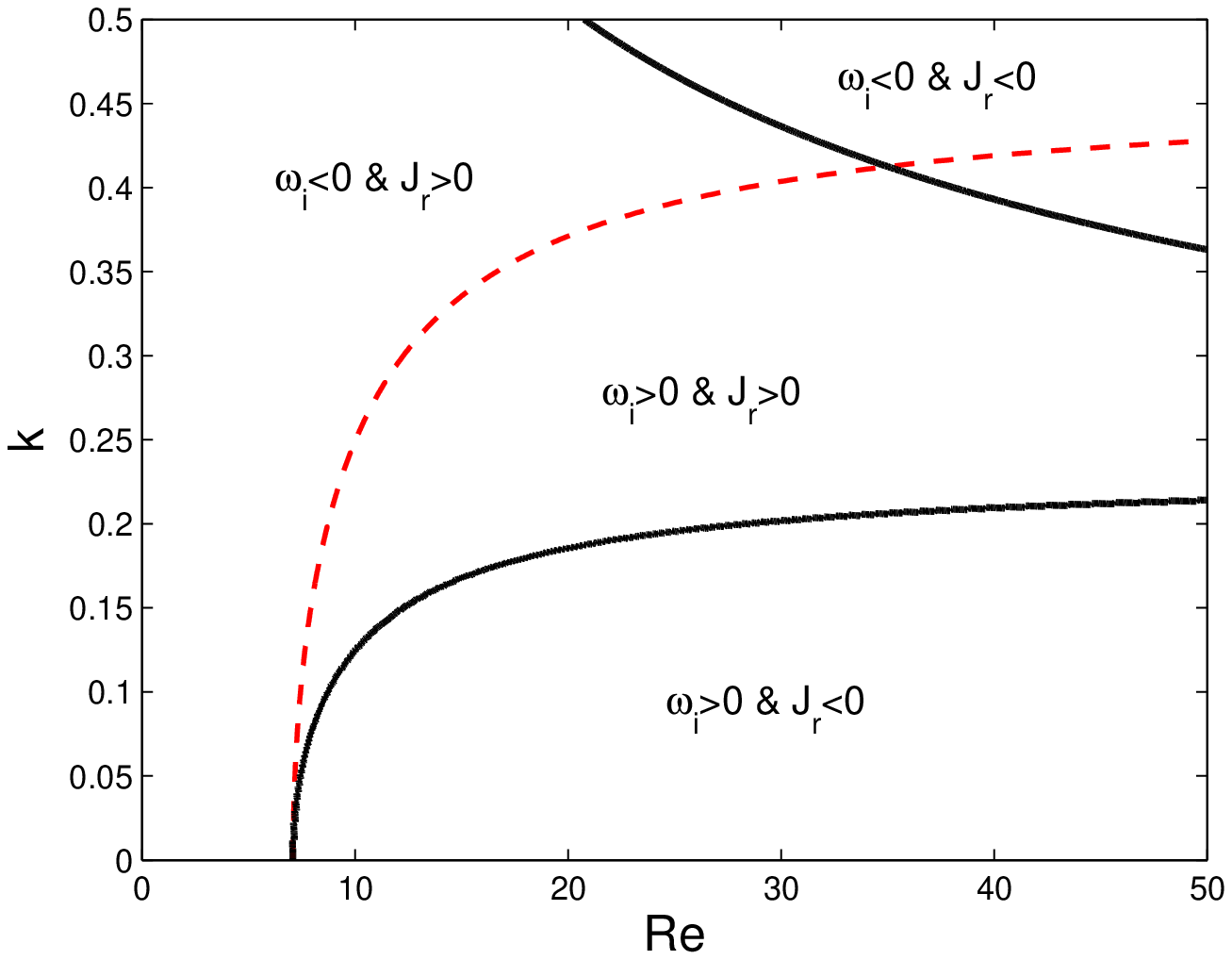}}
		\subfigure[$\beta=4000\epsilon^2$]{\label{f15b}\includegraphics*[width=4.4cm]{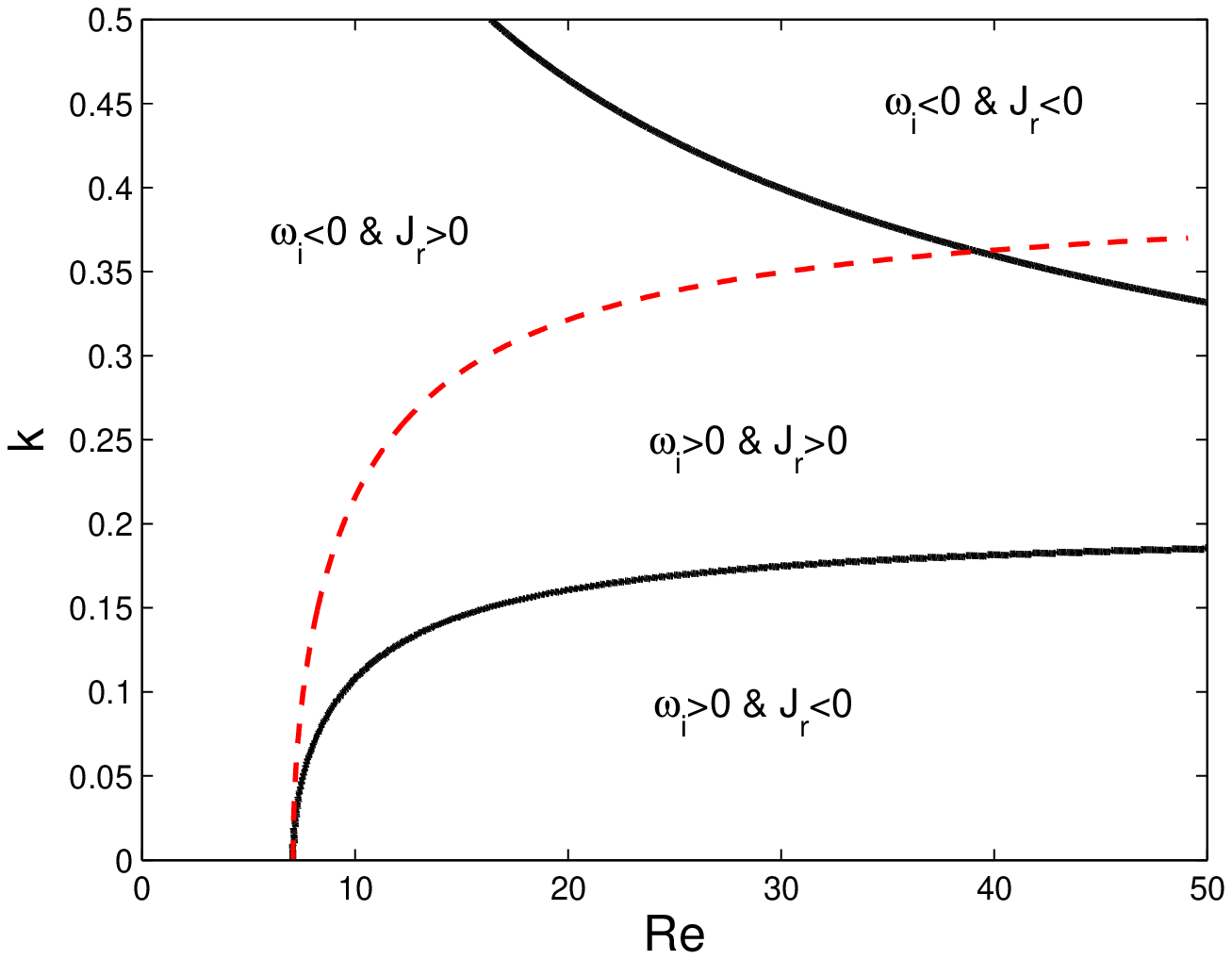}}
		\subfigure[$\beta=4496\epsilon^2$]{\label{f15c}\includegraphics*[width=4.4cm]{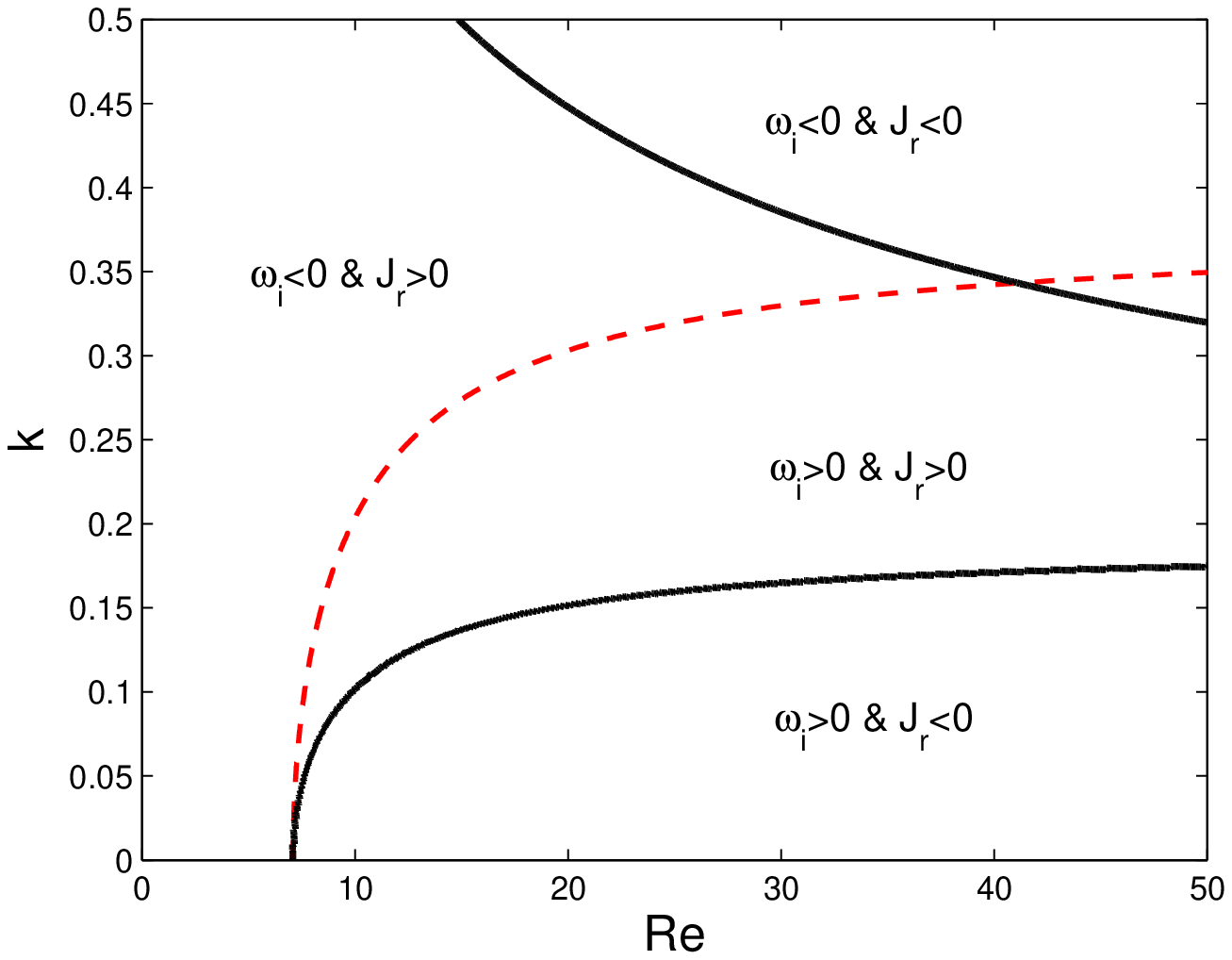}}
	\end{center}
	\caption{Neutral stability curves in the absence of floating elastic plate, showing the supercritical explosive region ($\omega_i>0$ \& $J_r<0$), supercritical stable region ($\omega_i>0$ \& $J_r>0$), sub-critical unstable region ($\omega_i<0$ \& $J_r<0$) and  sub-critical stable region ($\omega_i<0$ \& $J_r>0$), for different values surface tension $\beta$. The constant parameters are $\alpha=0$, $\epsilon=0.01$ and $\gamma=0$.}
	\label{f15}
\end{figure}

\begin{figure}
	\begin{center}
		\subfigure[$\alpha=0.1$]{\label{f16a}\includegraphics*[width=4.4cm]{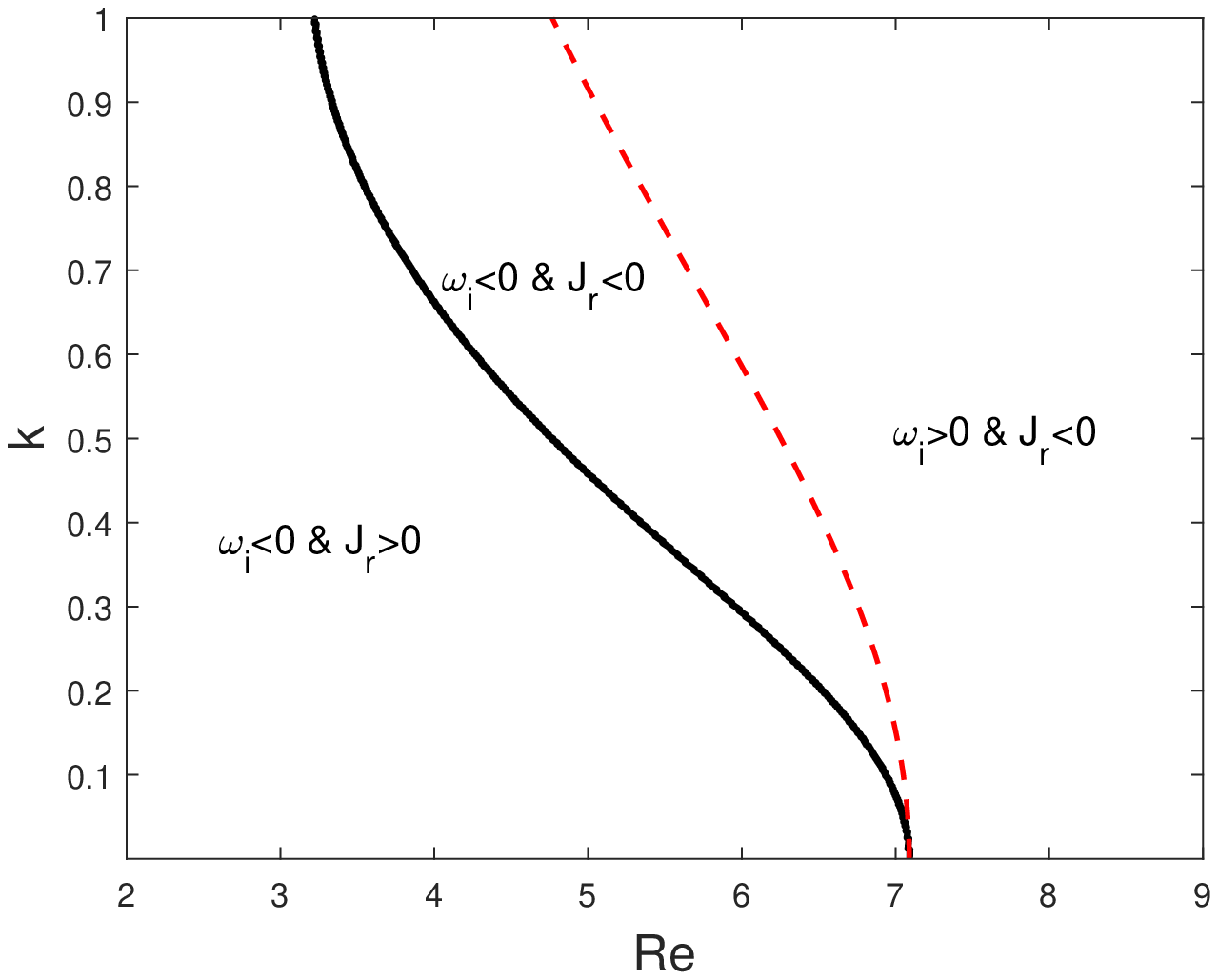}}
		\subfigure[$\alpha=0.5$]{\label{f16b}\includegraphics*[width=4.4cm]{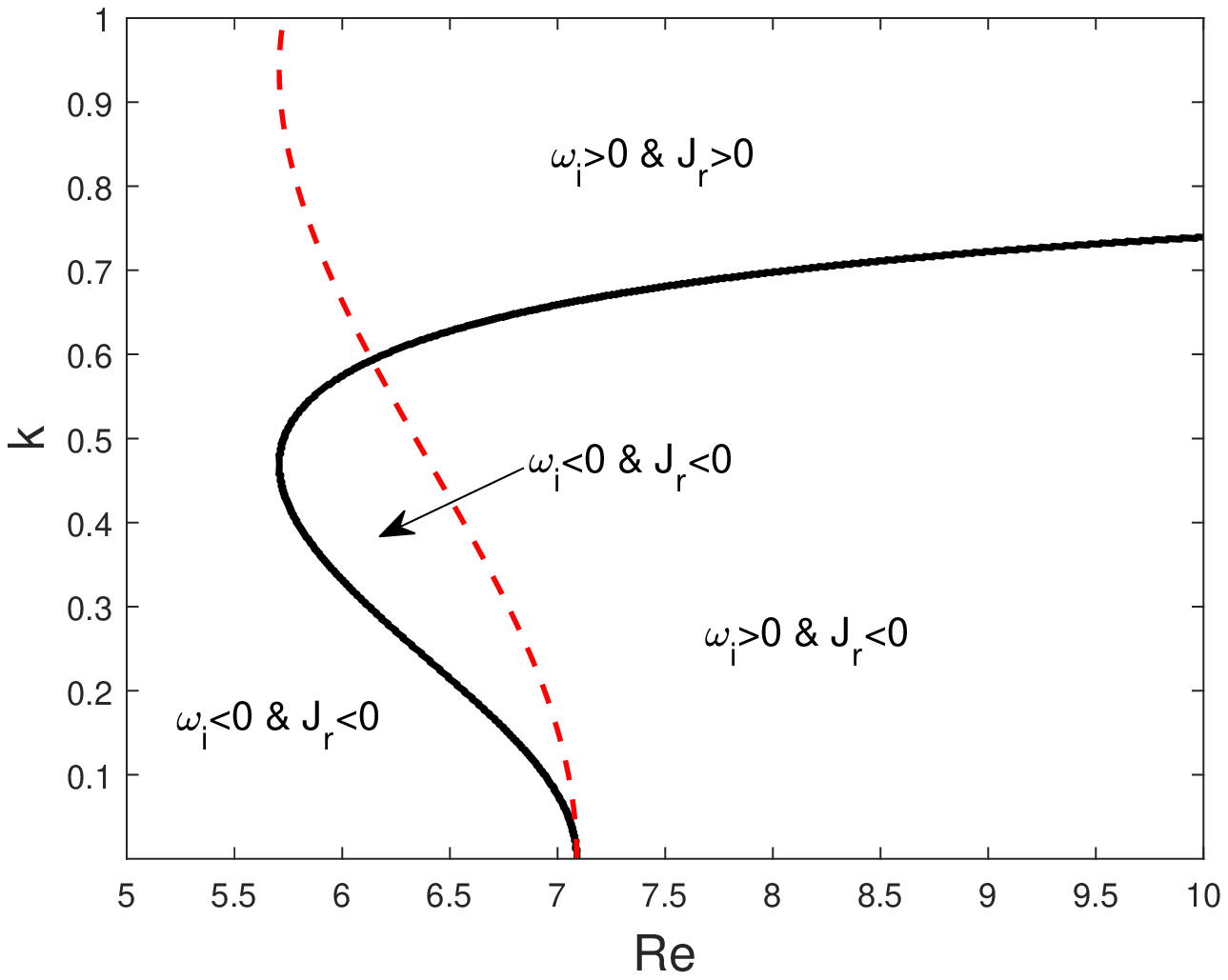}}
		\subfigure[$\alpha=1$]{\label{f16c}\includegraphics*[width=4.4cm]{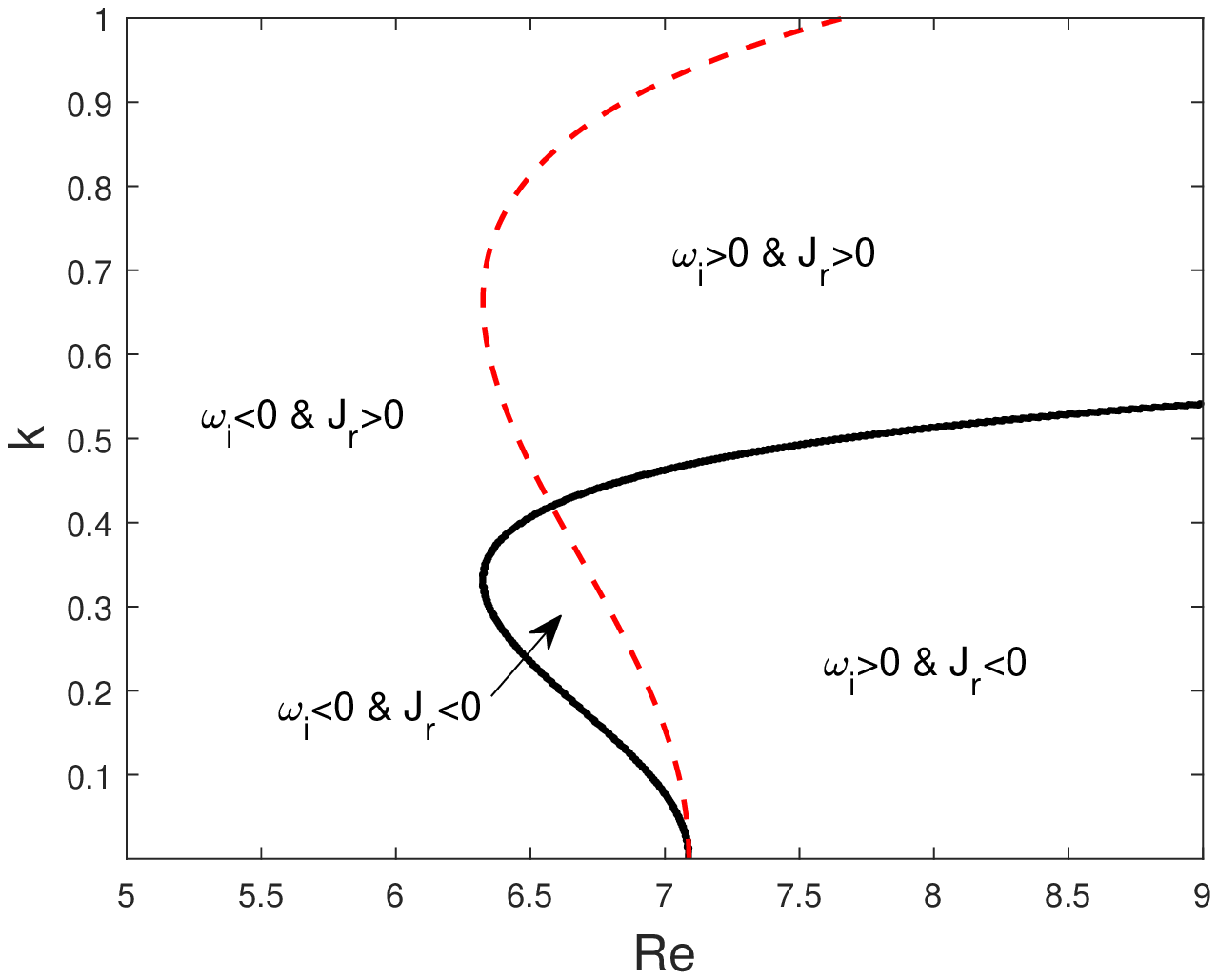}}
	\end{center}
	\caption{Neutral stability curves in the presence of floating elastic plate, showing the supercritical explosive region ($\omega_i>0$ \& $J_r<0$), supercritical stable region ($\omega_i>0$ \& $J_r>0$), sub-critical unstable region ($\omega_i<0$ \& $J_r<0$) and  sub-critical stable region ($\omega_i<0$ \& $J_r>0$), for different values of structural rigidity $\alpha$. The constant parameters are $\beta=1$ and $\gamma=0.03$.}
	\label{f16}
\end{figure}

\begin{figure}
	\begin{center}
		\subfigure[$\gamma=0.01$]{\label{f17a}\includegraphics*[width=4.4cm]{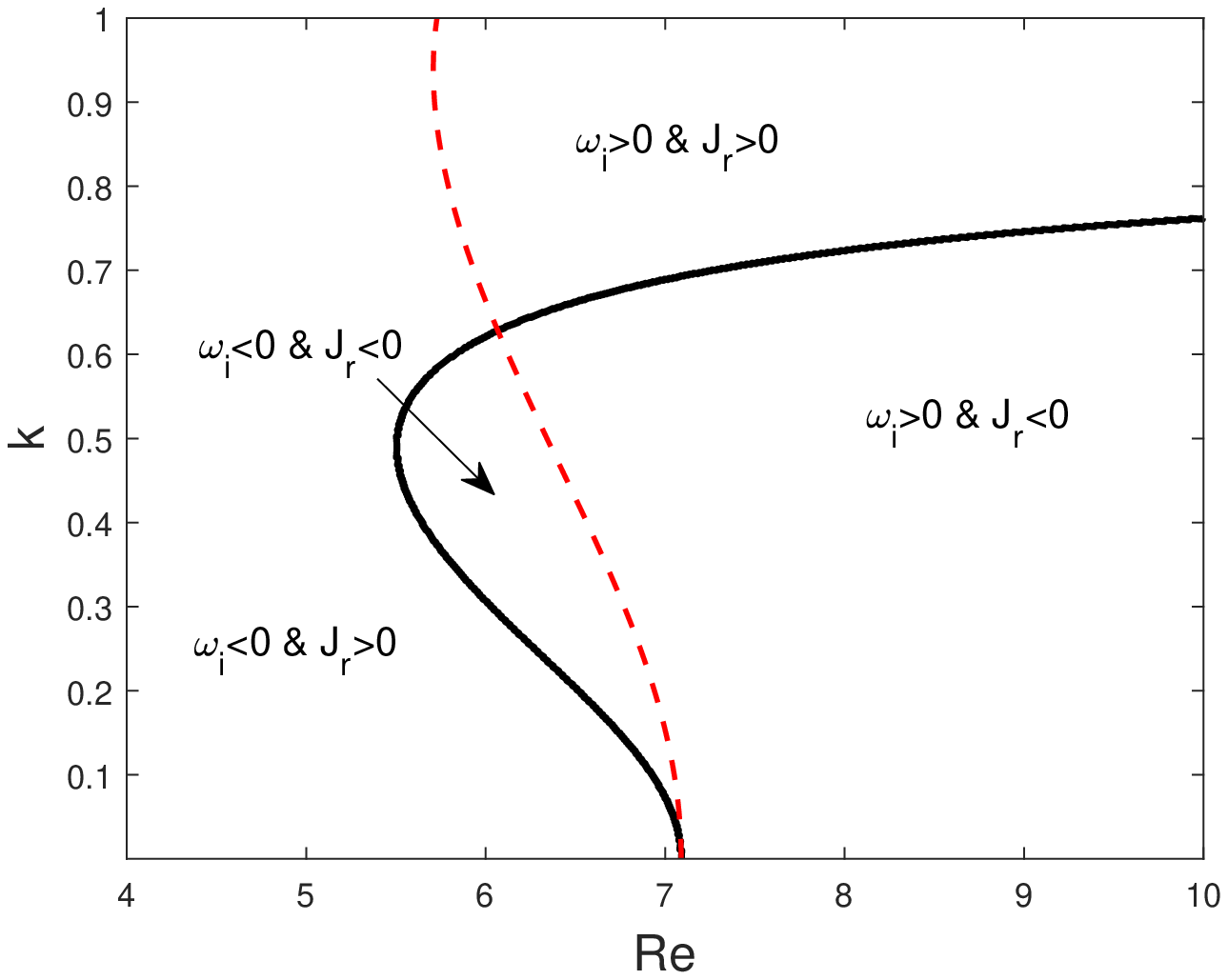}}
		\subfigure[$\gamma=0.03$]{\label{f17b}\includegraphics*[width=4.4cm]{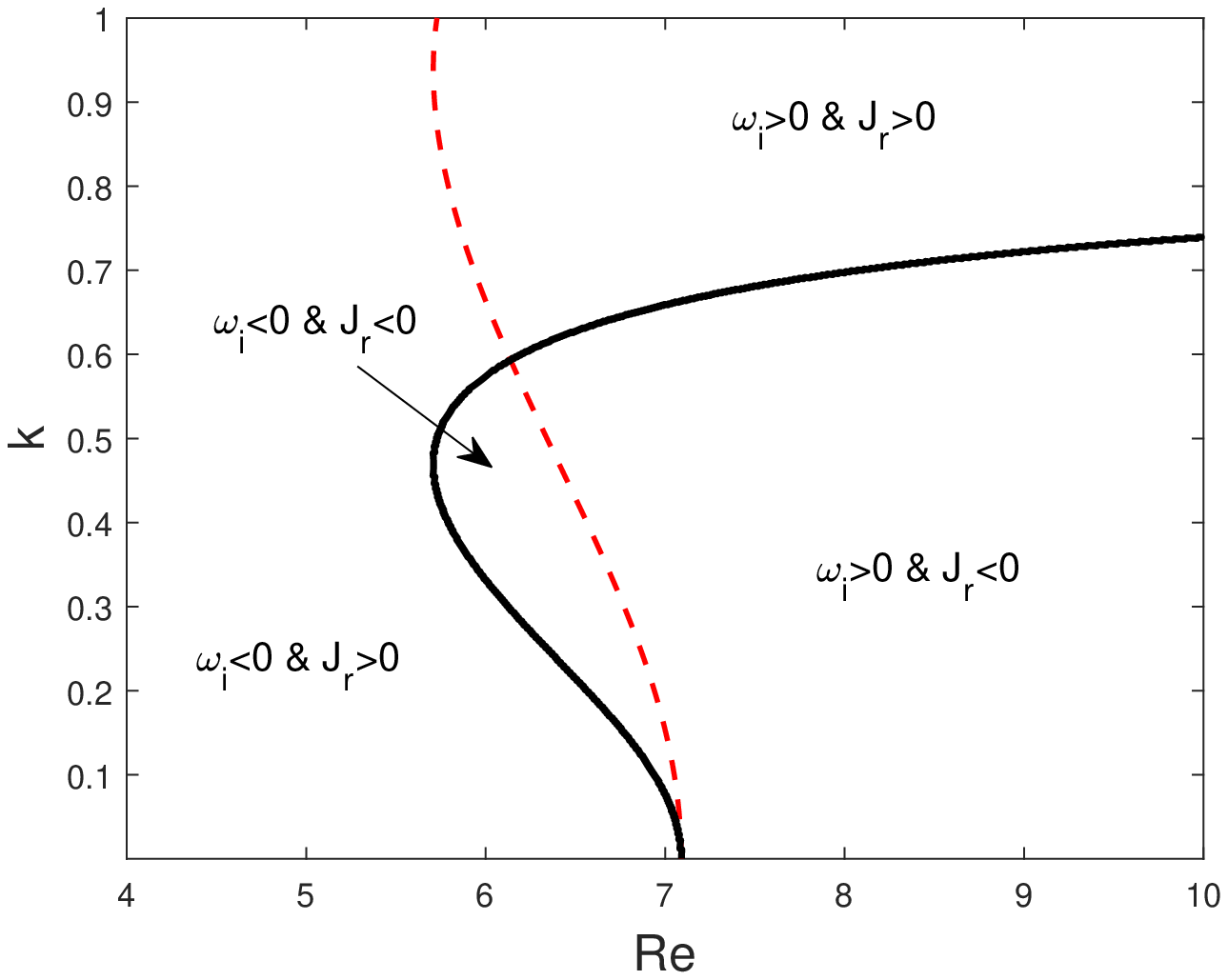}}
		\subfigure[$\gamma=0.05$]{\label{f17c}\includegraphics*[width=4.4cm]{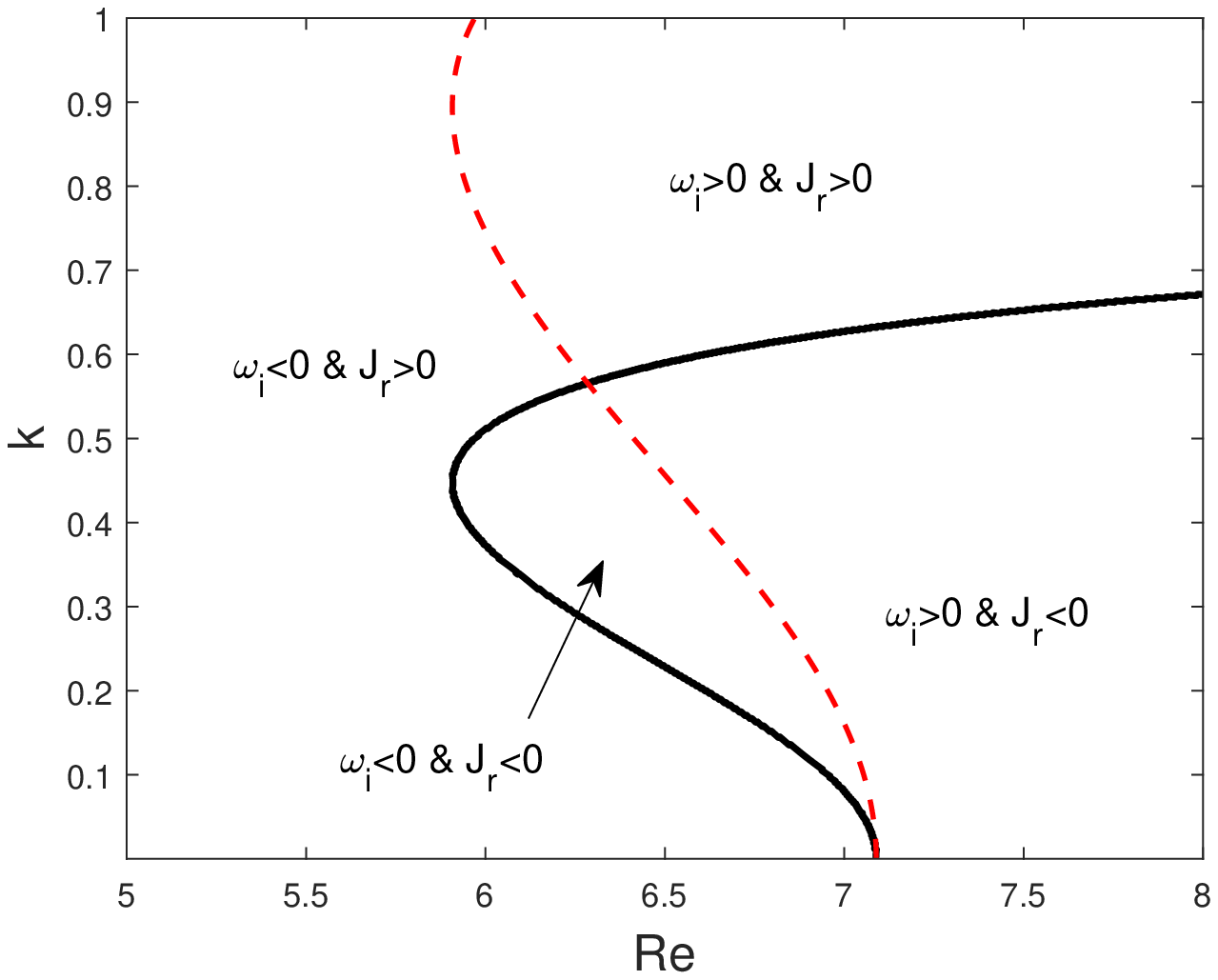}}
	\end{center}
	\caption{Neutral stability curves in the presence of floating elastic plate, showing the supercritical explosive region ($\omega_i>0$ \& $J_r<0$), supercritical stable region ($\omega_i>0$ \& $J_r>0$), sub-critical unstable region ($\omega_i<0$ \& $J_r<0$) and  sub-critical stable region ($\omega_i<0$ \& $J_r>0$), for different values of uniform mass per unit length $\gamma$. The constant parameters are $\alpha=0.5$ and $\beta=1$.}
	\label{f17}
\end{figure}

In section \ref{LWA}, analysis of the perturbed system is performed when the amplitude of the disturbance is small but finite, and the nonlinear effects become significant. In order to validate the result for free surface flow, the value of aspect ratio is considered as $\epsilon=0.05$ in this section. Results corresponding to the weakly nonlinear analysis are elaborated herewith. 
Fig.~\ref{f15} and Figs.~\ref{f16} \& \ref{f17} are respectively display the supercritical stable ($\omega_i>0$ \& $J_r>0$) and sub-critical unstable ($\omega_i<0$ \& $J_r<0$) regions in the vicinity of criticality for a free surface film flow and floating elastic plate. As a special case, the results of free-surface flow is validated with \cite{sadiq2008thin} in Fig.~\ref{f15}(c). The positive rate of nonlinear amplification in the unstable region stabilizes the nonlinear disturbance of film flow. Further, the bandwidth of the supercritical stability increases for an increase in the values of surface tension $\beta$. In that instance, for floating elastic plate as in Fig.~\ref{f16}, the supercritical stable region bandwidth is much smaller as compared to the free surface flow. With a decrease in the value of $\alpha$, the bandwidth of explosive state decreases and becomes closure to the criticality. Further, it vanishes for smaller values of $\alpha$. For this reason, the structural rigidity is fixed as $\alpha=0.5$ for the remainder of the study. On the other hand, bifurcation around the criticality is shown for varying values of $\gamma$ in Fig.~\ref{f17}. It is observed that the bandwidth of the supercritical region increases for an increase in the values of $\gamma$.

\begin{figure}
	\begin{center}
		\subfigure[Absence of elastic plate]{\label{f18a}\includegraphics*[width=4.4cm]{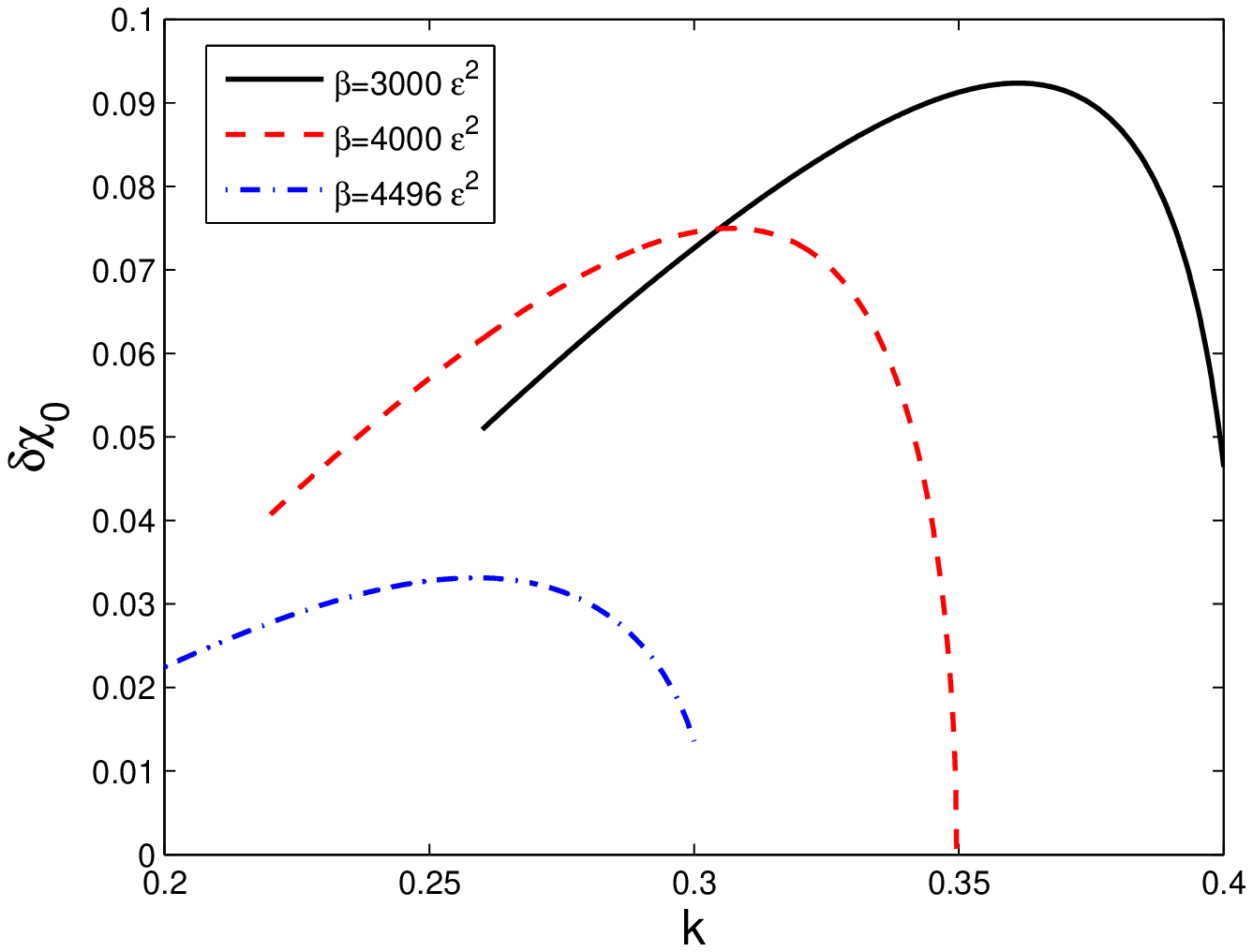}}
		\subfigure[Presence of elastic plate]{\label{f18b}\includegraphics*[width=4.4cm]{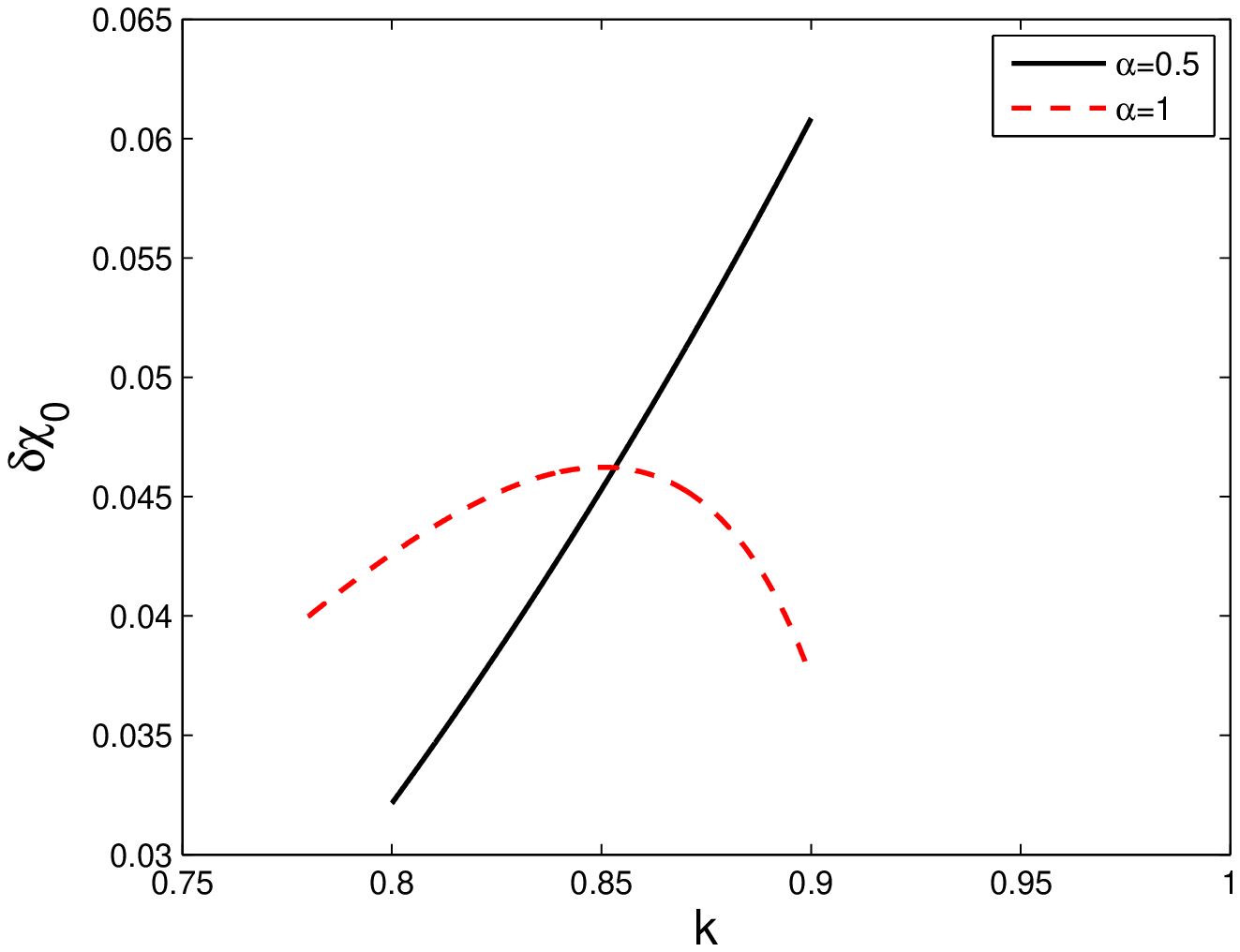}}
		\subfigure[Presence of elastic plate]{\label{f18c}\includegraphics*[width=4.4cm]{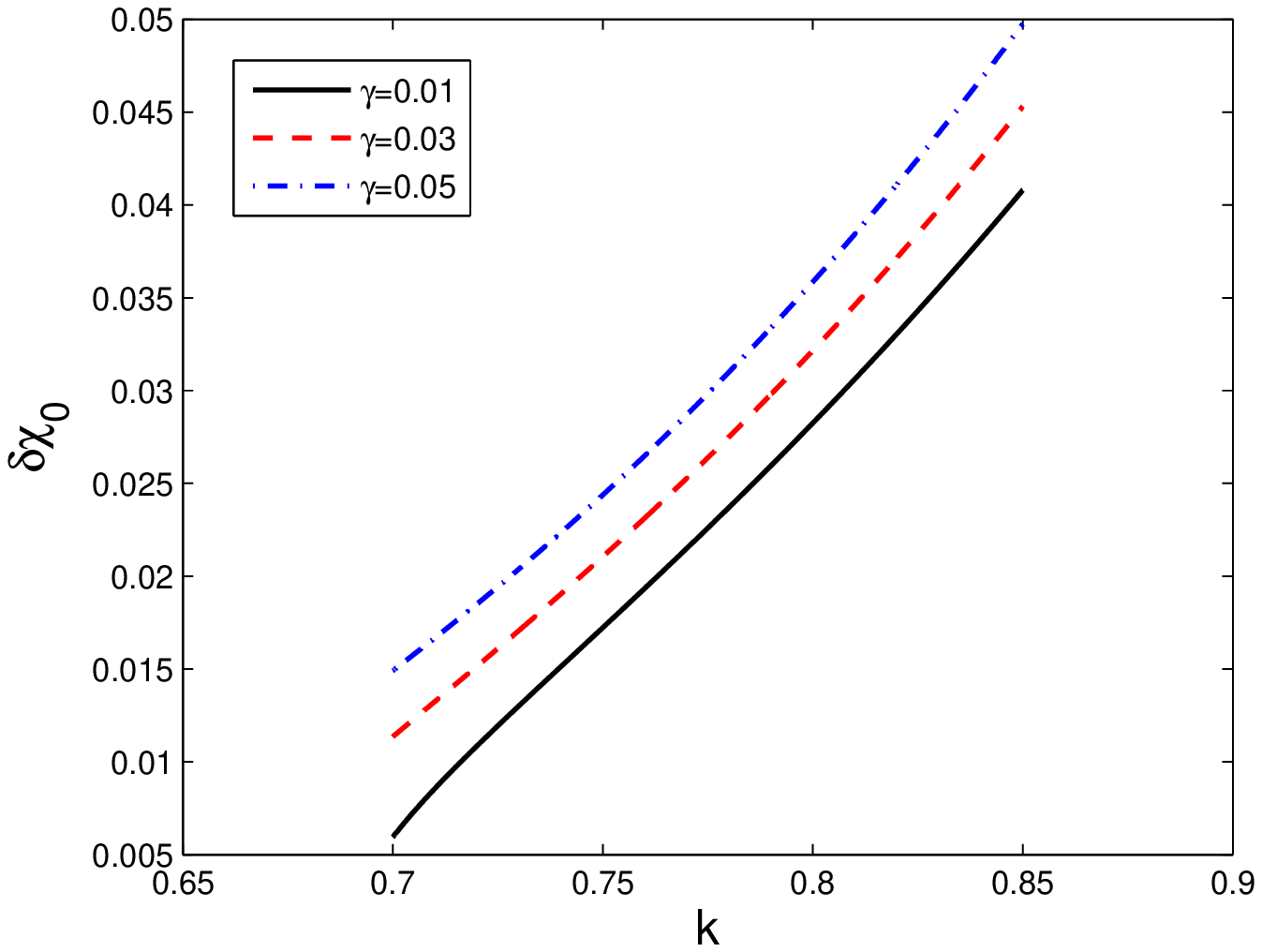}}
	\end{center}
	\caption{Threshold amplitude in the supercritical stable region showing the effect of (a) surface tension parameter $\beta_0$ with $Re=30$, $\alpha=0$ \& $\gamma=0$, (b) structural rigidity $\alpha$ with $Re=7$, $\beta=1$ \& $\gamma=0.03$ and (c) uniform mass per unit length $\gamma$ with $Re=7$, $\beta=1$ \& $\alpha=0.5$.}  \label{f18}
\end{figure}

In Fig.~\ref{f18}(a), the threshold amplitude $\delta\chi_0$ in the supercritical region is plotted with different values of surface tension $\beta$ for free-surface flow. It is observed that the threshold amplitude $\delta\chi_0$ increases with wave number ($k$) and attains a maximum, then decreases. In classical free surface flow (Fig.~\ref{f18}(a)), the threshold amplitude decreases for increasing value of $\beta$. On the other hand, in Figs.~\ref{f18}(b) and (c), the threshold amplitude $\delta\chi_0$ in the supercritical region is plotted with different structural rigidity $\alpha$ and uniform mass $\gamma$, respectively, for flow covered by a flexible plate. In general, the threshold amplitude decreases for larger values of $\alpha$. For smaller values of $\alpha$, the threshold amplitude varies linearly, whilst, the variation becomes two-fold for larger values of $\alpha$. This is due to the enlargement of the supercritical region's bandwidth for larger values of $\alpha$.
In Fig.~\ref{f18}(c), the threshold amplitude increases for an increase in the values of $\gamma$. Further, the threshold amplitude is of smaller in magnitude for the flow with a floating elastic plate as compared to that of the free-surface flow.
\begin{figure}
	\begin{center}
		\subfigure[Absence of elastic plate]{\label{f19a}\includegraphics*[width=4.4cm]{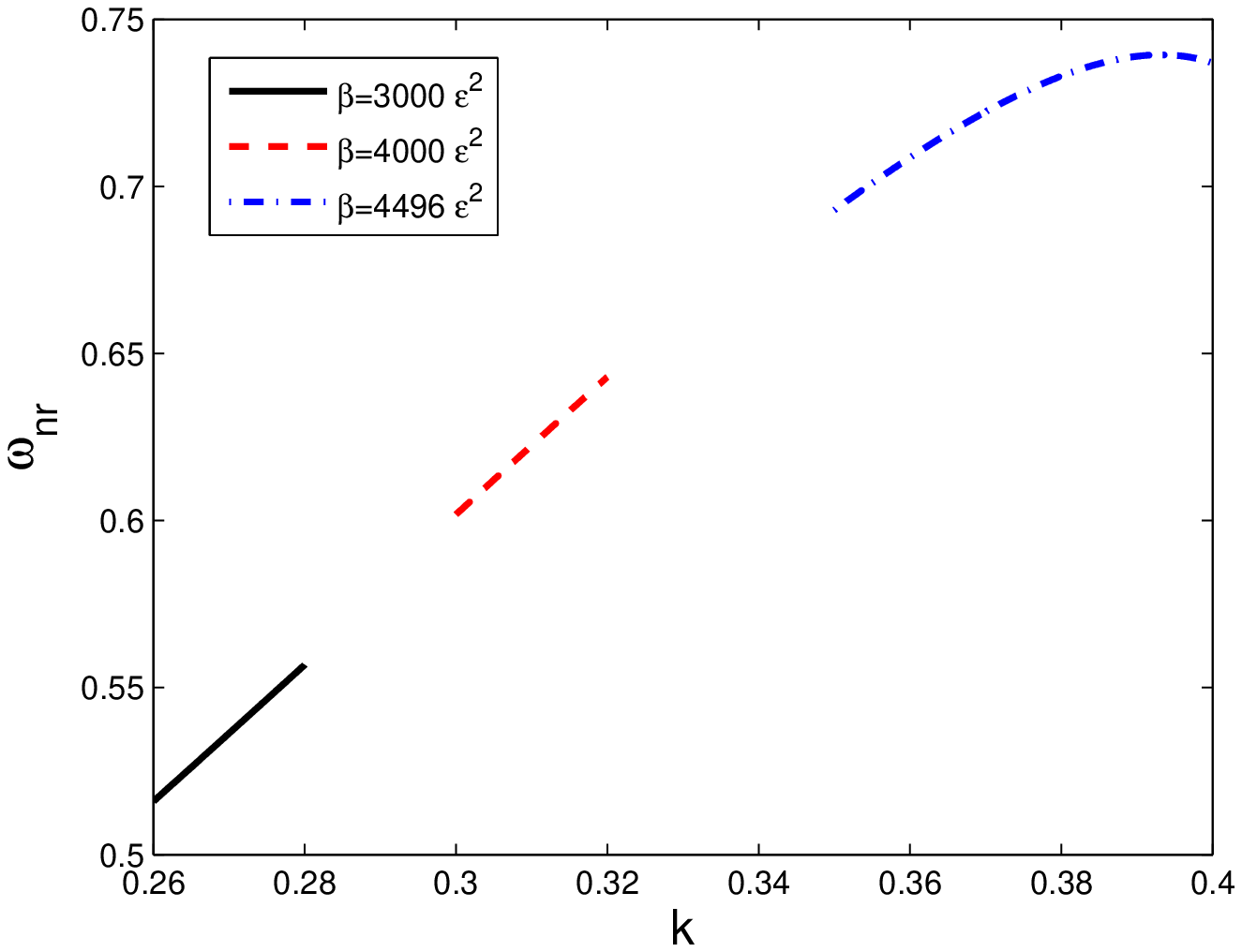}}
		\subfigure[Presence of elastic plate]{\label{f19b}\includegraphics*[width=4.4cm]{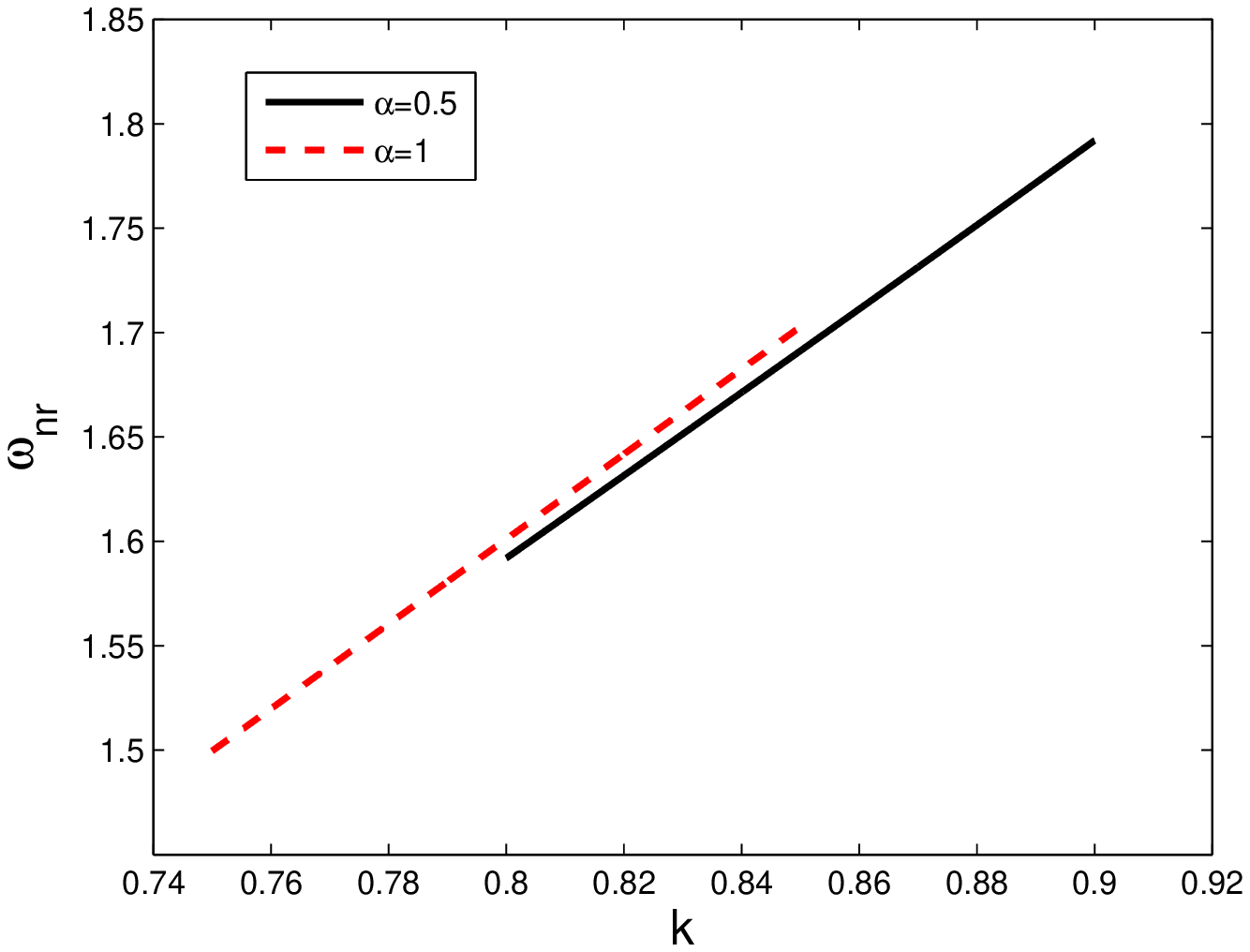}}
		\subfigure[Presence of elastic plate]{\label{f19c}\includegraphics*[width=4.4cm]{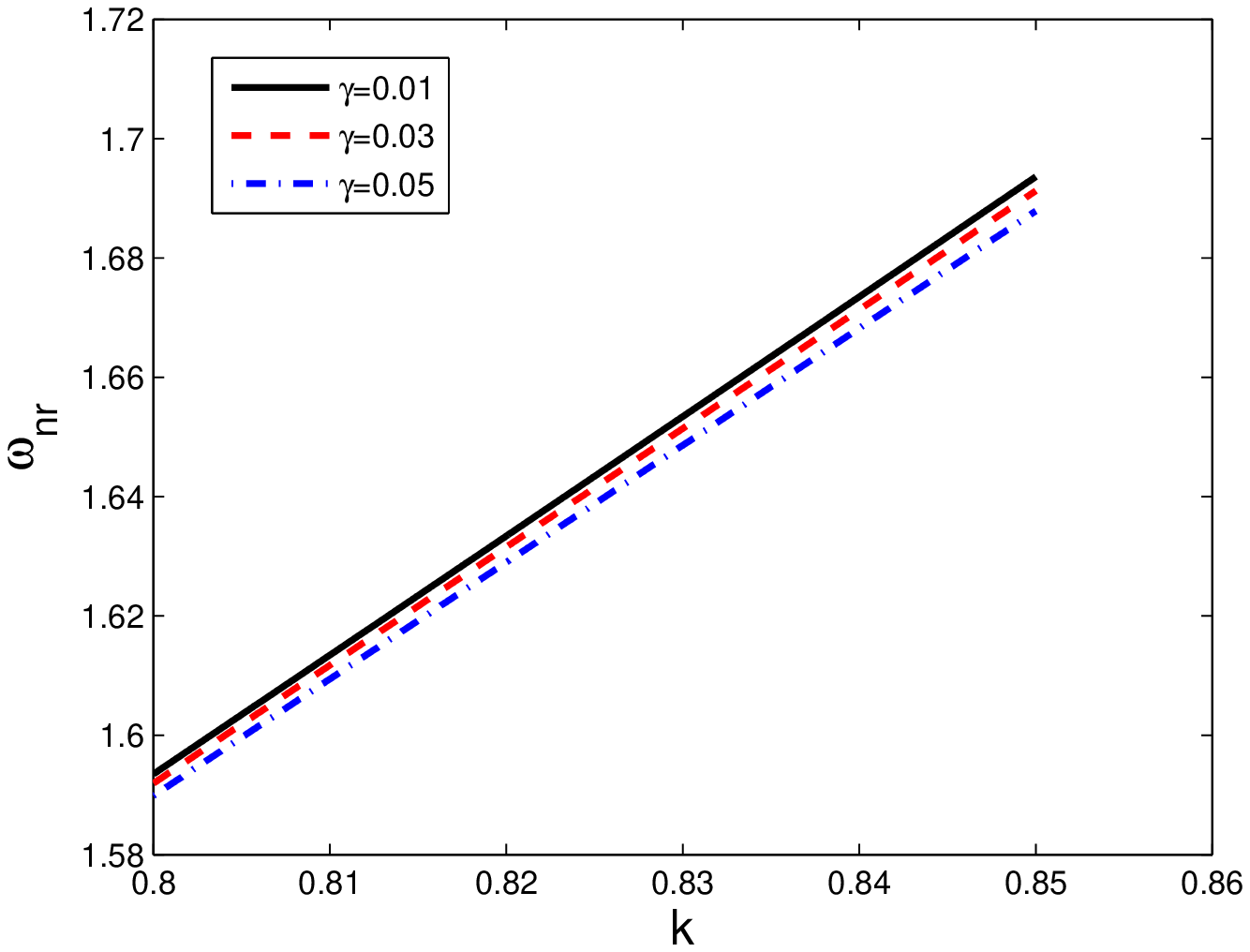}}
	\end{center}
	\caption{Nonlinear phase speed $\omega_{nr}$ in the supercritical stable region showing the effect of (a) surface tension parameter $\beta_0$ with $Re=30$, $\alpha=0$ \& $\gamma=0$, (b)structural rigidity $\alpha$ with $Re=7$, $\beta=1$ \& $\gamma=0.03$ and (c) uniform mass per unit length $\gamma$ with $Re=7$, $\alpha=0.5$ \& $\beta=1$.}
	\label{f19}
\end{figure}
\begin{figure}
	\begin{center}
		\subfigure[Absence of elastic plate]{\label{f20a}\includegraphics*[width=4.4cm]{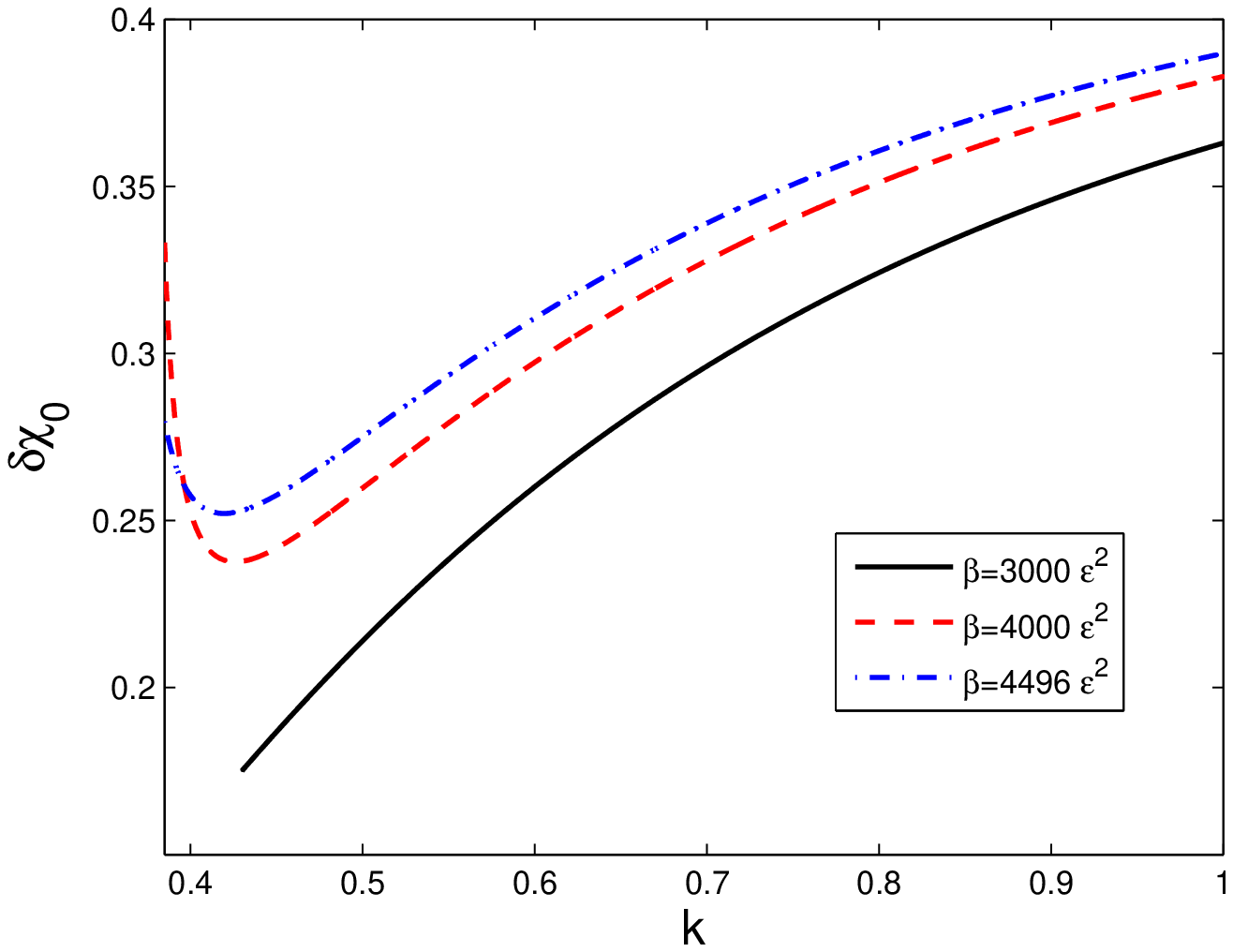}}
		\subfigure[Presence of elastic plate]{\label{f20b}\includegraphics*[width=4.4cm]{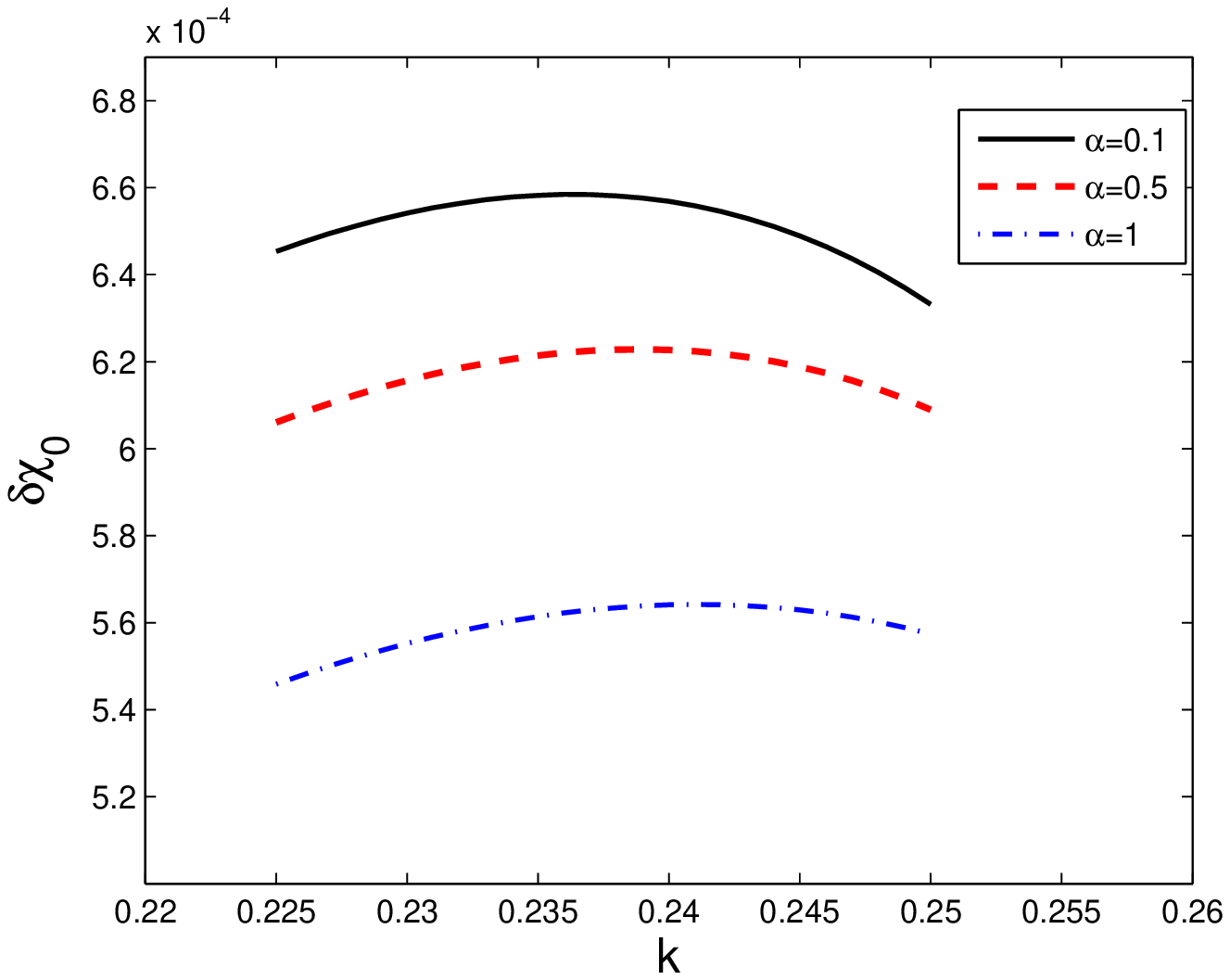}}
		\subfigure[Presence of elastic plate]{\label{f20c}\includegraphics*[width=4.4cm]{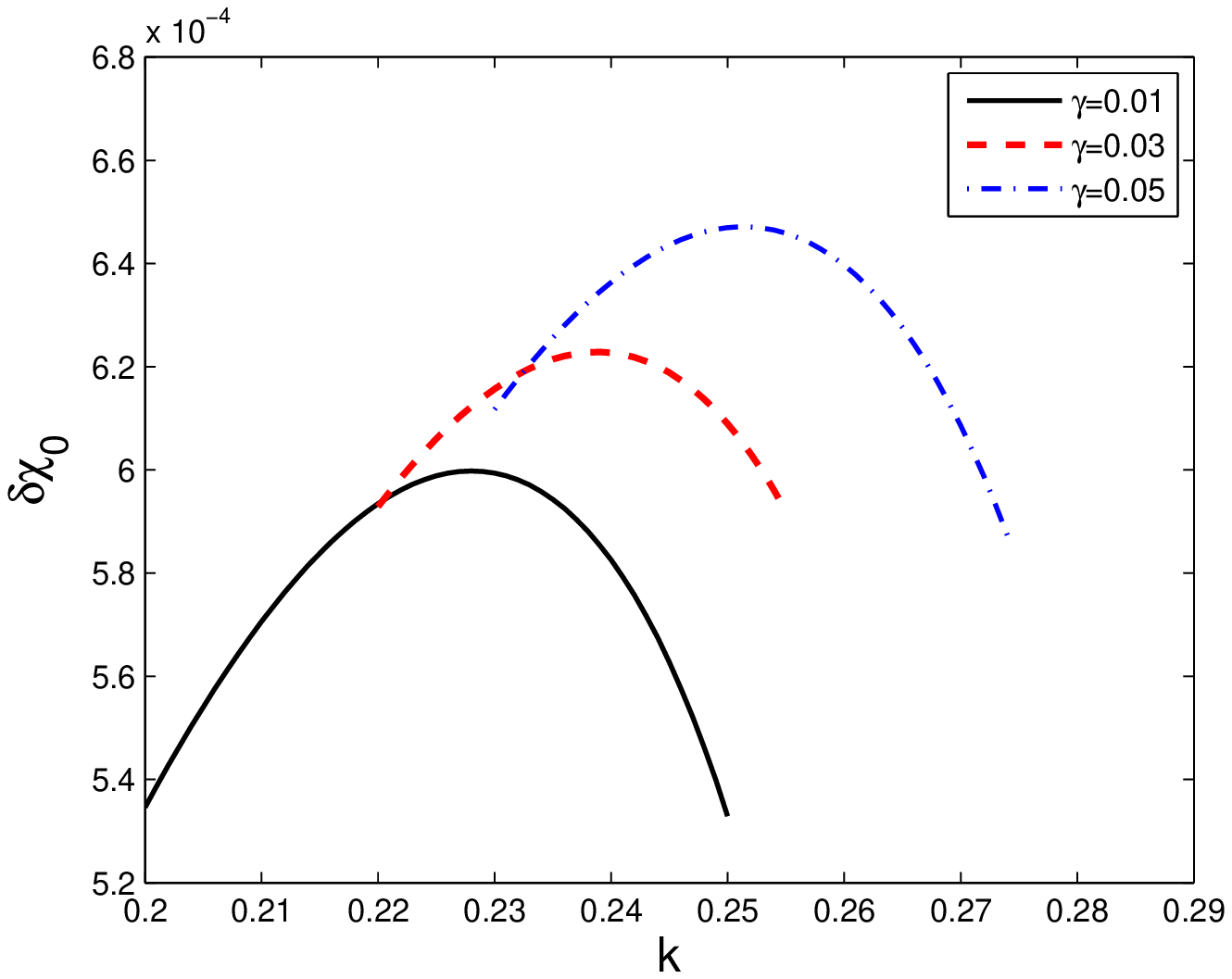}}
	\end{center}
	\caption{Threshold amplitude $\delta\chi_0$ in the sub-critical unstable region showing the effect of (a) surface tension parameter $\beta$ in the absence of floating elastic plate with $\alpha=0$, $Re=35$ \& $\gamma=0$, and (b) structural rigidity $\alpha$ with $\gamma=0.03$ and (c) uniform mass per unit length $\gamma$ with $\alpha=0.5$, in the presence of floating elastic plate with $\beta=0$ \& $Re=6.8$.}
	\label{f20}
\end{figure}
\begin{figure}
	\begin{center}
		\subfigure[Absence of elastic plate]{\label{f21a}\includegraphics*[width=4.4cm]{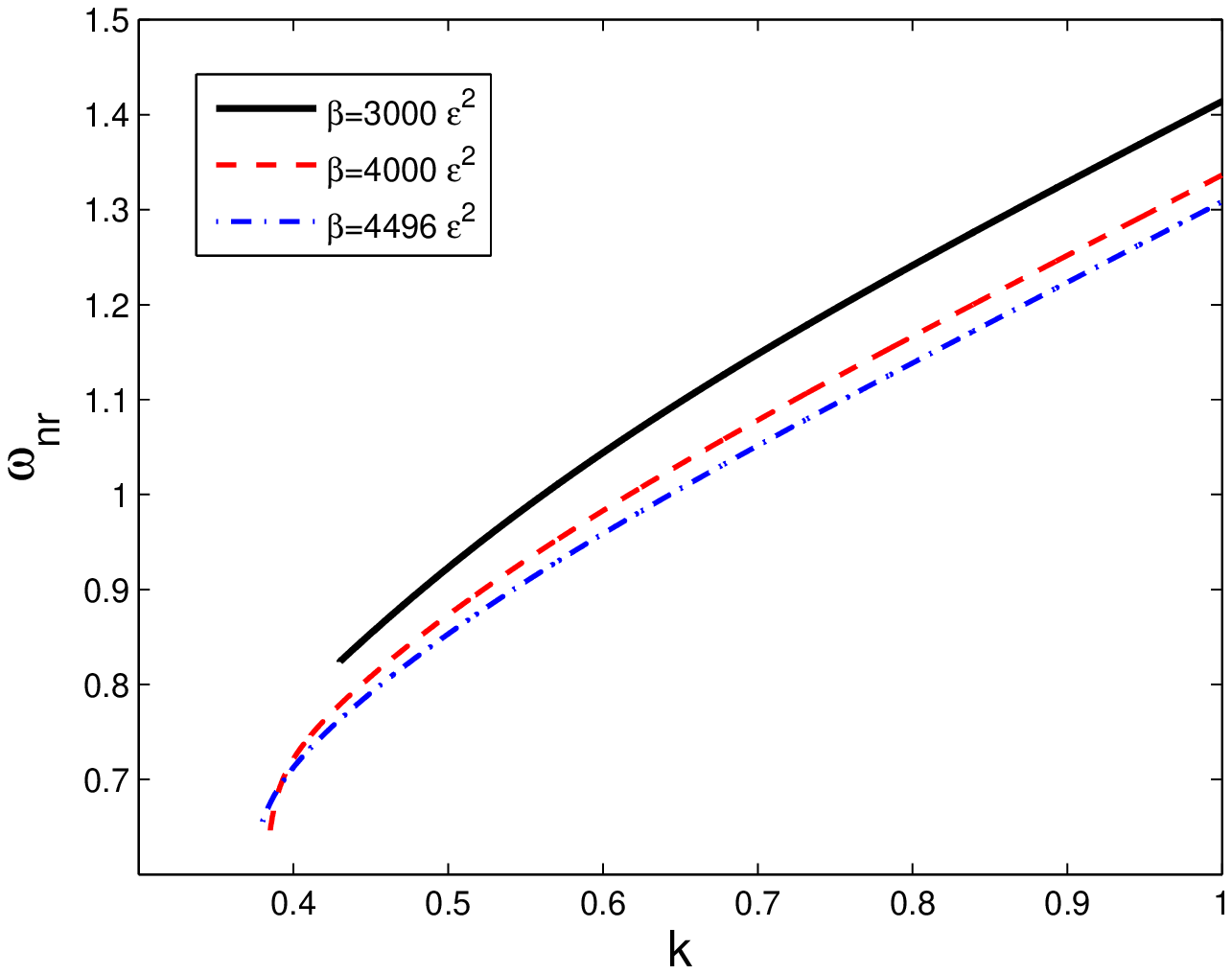}}
		\subfigure[Presence of elastic plate]{\label{f21b}\includegraphics*[width=4.4cm]{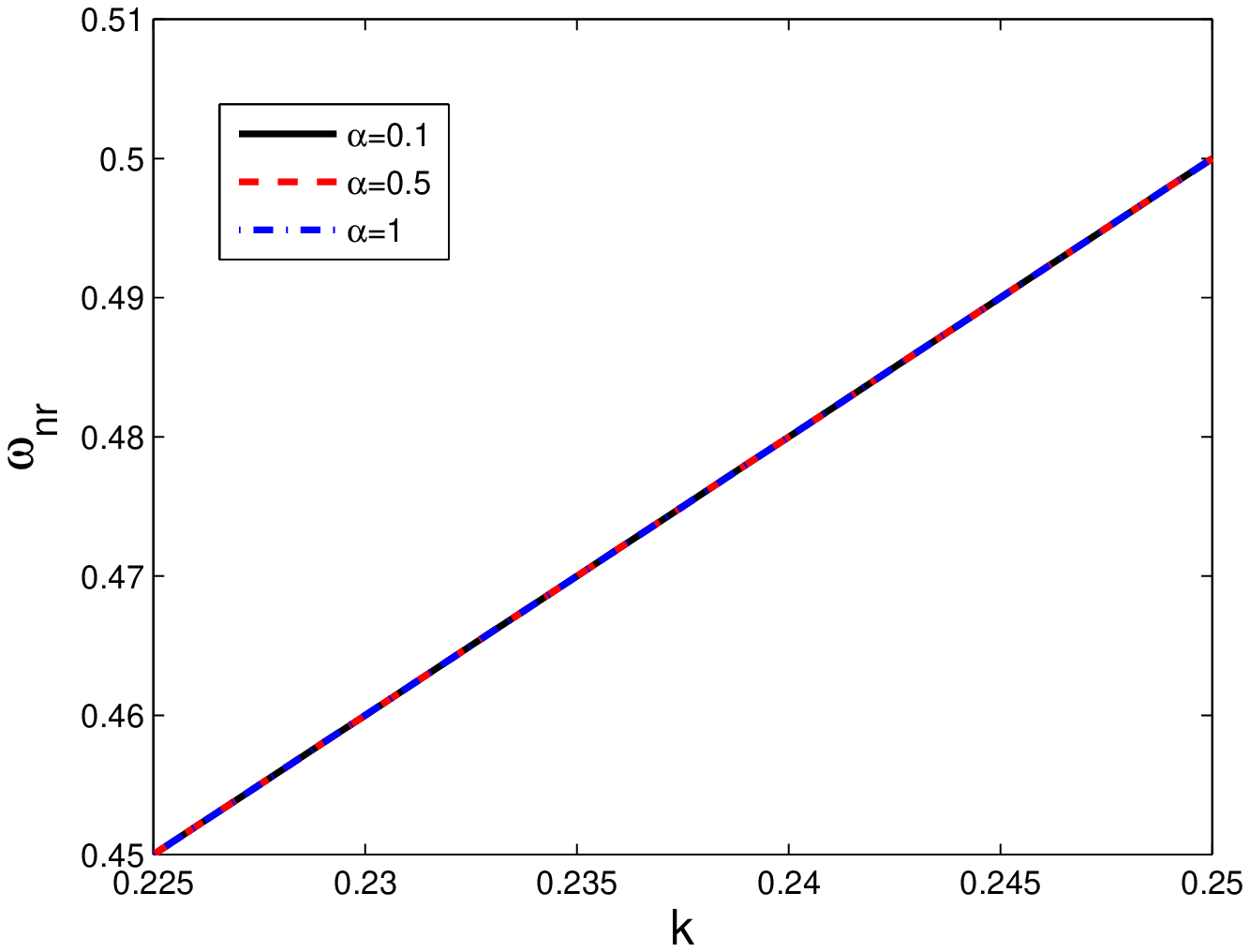}}
		\subfigure[Presence of elastic plate]{\label{f21c}\includegraphics*[width=4.4cm]{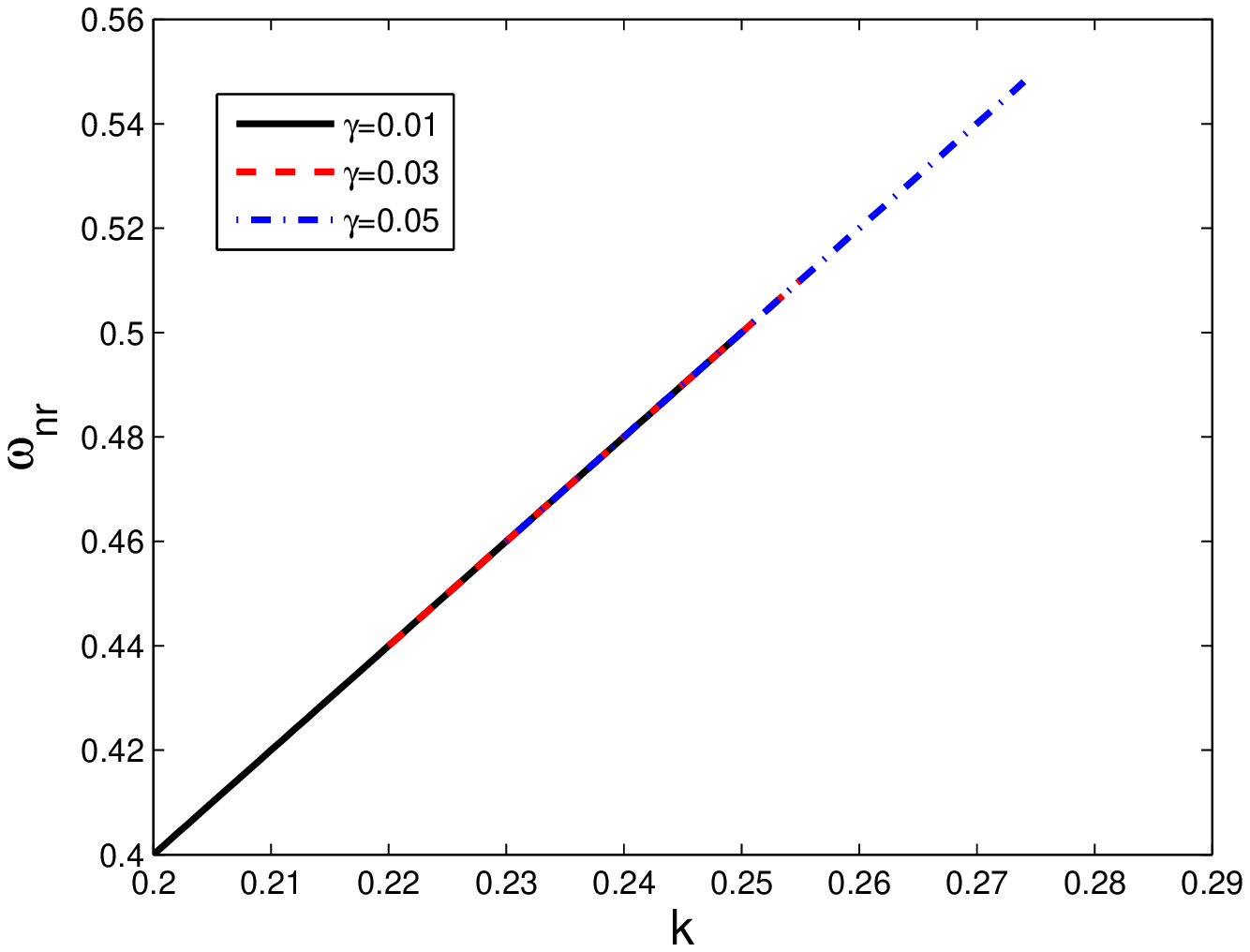}}
	\end{center}
	\caption{Nonlinear phase speed $\omega_{nr}$ in the sub-critical stable region showing the effect of (a) surface tension parameter $\beta$ in the absence of floating elastic plate with $\alpha=0$, $Re=35$ \& $\gamma=0$, and (b) structural rigidity $\alpha$ with $\gamma=0.03$ and (c) uniform mass per unit length $\gamma$ with $\alpha=0.5$, in the presence of floating elastic plate with $\beta=0$ \& $Re=6.8$.}
	\label{f21}
\end{figure}
Figs.~\ref{f19}(a), (b) and (c) exhibit the nonlinear phase speed $\omega_{nr}$ in the supercritical stable region for different values of $\beta$, $\alpha$ and $\gamma$, respectively. In the free surface flow as in Fig.~\ref{f19}(a), $\omega_{nr}$ increases for larger values of $\beta$ and it shifts towards higher $k$. A reverse trend is followed with superior $\alpha$ values in the presence of a floating elastic plate (Fig.~\ref{f19}(b)). Similarly, the nonlinear phase speed of perturbed flow decreases for increasing values of $\gamma$ as in Fig.~\ref{f19}(c).

Figs.~\ref{f20}(a), (b) and (c) depict the threshold amplitude in the sub-critical region for different values of surface tension $\beta_0$, structural rigidity $\alpha$ and uniform mass per unit length $\gamma$. In the sub-critical region, the threshold amplitude of free surface flow varies linearly against $k$ as in Fig.~\ref{f20}(a). Further, it increases for an increase in the values of surface tension $\beta$. On the other hand, in the case of flow in the flexible plate covered surface as in Figs.~\ref{f20}(b) and (c), the variation of threshold amplitude against the wavenumber is two-fold. As the value of $\alpha$ increases, the threshold amplitude decreases and reverse trends are observed while increasing the value of $\gamma$.

In Fig.~\ref{f21}, the nonlinear phase speed in the sub-critical region is plotted for both the classical free surface  (Fig.~\ref{f21}(a)) and flexural flow (Figs.~\ref{f21}(b) and (c)). It is found that the nonlinear phase speed decreases for an increase in the values of surface tension (in Fig.~\ref{f21}(a)). On contrary, for flexible plate flow, there is no major deviation in nonlinear phase speed for varying structural rigidity and uniform mass. However, the nonlinear phase speed is less as compared to the free surface flow. 

\section{Conclusions}
\label{con}

The effect of a flexible floating plate on the dynamics of fluid flow over a mild-slope is investigated using both linear and weakly nonlinear stability analysis. The linear stability analysis is performed using normal-mode analysis and results in an Orr-Sommerfeld system of equations, which is numerically solved using a spectral-collocation method. The long-wave analysis using the direct technique shows that the linear phase speed and growth rate are independent of the structural parameters. Further, the analysis with small film aspect ratio is carried out to estimate the critical parameters for instability. Such a procedure helps us to check analytically the influence of plate parameters on the instability of the flow. 
Moreover, the occurrence of supercritical and sub-critical regimes around the critical points are investigated using weakly nonlinear analysis. The numerical outcomes of the obtained Orr-Sommerfeld equations for the perturbed flow are explored for various structural and wave parameters. The long-wave approximation using a small film aspect ratio is validated through the neutral stability curve acquired by solving the Orr-Sommerfeld system. Further, the results show that the temporal growth rate of the dominant unstable mode diminishes in the presence of a floating elastic plate. Importantly, the unstable Yih mode for the classical free falling film is suppressed due to stabilizing influence by the structural parameters of the floating plate. The possible mechanism for this stabilization is the fall of acting pressure on the plate covered surface as the structural rigidity and uniform mass raise. Whereas, the weakly nonlinear analysis confirms that the bandwidth of the supercritical region increases, whilst the sub-critical region decreases for an increase in the value of these structural parameters. The threshold amplitude is lesser in the case of a plate covered surface when compared to that of a usual free surface flow. Therefore, the presence of a floating elastic plate over a falling flow can be used as a passive control option for suitable natural and mechanical applications.
\section*{Acknowledgment}
HB gratefully acknowledges the financial support from SERB, Department of Science and Technology, Government of India through “CRG” project, Award No. CRG/2018/004521. 

\section*{Declaration of interests}

The authors report no conflict of interest.

\bibliographystyle{unsrt}  
\bibliography{Ref}  

%
%
%
%
%

\end{document}